\documentclass[11pt,a4paper,preprintnumbers,nofootinbib]{revtex4}

\pdfoutput=1

\usepackage[T1]{fontenc}
\usepackage[utf8]{inputenc}
\usepackage{amsmath,amssymb,amsfonts,mathtools,bm}
\usepackage{hyperref}
\usepackage[final]{graphicx}
\usepackage{xcolor}
\usepackage{booktabs}
\usepackage{relsize}
\usepackage{lineno}
\usepackage{comment}

\def\ubar{\overline{u}}

\def\qbar{\overline{q}}
\def\lbar{\overline{\ell}}

\def\Babar{{\mbox{\slshape B\kern-0.1em{\smaller A}\kern-0.1em B\kern-0.1em{\smaller A\kern-0.2em R}}}}

\usepackage{cancel}

\usepackage{soul}
\setstcolor{red}

\begin{document}

\title{The New Physics Reach of  Null Tests with  \texorpdfstring{$D \to \pi \ell \ell $}{D to pill} and \texorpdfstring{$D_s \to K \ell \ell $}{Ds to Kll} Decays}

\author{Rigo Bause}
\email{rigo.bause@tu-dortmund.de}
\author{Marcel Golz}
\email{marcel.golz@tu-dortmund.de}
\author{Gudrun Hiller}
\email{ghiller@physik.uni-dortmund.de}
\author{Andrey Tayduganov}
\email{andrey.tayduganov@tu-dortmund.de}
\affiliation{Fakultät für Physik, TU Dortmund, Otto-Hahn-Str.4, D-44221 Dortmund, Germany}
\begin{abstract}
$|\Delta c|=|\Delta u|=1$ processes
are unique probes of flavor physics in the up-sector within and beyond the Standard Model (SM).
SM  tests  with rare semileptonic charm meson decays are based on an approximate CP--symmetry, a superior GIM--mechanism, angular distributions, lepton-universality and lepton flavor conservation.
We analyze the complete set of null test observables
in  $D \to \pi \ell \ell^{(\prime)}$ and $D_s \to K \ell \ell^{(\prime)}$ decays, $\ell^{(\prime)}=e, \mu$,
and find large room for new physics safely  above the SM contribution.
We identify signatures of  supersymmetry, leptoquarks and anomaly-free $U(1)^\prime$--models with generation-dependent charges, for which we provide explicit
examples. $Z^\prime$--effects in $c \to u \ell \ell^{(\prime)}$ transitions can be sizable if both left-handed and right-handed couplings to quarks are present.

\end{abstract}

\preprint{DO-TH 19/11}
\preprint{QFET 2019-11}

\maketitle

\section{Introduction}

New physics  (NP) can be probed in rare semileptonic charm decays 
\cite{Burdman:2001tf,Paul:2011ar,deBoer:2015boa,Fajfer:2015mia,Fajfer:2012nr,deBoer:2018buv,Cappiello:2012vg}. Null test observables, 
already key to ongoing and future  precision programs 
in rare semileptonic $b$-decays  \cite{Kou:2018nap,Cerri:2018ypt}, are indispensable when methods to achieve sufficient theoretical control on decay amplitudes are not available.
This is exactly the case in charm decays, subject to notorious resonance pollution  and poor convergence of the heavy quark expansion \cite{Feldmann:2017izn}.
Null tests are based on approximate symmetry limits, which follow from  parametric  or structural features  of the Standard Model (SM) that  do not
have to be present in general models beyond the SM.
Useful limits include the suppression of CP-violation, 
lepton-nonuniversality and lepton flavor violation (LFV), and  an efficient Glashow-Iliopoulos-Maiani (GIM)
mechanism in $|\Delta c|=|\Delta u|=1$ processes. 
Looking for a breakdown of such symmetries above the residual SM background is a useful strategy when there is a limit only, or even none, 
and  large room for NP. Null test searches with rare charm decays  provide hence a unique opportunity to test the SM  now.

Here we study the  complete catalog of null tests in $D \to \pi \ell \ell^{(\prime)}$ and $D_s \to K \ell \ell^{(\prime)}$ decays, $\ell^{(\prime)}=e, \mu$.
We closely follow   \cite{deBoer:2015boa}
and include experimental updates and recent lattice results for $D\to\pi$ form factors \cite{Lubicz:2017syv, Lubicz:2018rfs}.
While the decays $D \to \pi^+ \pi^- \mu^+ \mu^-$  and $D \to K^+ K^-  \mu^+ \mu^-$ have been observed at a level consistent with the 
SM \cite{Aaij:2017iyr,deBoer:2018buv,Cappiello:2012vg}, for the branching ratios of 
$D \to \pi \ell^+ \ell^- $ and $D_s \to K \ell^+ \ell^- $  only upper limits exist \cite{Tanabashi:2018oca}.
Experimental study is suitable for high luminosity flavor facilities LHCb \cite{Bediaga:2012py}, Belle II \cite{Aushev:2010bq} and BESIII \cite{Asner:2008nq}.
We also discuss signatures in  concrete BSM models, specifically leptoquarks, supersymmetry, and flavorful $Z^\prime$-models. We comment on the complementarity
between charm and $K, B$-flavor probes.

The plan of the paper is as follows:
In Sec.~\ref{sec:SM} we review the predictions in the SM, taking into account improved form factor computations.
In Sec.~\ref{sec:BSM} constraints on Wilson coefficients are obtained, and null test observables  for $D \to \pi \ell \ell^{(\prime)}$ and $D_s \to K \ell \ell^{(\prime)}$ decays, $\ell^{(\prime)}=e, \mu$, are analyzed. Signatures of concrete BSM models in  $c \to u \ell \ell^{(\prime)}$ transitions  are worked out in Sec.~\ref{sec:BSM_models}.
In Sec.~\ref{sec:summary} we conclude.
In App.~\ref{app:FF} we provide details on the $D \to \pi$ form factors.
Constraints from  \texorpdfstring{$D^0-\overline{D}^0$}{D-Dbar}   mixing are discussed in App.~\ref{sec:D_mixing}.
In App.~\ref{sec:anomalies} we give details on and explicit examples of anomaly-free $Z^\prime$--models with generation-dependent charges.

\section{Standard Model predictions \label{sec:SM}}

We discuss exclusive rare charm meson decays in the SM.
In Sec.~\ref{Sec:eff_SM} we review the SM contribution in an effective field theory framework at the charm mass scale. We discuss resonant contributions from intermediate mesons in Sec.~\ref{Sec:res_SM}. In Sec.~\ref{Sec:pheno_SM} we present the $D \to \pi  \ell^+ \ell^-$ and $D_s \to K \ell^+ \ell^-$ differential branching ratios  and discuss SM uncertainties.

\subsection{An effective field theory approach to charm physics \label{Sec:eff_SM}}

Rare $c\to u\ell^+\ell^-$ processes can be described by the effective Hamiltonian,
\begin{equation}
\mathcal{H}_{\rm eff} \supset -\frac{4G_F}{\sqrt2} \frac{\alpha_e}{4\pi} \biggl[ \sum_{i=7,9,10,S,P} \bigl( C_i O_i + C_i^\prime O_i^\prime \bigr) + \sum_{i=T,T5} C_i O_i +\sum_{q=d,s} V_{cq}^*V_{uq} \sum_{i=1}^2 C_i O_i^{q} \biggr]\,,
\label{eq:Heff}
\end{equation}
where the dimension 6 operators which can receive BSM contributions are defined as follows:
\begin{equation}
\begin{split}
O_7 &= {m_c \over e} (\ubar_L \sigma_{\mu\nu} c_R) F^{\mu\nu} \,, \\
O_9 &= (\ubar_L \gamma_\mu c_L) (\lbar \gamma^\mu \ell) \,, \\ 
O_{10} &= (\ubar_L \gamma_\mu c_L) (\lbar \gamma^\mu \gamma_5 \ell) \,, \\
O_S &= (\ubar_L c_R) (\lbar\ell) \,, \\
O_P &= (\ubar_L c_R) (\lbar \gamma_5 \ell) \,, \\
O_T &= \frac{1}{2} (\ubar \sigma_{\mu\nu} c) (\lbar \sigma^{\mu\nu} \ell) \,, \\
O_{T5} &= \frac{1}{2} (\ubar \sigma_{\mu\nu} c) (\lbar \sigma^{\mu\nu} \gamma_5 \ell) \,.
\end{split}
\label{eq:operators}
\end{equation}
The operators $O_i^\prime$ are obtained from the $O_i$ by interchanging left-handed $(L)$ and right-handed $(R)$ chiral fields, $L\leftrightarrow R$. As in $B$-decays, in the SM, $C_{S,P,T,T5}=0$,  and all $C_i^\prime$ are negligible. Unlike in $B$--physics, the GIM--mechanism kills penguin contributions to $C_{7,9,10}$  at the $W$-scale $\mu_W$ in $\mathcal{H}_{\rm eff}$.
Therefore, the SM contributions to $O_{7,9,10}$ are induced by the charged-current operators 
\begin{equation}
O_1^q = (\ubar_L \gamma_\mu T^a\, q_L)(\qbar_L \gamma^\mu T^a\, c_L) \,, \quad 
O_2^q = (\ubar_L \gamma_\mu q_L)(\qbar_L \gamma^\mu c_L) \,, \quad  q=d,\,s \, , \\
\end{equation}
by renormalization group running to the charm scale  $\mu_c$.
 After two-step matching at $\mu_W$ and the $b$-mass scale, see~\cite{deBoer:2015boa,deBoer:thesis} for details,  the effective coefficients $C_{7/9}^{\,\rm eff}$ 
 at $\mu_c$
can be estimated as \cite{deBoer:2015boa}
\begin{equation}
\begin{split}
C_7^{\,\rm eff}(q^2 \approx 0) &\simeq -0.0011-0.0041\,\text{i} \,, \\
C_9^{\,\rm eff} (q^2) &\simeq -0.021 [V_{cd}^*V_{ud}L(q^2,m_d,\mu_c) + V_{cs}^*V_{us}L(q^2,m_s,\mu_c)] \,,
\end{split}
\label{eq:CSM}
\end{equation}
where 
\begin{equation}
\begin{split}
L(q^2,m_q,\mu_c) &= {5\over3} - \ln{m_q^2 \over \mu_c^2} + x - (2+x) \sqrt{x-1}
\begin{cases}
2\,{\rm arctan}{1\over \sqrt{x-1}} \,, \quad~~ x>1 \\
\ln {1 + \sqrt{1-x} \over \sqrt{x}} - {\text i\pi \over 2} \,, \quad x\leq1
\end{cases}\,, \\
L(q^2,0,\mu_c) &= {5\over3} - \ln{q^2 \over \mu_c^2} + \text{i}\,\pi\,,
\end{split}
\label{eq:hq}
\end{equation}
with $x = 4m_q^2/q^2$ and the dilepton invariant mass squared $q^2$. 
Taking $\mu_c=m_c=1.275$~GeV one obtains $|C_9^{\,\rm eff}|\lesssim 0.01$. 
$\text{Im}[C_7^{\,\rm eff}]$ increases from -0.004 at $q^2=0$  to -0.001 at high $q^2$ at NNLL order~\cite{deBoer:thesis}. 
Importantly, $C_{10}^{\rm SM}=0$, which implies that at the charm scale effects from the V-A structure of the weak interaction are shut off.
Numerically, the short-distance  SM contributions are negligible in the $D \to P \ell^+ \ell^-$ decay rates  compared to the resonance-induced effects, discussed in Sec.~\ref{Sec:res_SM}.

\subsection{Resonance contributions \label{Sec:res_SM}}

The dominant process inducing the $O_{9,P}$ operators is $D\to P\gamma^*$ with $\gamma^*\to M \to\ell^+\ell^-$, which can be parametrized by a phenomenological shape, 
\begin{equation}
\begin{split}
C_9^R (q^2) &= a_\rho e^{\text{i}\,\delta_\rho} \biggl( {1 \over q^2 - m_\rho^2 + \text{i}\,m_\rho\Gamma_\rho} - {1\over3} {1 \over q^2 - m_\omega^2 + \text{i}\,m_\omega\Gamma_\omega} \biggr) + {a_\phi e^{\text{i}\,\delta_\phi} \over q^2 - m_\phi^2 + \text{i}\,m_\phi\Gamma_\phi} \,, \\
C_P^R (q^2) &= {a_\eta e^{\text{i}\,\delta_\eta} \over q^2 - m_\eta^2 + \text{i}\,m_\phi\Gamma_\eta} +{a_{\eta^\prime} \over q^2 - m_{\eta^\prime}^2 + \text{i}\,m_{\eta^\prime}\Gamma_{\eta^\prime}} \,,
\end{split}
\label{eq:C9_CP_res}
\end{equation}
with resonance parameters $a_M$, $M=\rho,\phi,\eta,\eta^\prime$ and $P=\pi,K$.
\begin{table}[!t]\centering
\caption{Phenomenological resonance parameters (in GeV$^2$) extracted from the experimental measurements of $\mathcal{B}(D\to\pi M)$ and $\mathcal{B}(D_s\to KM)$ with resonances $M=\rho,\phi,\eta,\eta^\prime$ decaying to $\mu^+\mu^-$.}
\label{tab:a_res}
\begin{tabular}{c||cc|cc|cc|cc}
&& $a_\rho$ && $a_\phi$ && $a_\eta$ && $a_{\eta^\prime}$ \\
\hline
$D^+\to\pi^+$  && $0.18\pm0.02$ && $0.23\pm0.01$ && $(5.7\pm0.4)\times10^{-4}$ && $\sim8\times10^{-4}$ \\
\hline
$D^0\to \pi^0$ && $0.86\pm 0.04$ && $0.25\pm 0.01$ && $(5.3\pm0.4)\times10^{-4}$ && $\sim8\times10^{-4}$ \\
\hline
$D_s\to K$ && $0.48\pm 0.04$ && $0.07\pm 0.01$ && $(5.9 \pm 0.7)\times10^{-4}$ && $\sim7\times10^{-4}$ \\
\end{tabular}
\end{table}
Here, $m_M$ and $\Gamma_M$ denote the mass and the total decay rate, respectively, of the resonance $M$. The moduli of the $a_M$ parameters can be extracted from measurements of  branching ratios $\mathcal{B}(D\to P M)$ and 
$\mathcal{B}(M \to \ell^+ \ell^-)$, and are given in Tab.~\ref{tab:a_res}. To reduce the number of input parameters and corresponding theoretical uncertainties we employ the isospin relation $a_\omega=a_\rho/3$ in (\ref{eq:C9_CP_res}).
This is justified by the measurement  of $\mathcal{B}(D^+\to\pi^+\omega)$, which returns $a_\omega^{D\pi^+ }\simeq0.03 \, \mbox{GeV}^{2}$ and $a_\omega^{D\pi^+}/a_\rho^{D\pi^+} \simeq  0.2$, somewhat smaller than the isospin limit;  for $D^0 \to \pi^0 \omega$, 
$a_\omega^{D\pi^0}\simeq0.04 \, \mbox{GeV}^{2}$ and $a_\omega^{D\pi^0}/a_\rho^{D\pi^0} \simeq  0.05$.
From the presently only available upper limit on $\mathcal{B}(D_s\to K\omega)$ one obtains $a_\omega^{D_s K}\lesssim 0.13 \, \mbox{GeV}^{2}$, yielding $a_\omega^{D_s K}/a_\rho^{D_sK}\lesssim 0.27 $, consistent with the isospin limit.
The strong phases $\delta_{\rho,\,\phi,\,\eta}$ remain undetermined by this procedure and are varied within $-\pi$ and $\pi$ in the numerical analyses.

\begin{figure}[!t]\centering
\includegraphics[width=0.49\textwidth]{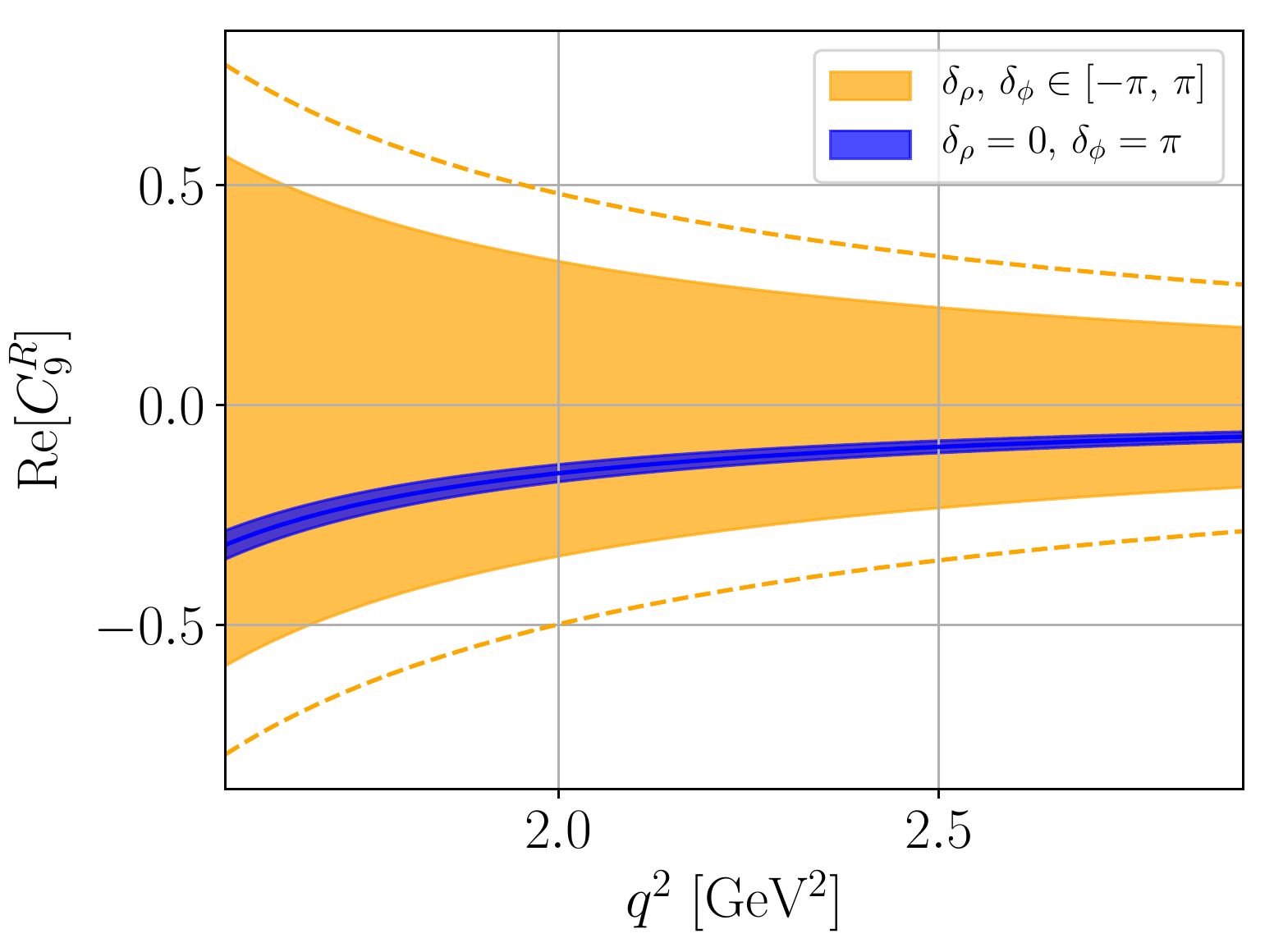}
\includegraphics[width=0.49\textwidth]{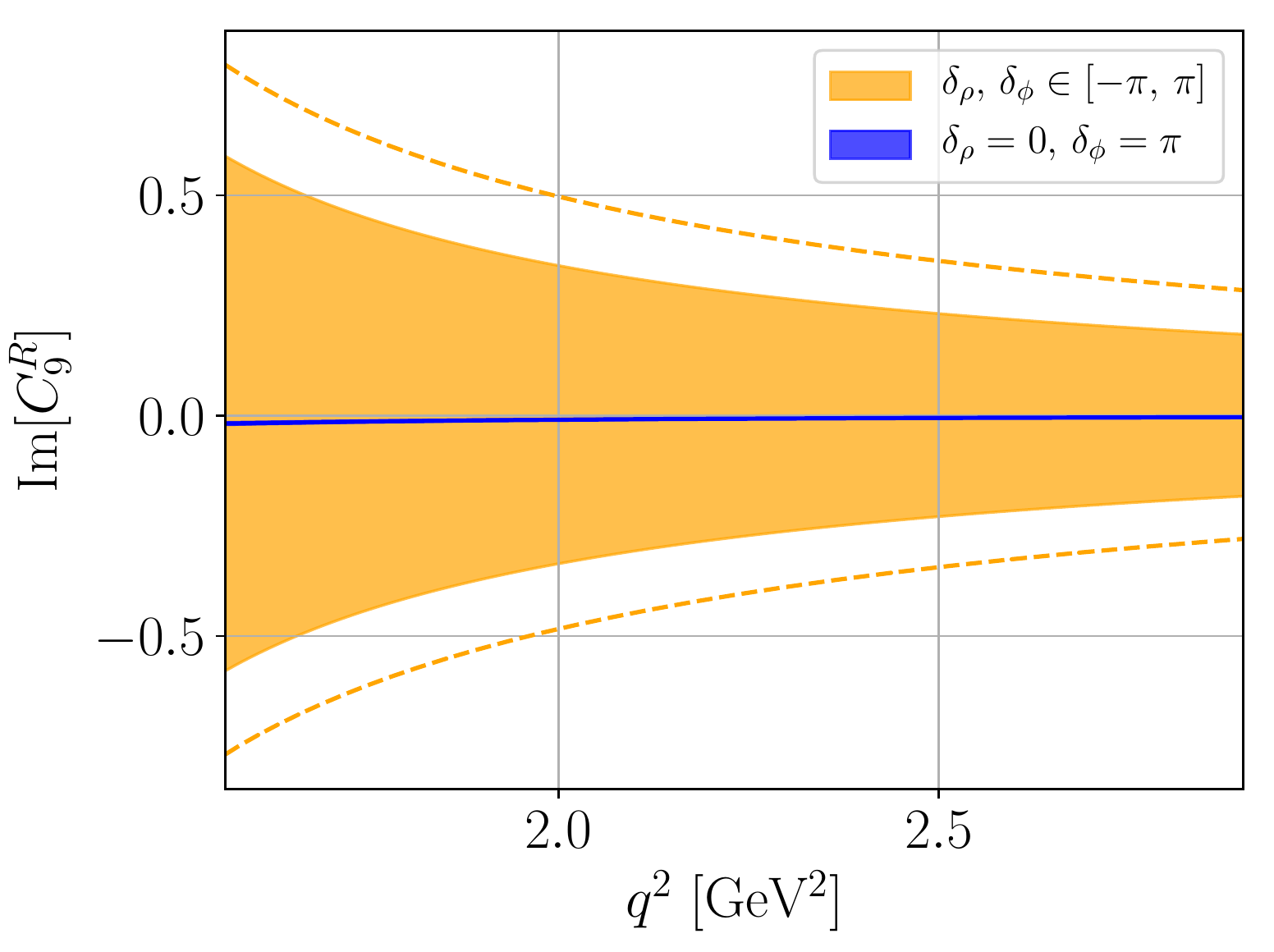}
\caption{Resonance contributions to the real part (plot to the left) and the imaginary part (plot to the right)  of $C_9^R$ in the high $q^2$--region using the $D^+\to\pi^+$ parameters from Tab.~\ref{tab:a_res}.
The wider yellow bands correspond to the uncertainties from strong phases $\delta_\rho, \delta_\phi$ varied within $[-\pi, +\pi]$, whereas the smaller blue bands correspond to fixed strong phases $\delta_\rho=0, \delta_\phi=\pi$.
Additional uncertainties arise from $a_\rho, a_\phi$, and are included in the plots. The dashed curves illustrate the uncertainties in the case of $D^0\to\pi^0$ parameters.}
\label{fig:C9R}
\end{figure}

Fig.~\ref{fig:C9R} illustrates the impact of the uncertainties from the unknown phases on real and imaginary parts of $C_9^R$ in the high $q^2$--region ($\sqrt{q^2}\geq1.25$~GeV) for $D \to \pi \mu^+ \mu^-$ decays.
The  phases  $\delta_{\rho,\,\phi}$ give rise to huge uncertainties, shown by the yellow wider bands. As a result, even the sign of $C_9^R$ cannot be predicted.
Once the phases are fixed, the residual uncertainties from $a_\rho, a_\phi$, shown by the blue bands, are small. In addition, they could be reduced by improved measurements of $D \to P M$ and $M \to \ell^+ \ell^-$ branching ratios. Darker shaded solid  lines correspond to central values of input with 
strong phases fixed to $\delta_\rho=0$, $\delta_\phi=\pi$, consistent with the $SU(3)_F$ limit $\delta_\rho - \delta_\phi=\pi$.

\subsection{Phenomenological \texorpdfstring{$q^2$}{q2}--distributions \label{Sec:pheno_SM}}

The differential decay distribution of $D \to P \ell^+\ell^-$ can be written as \cite{Bobeth:2007dw}
\begin{equation}
\begin{split}
{\text{d}\Gamma \over \text{d}q^2} &= {G_F^2 \alpha_e^2 \over 1024\pi^5 m_D^3} \sqrt{\lambda_{DP} \biggl( 1 - {4m_\ell^2 \over q^2} \biggr)}\, \biggl\{\biggr. \\
& \quad {2\over3} \biggl| C_9 + C_9^R + C_7 {2m_c \over m_D + m_P} {f_T \over f_+} \biggr|^2 \left( 1 + {2m_\ell^2 \over q^2} \right) \lambda_{DP} f_+^2 \\
& \quad + |C_{10}|^2 \left[ {2\over3} \left( 1 - {4m_\ell^2 \over q^2} \right) \lambda_{DP} f_+^2 + {4m_\ell^2 \over q^2} (m_D^2-m_P^2)^2 f_0^2 \right]\\
& \quad + \left[ |C_S|^2 \left (1 - {4m_\ell^2 \over q^2} \right) + |C_P + C_P^R|^2 \right] {q^2 \over m_c^2} (m_D^2-m_P^2)^2 f_0^2 \\
& \quad+ {4\over 3} \left[ |C_T|^2 + |C_{T5}|^2 \right] \left( 1 - {4m_\ell^2 \over q^2} \right) {q^2 \over (m_D + m_P)^2} \lambda_{DP} f_T^2 \\
& \quad + 8 \,\text{Re}\left[ \left( C_9 + C_9^R + C_7 {2m_c \over m_D + m_P} {f_T \over f_+} \right)C_T^*\right]  {m_\ell \over m_D + m_P} \lambda_{DP} f_+ f_T\\
& \quad + 4\,\text{Re}\left[ C_{10} \left( C_P + C_P^R \right)^* \right] {m_\ell \over m_c} (m_D^2-m_P^2)^2 f_0^2 +16\,|C_T|^2 \frac{m_\ell^2}{(m_D+m_P)^2}\lambda_{DP}f_T^2 \biggr.\biggr\}
\end{split}
\label{eq:dGamma}
\end{equation}
with $\lambda_{DP}=m_D^4+m_P^4+q^4-2m_D^2\,m_P^2-2m_D^2\,q^2-2m_P^2q^2\,$. To ease notation, here and in the following, all Wilson coefficients $C_i$, with the exception of the tensor ones, are understood as
\begin{equation}
C_i \to C_i+C_i^\prime \, . 
\end{equation}
We neglect the up-quark mass. 
The  corresponding $D_s \to K \ell^+\ell^-$ decay distribution can be obtained by replacing $m_D\to m_{D_s}$, using 
$D_s\to K$ form factors $f_{+,0,T}$ and resonance parameters $a_{\rho,\,\phi,\,\eta,\,\eta^\prime}$ from Tab.~\ref{tab:a_res}. 
A detailed discussion of the form factors can be found in App.~\ref{app:FF}.

 In this work we give predictions  for $D^+ \to \pi^+ \ell \ell^{(\prime)}$ observables.
 Predictions for $D^0 \to \pi^0 \ell \ell^{(\prime)}$ decays
 are, within  uncertainties, the same as for $D^+ \to \pi^+ \ell \ell^{(\prime)}$ decays except for branching ratios, which need to be rescaled by the ratio of lifetimes,
 $\tau(D^+)/\tau(D^0) =2.54$ \cite{Tanabashi:2018oca}.
 
\begin{figure}[!t]\centering
\includegraphics[width=0.49\textwidth]{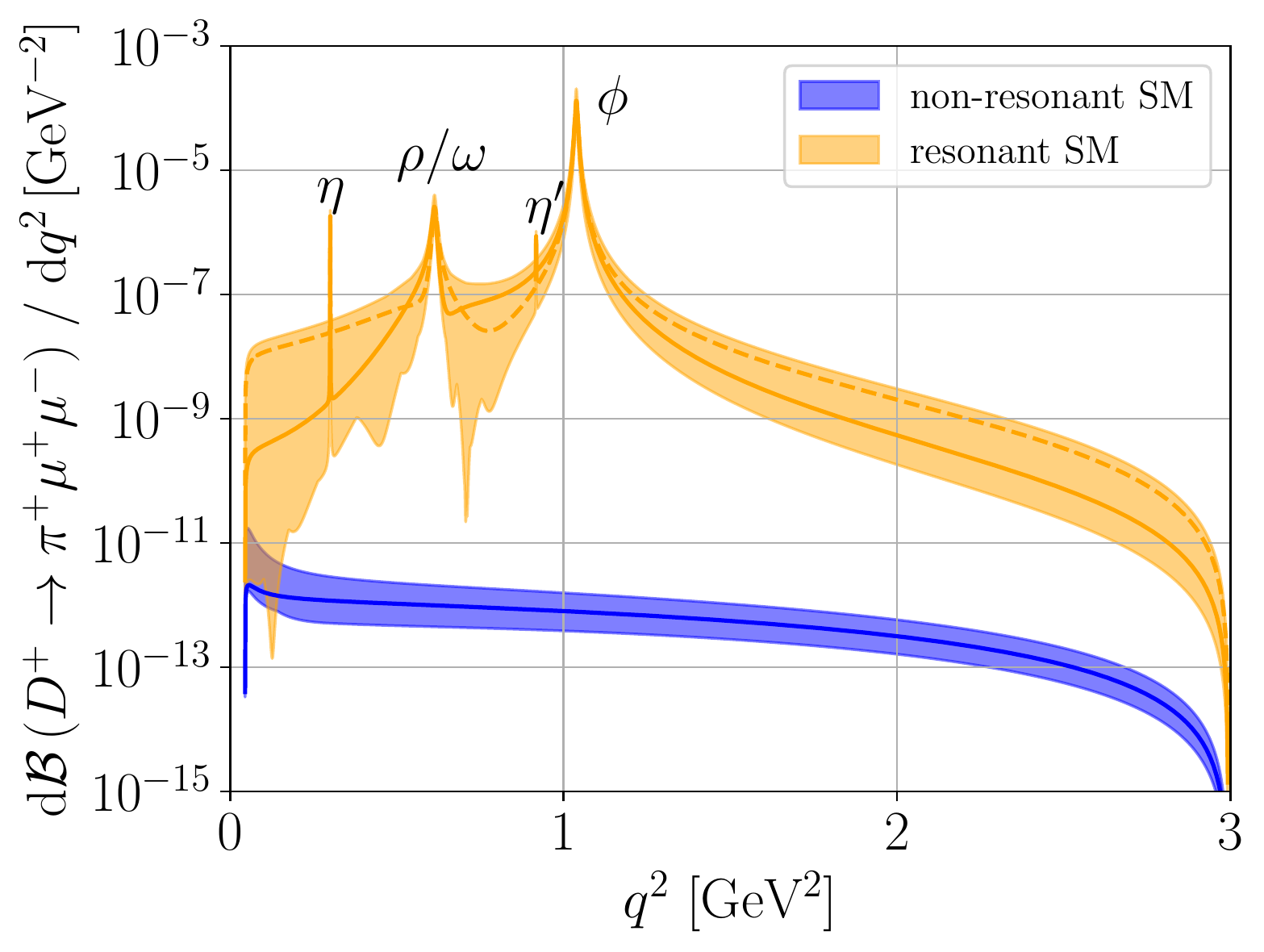}
\includegraphics[width=0.49\textwidth]{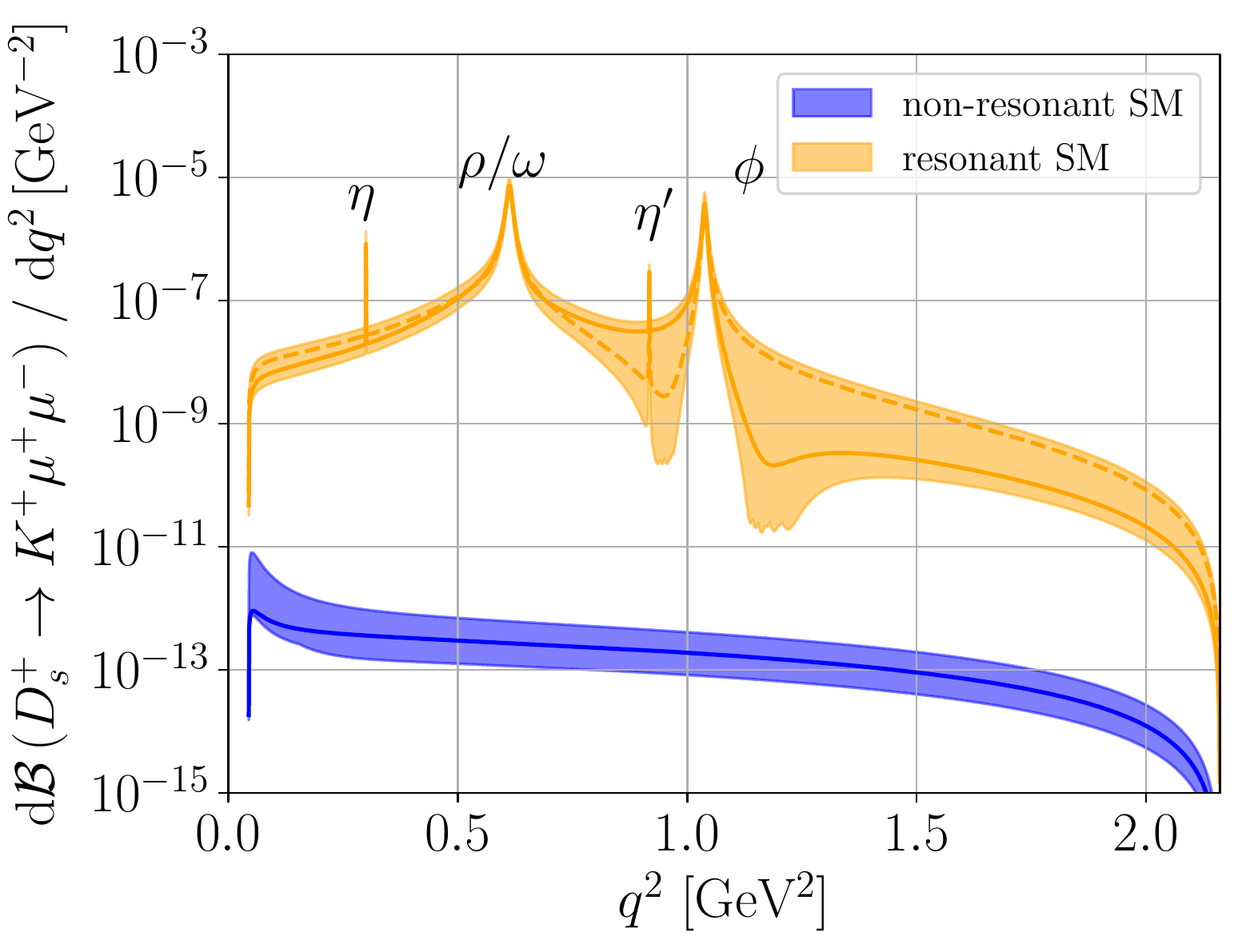}
\caption{The differential SM branching fractions of $D^+\to\pi^+\mu^+\mu^-$ (plot to the left) and $D_s^+\to K^+\mu^+\mu^-$ (plot to the right) decays. Yellow (blue) bands show pure resonant (short-distance) contributions. The band widths represent theoretical uncertainties of hadronic form factors, resonance parameters and $\mu_c$. 
Darker shaded thin curves represent all parameters set to their central values including $\delta_\rho=0$ and $\delta_\phi=\pi$ (solid) 
and $\delta_\rho=\delta_\phi=0$ (dashed). For the experimental limit by LHCb \cite{Aaij:2013sua} see Eq.~\eqref{eq:explimit}.}
\label{fig:dGamma_SM}
\end{figure}
 
In Fig.~\ref{fig:dGamma_SM} we show the pure non-resonant $q^2$--spectra determined by the SM short-distance $C_{7,9}^{\rm eff}$ coefficients (lower, blue curves) and the pure resonant ones determined by $C_{9,P}^R$ (yellow, wiggly curves). The non-resonant contribution is suppressed by several orders of magnitude with respect to the resonant one  and negligible in the SM. The band width represents theoretical uncertainties of the  form factors, the resonance parameters $a_M$ with corresponding phases $\delta_M$, see Eq.~\eqref{eq:C9_CP_res}, and the scale $\mu_c$. The largest uncertainty stems from the  unknown phases $\delta_M$, an effect which is pronounced through resonance interference.

The uncertainties from the strong phases $\delta_M$ are highly dependent on the values of the $a_M$. One can infer from Fig.~\ref{fig:dGamma_SM} that the $D$--mode has huge uncertainties at low $q^2$ and around the $\rho/\omega$ peak, and it has much smaller uncertainties and more stable behavior above the $\phi$ mass, while for the $D_s$--mode the situation is the opposite. The reason for this difference is the numerical values and hierarchy of $a_\rho$ and $a_\phi$ (see Tab.~\ref{tab:a_res}). Although the tiny uncertainty band in the low $q^2$--region of the $D_s$--mode looks promising for testing new physics, we stress that the uncertainty band width can change with future measurements of the $D_s\to KM$ and $M\to \ell^+ \ell^-$ branching ratios.

\section{Beyond the Standard Model \label{sec:BSM}}

BSM effects in the $c\to u \ell\ell^{(\prime)}$ transition are discussed in a model-independent way. In Sec.~\ref{sec:BSM_constraints} we update the constraints on the Wilson coefficients coming from $D^+\to \pi^+ \mu^+\mu^-$ and $D^0\to\mu^+\mu^-$ branching ratios and discuss the BSM sensitivity of $\mathcal{B}(D \to P \ell^+\ell^-$) at high $q^2$. In Sec.~\ref{sec:BSM_lfu}  lepton universality tests are considered. Null tests of the SM based on the angular distributions are discussed in Sec.~\ref{sec:BSM_nulltests}. CP--asymmetries and their sensitivity to BSM benchmark values is analyzed in Sec~\ref{sec:BSM_cpasym}. In Sec.~\ref{sec:BSM_lfv} lepton flavor violating $c \to u e^\pm \mu^\mp $ decays are discussed.

\subsection{Updated constraints on Wilson coefficients \label{sec:BSM_constraints}}

Using the experimental limits on the branching fraction of $D^+\to\pi^+\mu^+\mu^-$ in high and full $q^2$--regions at 90\%~CL~\cite{Aaij:2013sua}\footnote{We use the full $q^2$--region as given in Eq.~\eqref{eq:explimit} in order to present the strongest upper limit on the Wilson coefficients in Eq.~\eqref{eq:NP_constraints_full}. Since an actual measurement is only performed in $\sqrt{q^2}\in[250,525]~{\rm MeV}~{\rm and}~\sqrt{q^2}\geq1.25~{\rm GeV}$ and extrapolated in~\cite{Aaij:2013sua} in between, we prefer to use the bounds from the high $q^2$--region given in Eq.~\eqref{eq:NP_constraints_high}. },
\begin{equation}
\label{eq:explimit}
\begin{split}
\mathcal{B}(D^+\to\pi^+\mu^+\mu^-)\vert_{{\rm full}~q^2} &< 7.3 \times 10^{-8} \quad (250\,{\rm MeV}\leq \sqrt{q^2}\leq m_{D^+}-m_{\pi^+})\,,\\
\mathcal{B}(D^+\to\pi^+\mu^+\mu^-)\vert_{{\rm high}~q^2} &< 2.6 \times 10^{-8} \quad (\sqrt{q^2}\geq1.25~{\rm GeV}) \,, 
\end{split}
\end{equation}
and neglecting the SM contributions, we obtain the following constraints on the BSM Wilson coefficients in the full $q^2$--region,
\begin{equation}
\begin{split}
& 1.2|C_7|^2 + 1.2|C_9|^2 + 1.2|C_{10}|^2 + 2.4|C_S|^2 + 2.5|C_P|^2 + 0.4|C_T|^2 + 0.3|C_{T5}|^2 \\
& + 0.3\,\mathrm{Re}[C_9 C_T^*] + 1.0\,\mathrm{Re}[C_{10} C_P^*] + 2.4\,\mathrm{Re}[C_7 C_9^*] + 0.6\,\mathrm{Re}[C_7 C_T^*] \lesssim1 \,.
\end{split}
\label{eq:NP_constraints_full}
\end{equation}
and in the high $q^2$--region,
\begin{equation}
\begin{split}
& 0.6|C_7|^2 + 0.7|C_9|^2 + 0.8|C_{10}|^2 + 4.4|C_S|^2 + 4.5|C_P|^2 + 0.4|C_T|^2 + 0.4|C_{T5}|^2 \\
& + 0.3\,\mathrm{Re}[C_9 C_T^*] + 1.1\,\mathrm{Re}[C_{10} C_P^*] + 1.4\,\mathrm{Re}[C_7 C_9^*] + 0.3\,\mathrm{Re}[C_7 C_T^*]\lesssim1 \,,
\end{split}
\label{eq:NP_constraints_high}
\end{equation}

We find good agreement between Eq.~\eqref{eq:NP_constraints_high}, in the limit when only one $C_i$ is present, and Ref.~\cite{Fajfer:2015mia}. 
The limits \eqref{eq:NP_constraints_full}, \eqref{eq:NP_constraints_high} are also consistent with the fact that in all $D \to P \ell^+ \ell^-$ distributions the leptonic vector current is probed only through the combination
\begin{equation} \label{eq:U}
C_9 + C_9^R + C_9^\prime +  \gamma (C_7+C_7^\prime) \, ,
\end{equation}
with $\gamma={2m_c \over m_D + m_P} {f_T \over f_+} $, which numerically is around one in the full $q^2$--region, $\gamma \approx 1$, as shown in App.~\ref{app:FF}. 
Present bounds on dipole operators from $D \to \rho  \gamma$~\cite{Abdesselam:2016yvr,deBoer:2017que} are in agreement with \eqref{eq:NP_constraints_full}, \eqref{eq:NP_constraints_high}.

The numerical coefficients in Eqs.~\eqref{eq:NP_constraints_full} and \eqref{eq:NP_constraints_high} are smaller than the corresponding ones in Eqs.~(29) and (30) of Ref.~\cite{deBoer:2015boa}. This difference is caused by the $D\to\pi$ form factors, for which we use new lattice results~\cite{Lubicz:2017syv, Lubicz:2018rfs} (for details, see App.~\ref{app:FF}). This in particular affects $f_T$ and hence the available space for NP in $C_{T,T5}$ and $C_7^{(\prime)}$. 

Note that in Eqs.~\eqref{eq:NP_constraints_full} and \eqref{eq:NP_constraints_high} we neglected the contributions from $C_{9,\,P}^R$. 
Taking them into account, we obtain one additional constant term plus several interference terms proportional to Re$[C_i]$ and Im$[C_i]$ on the left-hand side of Eqs.~\eqref{eq:NP_constraints_full} and \eqref{eq:NP_constraints_high}. 
When varying strong phases, the constant term turns out to be smaller than 0.1. 
The largest numerical coefficients of the interference terms can vary in the full $q^2$--region within [-0.1, 0.1] and [-0.2, 0.2] for $C_9$ and $C_7$, respectively, whereas at high $q^2$ the corresponding ranges turn out to be larger, about a factor $\sim 2$ to 5.
Despite being less tight we use in the following the constraint from  the $D^+ \to \pi^+ \mu^+ \mu^-$ branching fraction in the high $q^2$--region as it is the best available bound we trust, since no extrapolations of signal have been included, in contrast to the full $q^2$--region.
One may wonder whether Wilson coefficients as large as order one of $|\Delta c|= | \Delta u|=1$ four-fermion operators are constrained from high-$p_T$ searches.
While a dedicated analysis for up-sector flavor changing neutral currents (FCNCs) is currently not available, constraints derived from $36.1 \, \rm{fb}^{-1}$ ATLAS data on semileptonic operators with two charm singlets or two second generation quark doublets \cite{Greljo:2017vvb} are close to probing the range allowed by charm decays into muons, and are somewhat stronger for electrons.

Further constraints on $C_{10,S,P}^{(\prime)}$ can be obtained from the $D^0\to\ell^+\ell^-$ branching ratio
\begin{equation}
\begin{split}
\mathcal{B}(D^0\to\ell^+\ell^-) &=\tau_{D} {G_F^2 \alpha_e^2 m_D^5 f_D^2 \over 64\pi^3 m_c^2} \sqrt{1 - {4m_\ell^2 \over m_D^2}} \left[ \left( 1 - {4m_\ell^2 \over m_D^2} \right) |C_S - C_S^\prime|^2 \right.\\
& \quad \left.+\left|C_P - C_P^\prime + {2m_\ell m_c \over m_D^2} (C_{10} - C_{10}^\prime) \right|^2 \right] \,.
\end{split}
\label{eq:Br_D-ll}
\end{equation}
The upper limit $\mathcal{B}(D^0\to\mu^+\mu^-)<6.2\times10^{-9}$ at 90\% CL \cite{Aaij:2013cza} gives
\begin{equation}
|C_S-C_S^\prime|^2 + |C_P - C_P^\prime + 0.1(C_{10}-C_{10}^\prime)|^2 \lesssim 0.007 \,.
\end{equation}
In addition, tensor Wilson coefficients are constrained by the leptonic decays  as they induce scalar ones from renormalization group running \cite{Gonzalez-Alonso:2017iyc}.
In the subsequent analysis we use $|C_{S,P}|\simeq0.1$ and $|C_i|\simeq0.5$ for all other NP Wilson coefficients as benchmark values. An exception is 
Sec.~\ref{sec:BSM_cpasym} on CP-violating effects, which are subject to model-dependent, generically stronger $D$-mixing constraints, see App.~\ref{sec:D_mixing} for details.
In the following $C_{i (j)}$ stands for "$C_i$ or $C_j$".

\begin{figure}[!t]\centering
\includegraphics[width=0.49\textwidth]{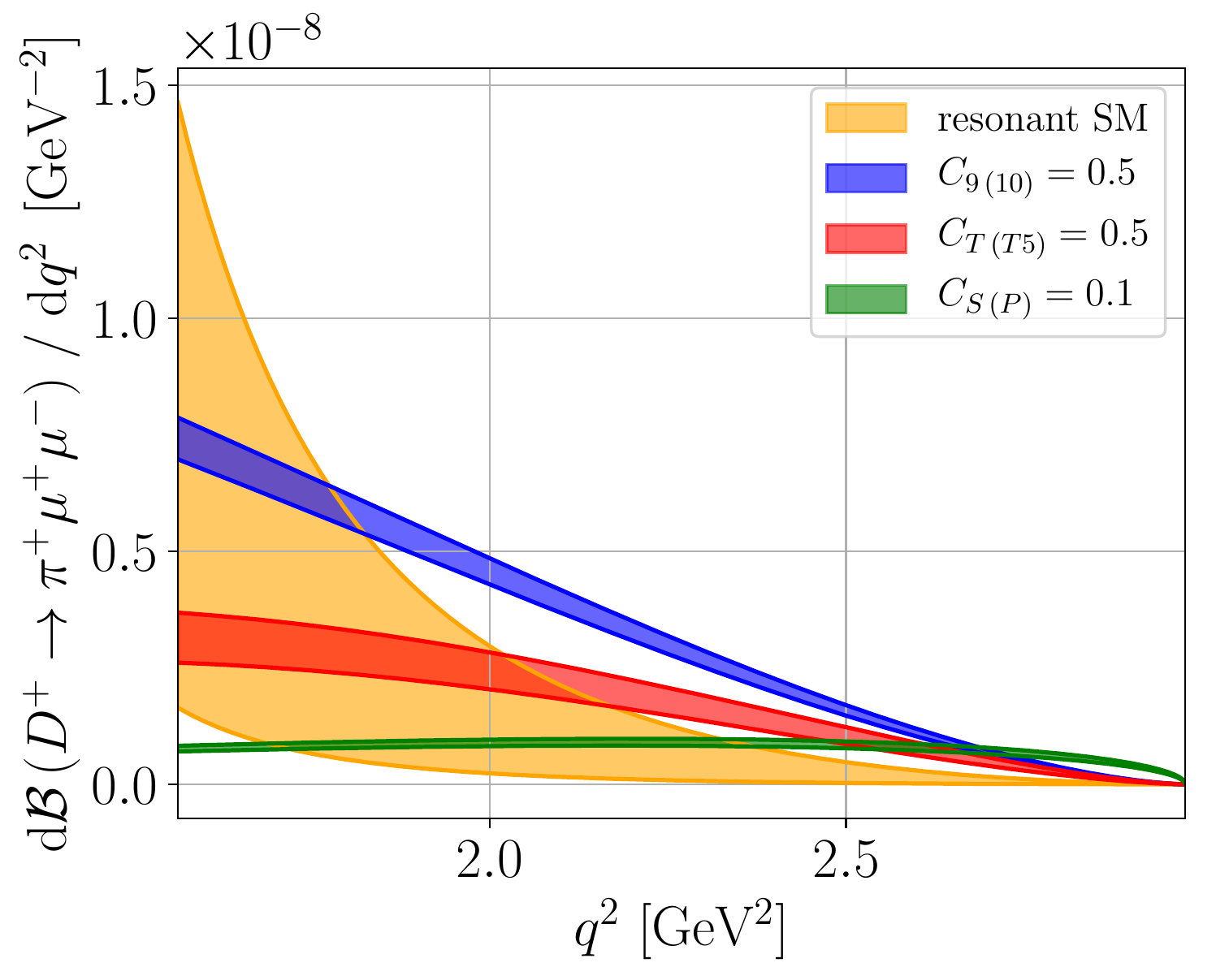}
\includegraphics[width=0.48\textwidth]{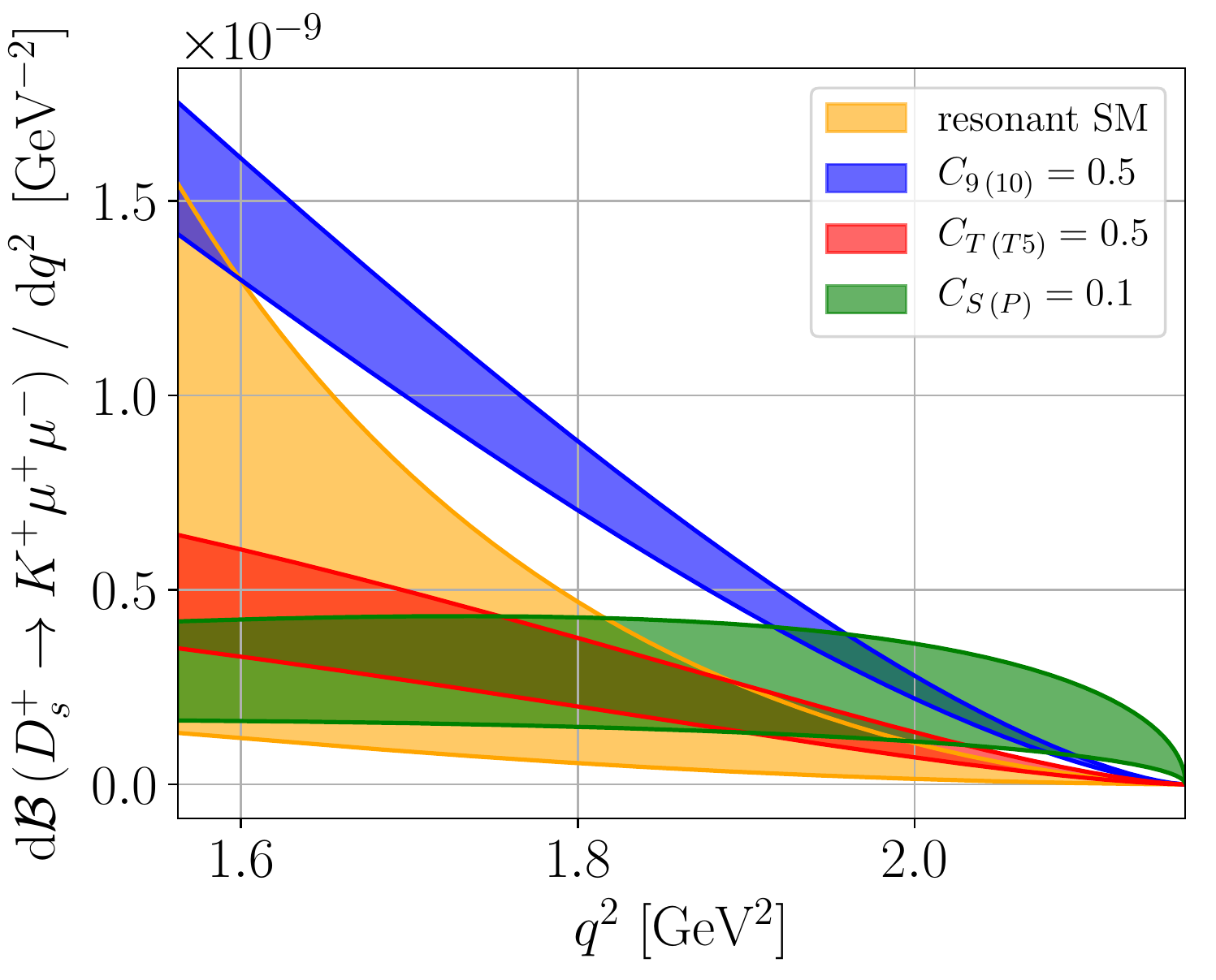}
\includegraphics[width=0.49\textwidth]{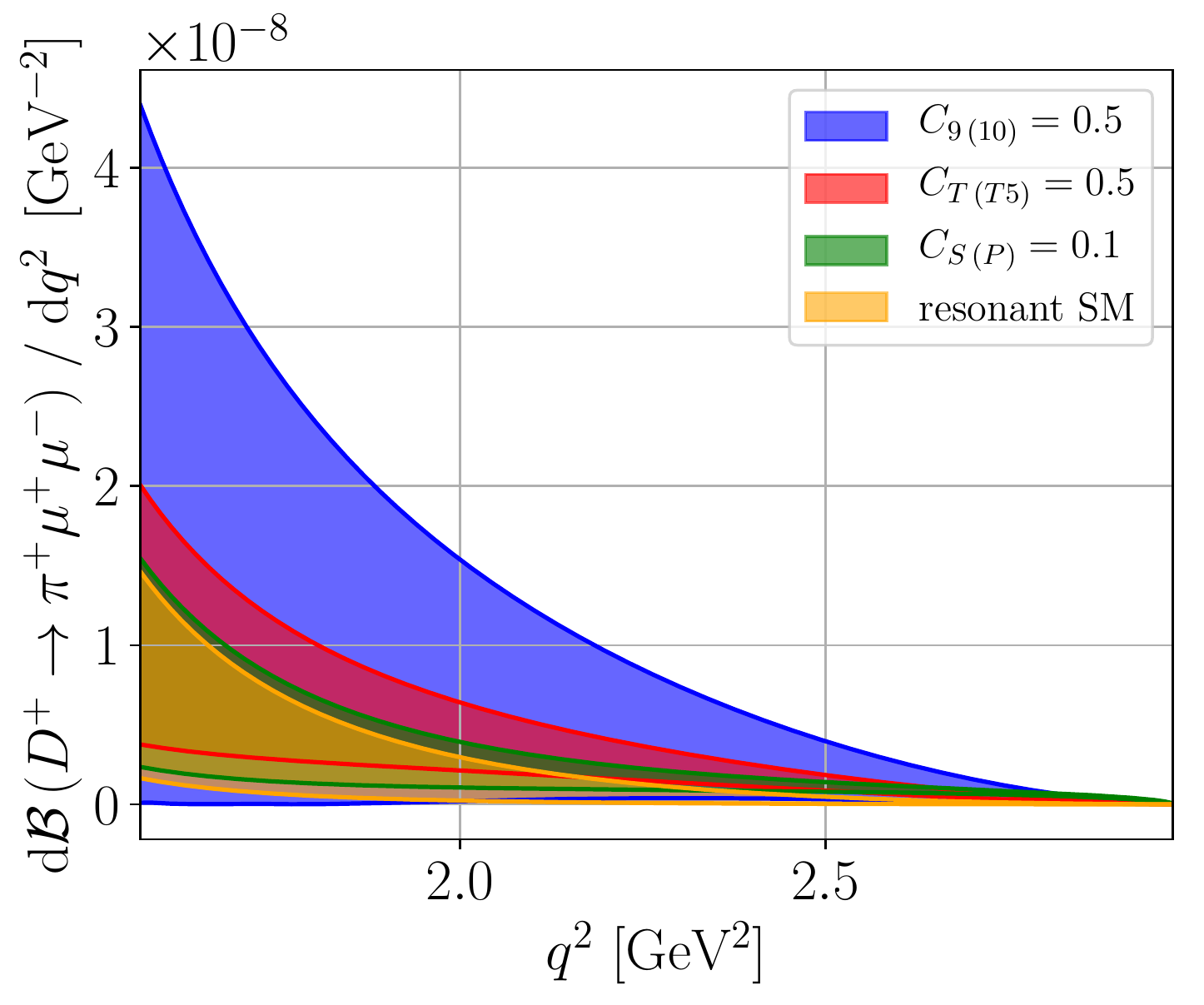}
\includegraphics[width=0.48\textwidth]{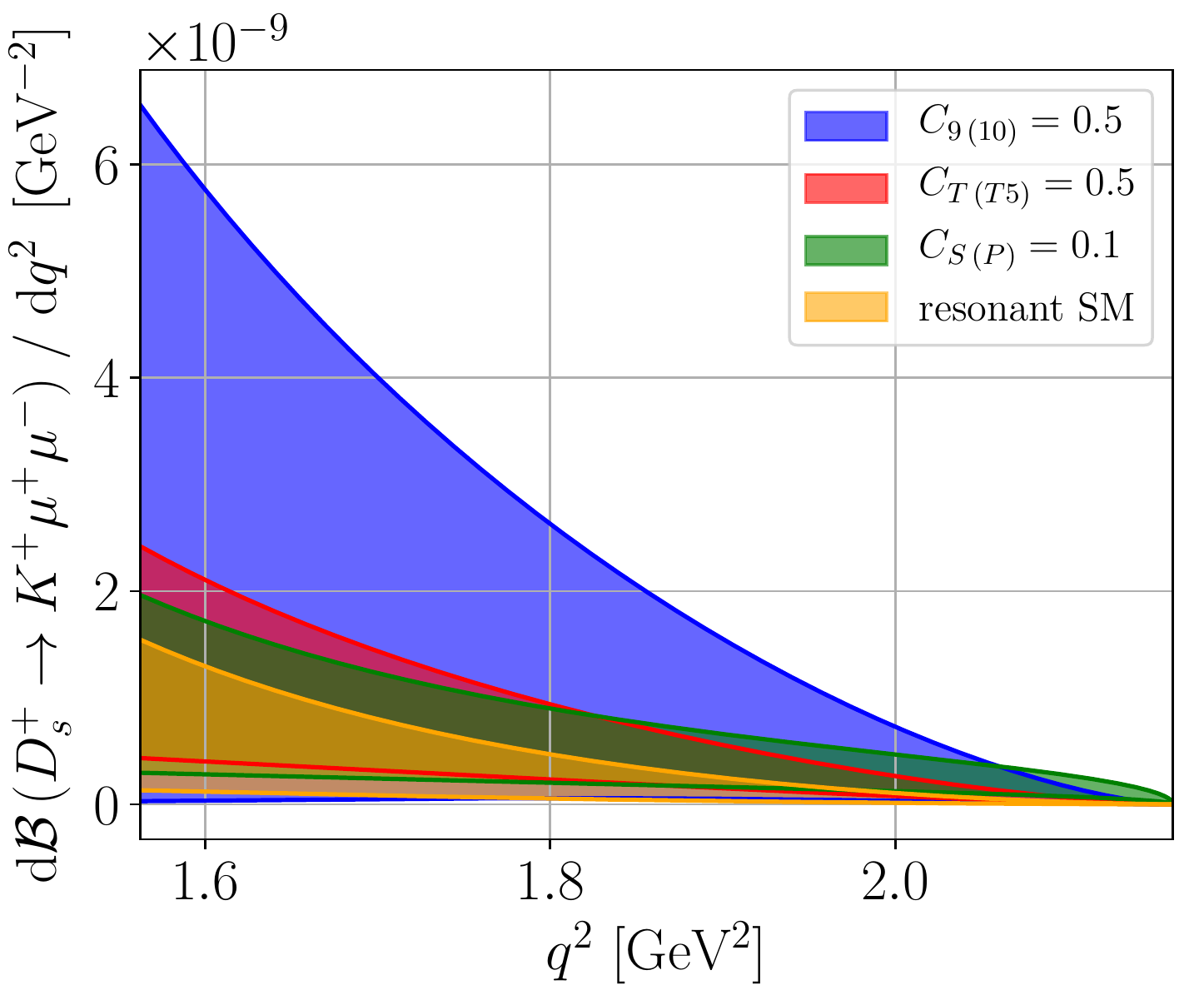}
\caption{The differential branching fractions in the SM and in three BSM scenarios in the high $q^2$--region for $D^+\to\pi^+\mu^+\mu^-$ (plots to the left) and $D_s^+\to K^+\mu^+\mu^-$ (plots to the right). 
Orange bands correspond to  the  pure resonant contributions. 
The band width represents theoretical uncertainties, where we distinguish between pure BSM contributions (upper plots) and BSM benchmarks including interference with the SM resonance contributions, such that the variation of the strong phases also effects the BSM predictions (lower plots). 
Predictions for BSM coefficients $C_7^{(\prime)}$ are similar to those of $C_9^{(\prime)}$ (cf. Eq.~\eqref{eq:U}), and not shown.}
\label{fig:Br_high}
\end{figure}

In Fig.~\ref{fig:Br_high} and Tab.~\ref{tab:integrated_BR} we illustrate the NP  sensitivity of the $D_{(s)}^+\to\pi(K)^+\mu^+\mu^-$ differential and integrated branching fractions, respectively,  at high $q^2$.
Results for the intermediate resonance region and at  low $q^2$ are not given due to sizable theoretical uncertainties, see, for instance, Fig.~\ref{fig:dGamma_SM}.
The largest deviations from the SM in the branching ratios are possible with (axial)-vector operators $O_{9,10}^{(\prime)}$, followed by tensors $O_{T,T5}$ and  (pseudo)-scalar operators $O_{S,P}^{(\prime)}$.
An exception to this arises close to the endpoint, where  $O_{S,P}^{(\prime)}$ effects can become the largest as they do have one power
of the K\"allen function $\lambda_{DP}$ less in Eq.~\eqref{eq:dGamma}.
While there can be NP distributions in excess of the SM one shown by the yellow band,  the window for NP is small. In view 
of the sizable hadronic uncertainties NP branching ratios
need to be close to  the present experimental upper limit to support a BSM interpretation without further input or data.
Taking the resonances into account enhances the BSM branching ratios and makes them more uncertain, 
see the lower plots in Fig.~\ref{fig:Br_high} (resonances plus NP), in comparison to the 
upper plots  of Fig.~\ref{fig:Br_high} (pure NP).
The sensitivity to BSM dipole operators $O_7^{(\prime)}$ is similar to the one of $O_9^{(\prime)}$, see Eq.~\eqref{eq:U}, and therefore not shown.

\begin{table}[!t]
\centering
\caption{Integrated branching fractions in the  high $q^2$--bin ($\sqrt{q^2}\geq1.25$~GeV) in the SM and in the NP benchmark scenarios as in Fig.~\ref{fig:Br_high}. 
In the third to sixth column, upper entries  correspond to NP-only branching ratios while for the lower entries the resonance contributions are taken into account.
}\label{tab:integrated_BR}
\begin{tabular}{cc||ccc|ccc|ccc|ccc|ccc}
$\mathcal{B}\vert_{\text{high}\, q^2}\,\times10^{9}$&&&SM&&&$C_{9\,(10)}=0.5$&&&$C_{S\,(P)}=0.1$&&& $C_{T\,(T5)}=0.5$&&& $C_9=\pm C_{10}=0.5$ & \\
\hline
$D^+\to\pi^+\mu^+\mu^-$&&&$0.3\ldots3.0$&&&$4.5\pm0.3$&&&$1.1\pm0.1$&&&$2.6\pm0.4$&&&$9.4\pm0.6$& \\
&&& &&&$7.8\pm7.4$&&&$2.9\pm1.4$&&&$5.0\pm2.9$&&&$12.6\pm7.7$\\
\hline
$D_s^+\to K^+\mu^+\mu^-$&&&$0.03\ldots0.3$&&&$0.40\pm0.05$&&&$0.15\pm0.07$&&&$0.15\pm0.05$&&&$0.8\pm0.1$&\\
&&& &&&$0.8\pm0.7$&&&$0.3\pm0.2$&&&$0.4\pm0.3$&&& $1.2\pm0.8$\\
\end{tabular}
\end{table}

\subsection{Testing lepton universality \label{sec:BSM_lfu}}

Lepton universality can be probed in semileptonic decays with the ratios \cite{Fajfer:2015mia,deBoer:2018buv}
\begin{equation} \label{eq:LNU}
R_P^D = {\displaystyle{\int_{q_{\rm min}^2}^{q_{\rm max}^2} {\text{d}\mathcal{B}(D\to P\mu^+\mu^-) \over \text{d}q^2} \text{d}q^2} \over \displaystyle{\int_{q_{\rm min}^2}^{q_{\rm max}^2} {\text{d}\mathcal{B}(D\to Pe^+e^-) \over \text{d}q^2} \text{d}q^2} } \,.
\end{equation}
Here, $q_{\rm min}^2$ ($q_{\rm max}^2$) denotes the lower (upper) dilepton mass cut; to ensure maximal cancellation of hadronic uncertainties and hence a controlled SM prediction near unity it is crucial that the cuts for electron and muon modes are identical~\cite{Hiller:2003js}.
Assuming that NP contributes only to the muon mode, we present in Tabs.~\ref{tab:R_ratios} and \ref{tab:R_ratios_Ds} the predicted ranges of $R_\pi^D$ and $R_K^{D_s}$, respectively,  in the full, low and high $q^2$--integrated intervals. 
As for the resonance parametrization, we use the same $C_{9,P}$ for electron and muon modes. 
Due to poor knowledge of the resonances' behavior, the predictions for the low $q^2$--region ($\sqrt{q^2}\in[250,525]$~MeV) have sizable uncertainties and we only give the order of magnitude of the largest values found. 
The main uncertainty comes from the  phases $\delta_{\rho,\,\phi,\,\eta}$ which are varied within $-\pi$ and $\pi$, while the uncertainties due to hadronic form factors are  of the order of few percent. 
Integration over the full $q^2$--interval gives ratios $R_\pi^D$ and $R_K^{D_s}$ which are insensitive to NP. 
On the other hand, in the high $q^2$--region NP  can induce significant effects. 
BSM effects in the low $q^2$--region can be huge in the $D$--mode, however, there are large uncertainties.

\begin{table}[!ht]
 \centering
  \caption{$R_\pi^D$~\eqref{eq:LNU} in the SM and in NP scenarios for different $q^2$--bins. Ranges correspond to uncertainties from form factors and resonance parameters. Due to large uncertainties at low $q^2$ in some cases only  the order of magnitude of the largest values is given.}
 \label{tab:R_ratios}
 \begin{tabular}{l||c|c|c|c|c|c|c}
&SM & $\vert C_9 \vert=0.5$& $\vert C_{10}\vert=0.5$ & $\vert C_{9}\vert=\pm\vert C_{10}\vert=0.5$  & $\vert C_{S\,(P)}\vert=0.1$ & $\vert C_T \vert=0.5$&$\vert C_{T5}\vert=0.5$\\
 \hline
  full $q^2$ & $1.00 \pm \mathcal O(\%)$ &  SM-like	& SM-like	& SM-like& SM-like& SM-like& SM-like\\
  low $q^2$ & $0.95 \pm \mathcal O(\%)$ & $\mathcal{O}(100)$		& $\mathcal{O}(100)$		& $\mathcal{O}(100)$&$0.9\ldots1.4$& $\mathcal{O}(10)$&$1.0\ldots5.9$\\
  high $q^2$ & $1.00 \pm \mathcal O(\%)$ & $0.2\ldots11$		& $3\ldots7$		& $2\ldots17$& $1\ldots2$&$1\ldots5$&$2\ldots4$\\
 \end{tabular}
\end{table}

\begin{table}[!ht]
 \centering
  \caption{$R_K^{D_s}$~\eqref{eq:LNU} in the SM and in NP scenarios for different $q^2$--bins, see Tab.~\ref{tab:R_ratios} and text.}
 \label{tab:R_ratios_Ds}
 \begin{tabular}{l||c|c|c|c|c|c|c}
  &SM & $\vert C_9 \vert=0.5$& $\vert C_{10}\vert=0.5$ & $\vert C_{9}\vert=\pm\vert C_{10}\vert=0.5$  & $\vert C_{S\,(P)}\vert=0.1$ & $\vert C_T \vert=0.5$&$\vert C_{T5}\vert=0.5$\\
 \hline
  full $q^2$ & $1.00 \pm \mathcal O(\%)$ &  SM-like	& SM-like	& SM-like& SM-like& SM-like& SM-like\\
  low $q^2$ & $0.94 \pm \mathcal O(\%)$ & $0.1\ldots3.0$		& $1.3\ldots1.5$		& $0.5\ldots3.6$&SM-like& $0.7\ldots1.2$& SM-like\\
  high $q^2$ & $1.00 \pm \mathcal O(\%)$ & $0.2\ldots16$		& $3\ldots11$		& $2\ldots26$& $1.5\ldots3.7$&$1\ldots6$&$1.6\ldots4.1$\\
 \end{tabular}
\end{table}

\subsection{Angular observables\label{sec:BSM_nulltests}}

\begin{figure}[!t]\centering
\includegraphics[width=0.49\textwidth]{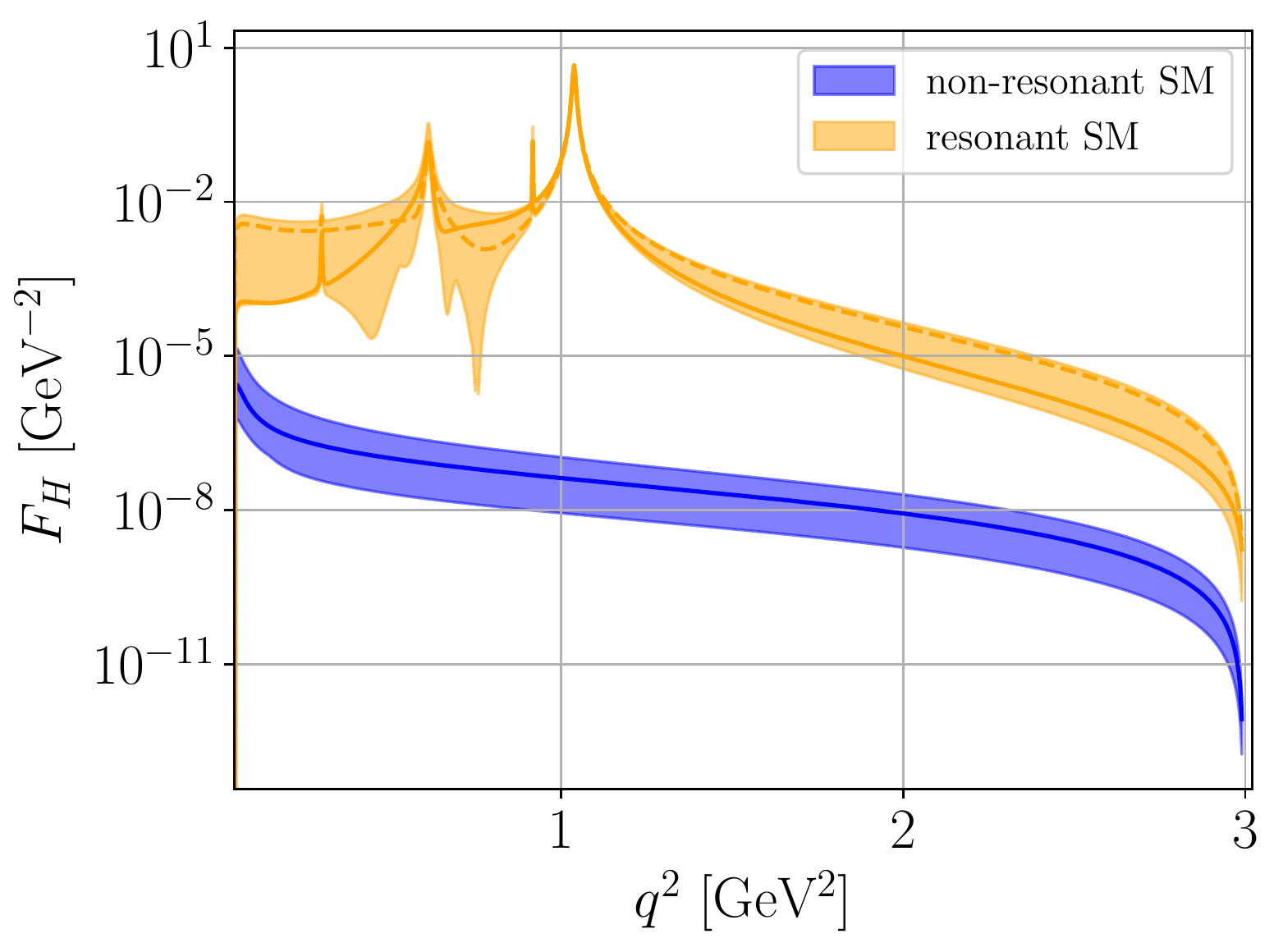}
\includegraphics[width=0.49\textwidth]{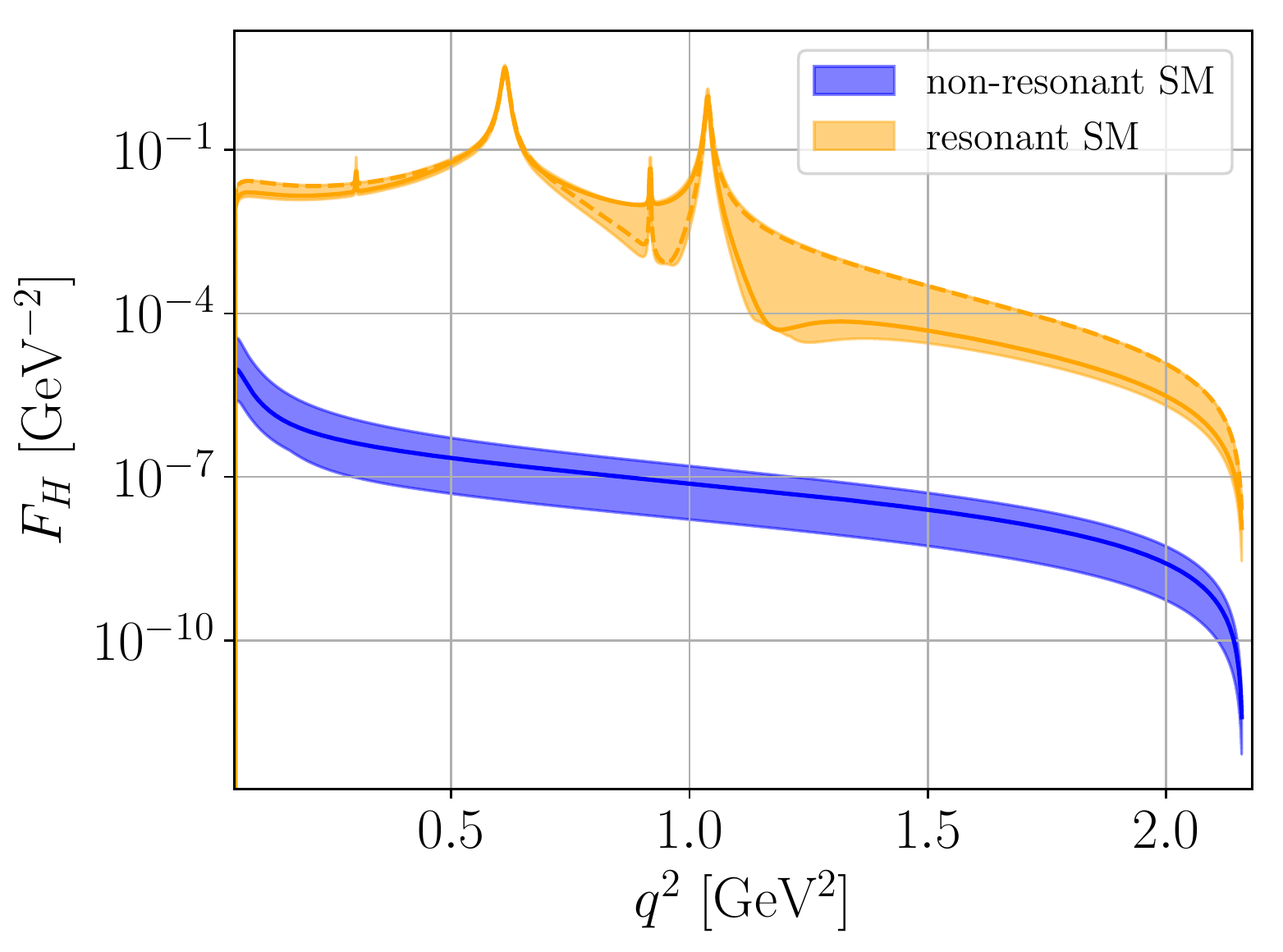}
\caption{The SM contributions to  $F_H(q^2)$ given in Eq.~(\ref{eq:FH}) for $D^+\to\pi^+\mu^+\mu^-$ (plot to the left) and $D_s^+\to K^+\mu^+\mu^-$ (plot to the right) decays, see Fig.~\ref{fig:dGamma_SM}.}
\label{fig:FH-SM}
\end{figure}

The double differential distribution of $D \to P \ell^+ \ell^-$ decays~\cite{Bobeth:2007dw}
\begin{equation}
{\text{d}^2\Gamma \over \text{d}q^2\,\text{d}\cos\theta} = a(q^2) + b(q^2) \cos\theta + c(q^2) \cos^2\theta \,,
\end{equation}
where  $\theta$ denotes the angle between the $\ell^-$--momentum and the $P$--momentum in the dilepton rest frame,
offers the measurement of two angular observables:
The  lepton forward-backward asymmetry,
\begin{equation}
\begin{split}
A_{\rm FB}(q^2) &= {1 \over \Gamma} \left[ \int_0^1 - \int_{-1}^0 \right] {\text{d}^2\Gamma \over \text{d}q^2 \text{d}\cos\theta} \text{d}\cos\theta = {b(q^2) \over \Gamma} = {1\over \Gamma} {G_F^2 \alpha_e^2 \over 512\pi^5 m_D^3} \lambda_{DP} \biggl( 1 - {4m_\ell^2 \over q^2} \biggr) \biggl\{\biggr. \\
& \quad \text{Re}\left[ \left( C_9 + C_9^R + C_7 {2m_c \over m_D + m_P} {f_T \over f_+} \right) C_S^* \right] {m_\ell \over m_c} f_+ + 2\text{Re}\left[ C_{10} C_{T5}^* \right] {m_\ell \over m_D  + m_P} f_T \\
& \quad + \text{Re}\left[ C_S C_T^*  + \left( C_P + C_P^R \right) C_{T5}^* \right] {q^2 \over m_c (m_D + m_P)} f_T \biggl.\biggr\} (m_D^2-m_P^2) f_0 \,,
\end{split}
\label{eq:AFB}
\end{equation}
and the ``flat'' term,
\begin{equation}
\begin{split}
F_H(q^2) &= {2\over \Gamma} [a(q^2) + c(q^2)] = {1 \over \Gamma} {G_F^2 \alpha_e^2 \over 1024\pi^5 m_D^3} \sqrt{\lambda_{DP} \biggl( 1 - {4m_\ell^2 \over q^2} \biggr)}\, \biggl\{\biggr. \\
& \quad \biggl| C_9 + C_9^R + C_7 {2m_c \over m_D+m_P} {f_T \over f_+} \biggr|^2 {4m_\ell^2 \over q^2} \lambda_{DP} f_+^2 + |C_{10}|^2 {4m_\ell^2 \over q^2} (m_D^2-m_P^2)^2 f_0^2 \\
& \quad + \left[ |C_S|^2 \left( 1 - {4m_\ell^2 \over q^2} \right) + |C_P + C_P^R|^2 \right] {q^2 \over m_c^2} (m_D^2-m_P^2)^2 f_0^2 \\
& \quad + 4\,\left[ |C_T|^2  + |C_{T5}|^2 \right] \left( 1 - {4m_\ell^2 \over q^2} \right) {q^2 \over (m_D + m_P)^2} \lambda_{DP} f_T^2 \\
& \quad + 8\,\text{Re}\left[ \left( C_9 + C_9^R + C_7 {2m_c \over m_D + m_P} {f_T \over f_+} \right) C_T^*\right] {m_\ell \over m_D + m_P} \lambda_{DP} f_+ f_T\\
& \quad + 4\,\text{Re}\left[ C_{10} \left( C_P + C_P^R \right)^* \right] {m_\ell \over m_c} (m_D^2-m_P^2)^2 f_0^2  +16\,|C_T|^2 \frac{m_\ell^2}{(m_D+m_P)^2}\lambda_{DP} f_T^2 \biggr.\biggr\} \,.
\end{split}
\label{eq:FH}
\end{equation}
both with normalization to the
$q^2$-integrated decay rate,
\begin{align} \label{eq:G}
\Gamma=\Gamma(q^2_{\rm min},q^2_{\rm max})= \int_{q^2_{\rm min}}^{q^2_{\rm max}} {\text{d}\Gamma \over \text{d}q^2}\text{d}q^2 = 2  \int_{q^2_{\rm min}}^{q^2_{\rm max}} \left(a(q^2) + \frac{ c(q^2)}{3}\right) \,  \text{d}q^2 \, , 
\end{align} with integration limits depending on the $q^2$--bin. 
Since the scalar and pseudotensor operators are absent in the SM, $A_{\rm FB}$ constitutes a null test~\footnote{Higher order corrections, which can induce a finite $A_{\rm FB}$, are suppressed by either powers of $m_D$ over the $W$-mass (higher dimensional operators) or by $\alpha_e/(4\pi)$ ($D\to\pi\gamma\gamma\to\pi\ell^+\ell^-$) \cite{Bobeth:2007dw,deBoer:thesis}.}.
We find that in the SM $F_H(D \to \pi \mu^+ \mu^-)$ is $\mathcal{O}(10^{-3})$ at low $q^2$, whereas $F_H(D_s \to K \mu^+ \mu^-)$ is $\mathcal{O}(10^{-2})$ and both are even smaller at high $q^2$, see Fig.~\ref{fig:FH-SM}. 
$F_H(D_{(s)} \to \pi(K) e^+ e^-)$ in the SM is even further suppressed.
We learn that $F_H$ is yet another a very promising null test of the SM, sensitive to  NP in (pseudo)-scalar and tensor operators.

\begin{figure}[!t]\centering
\includegraphics[width=0.49\textwidth]{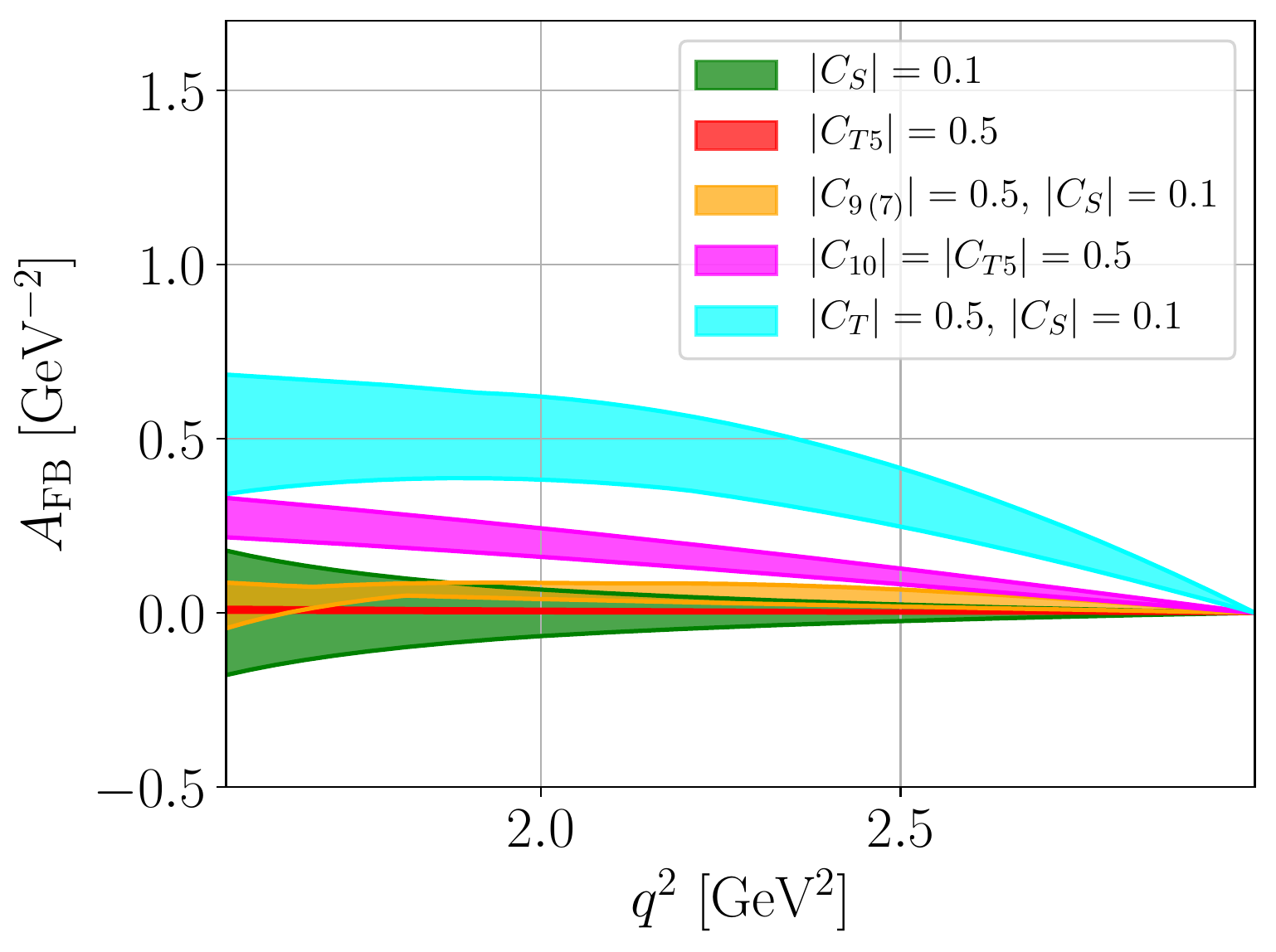}
\includegraphics[width=0.49\textwidth]{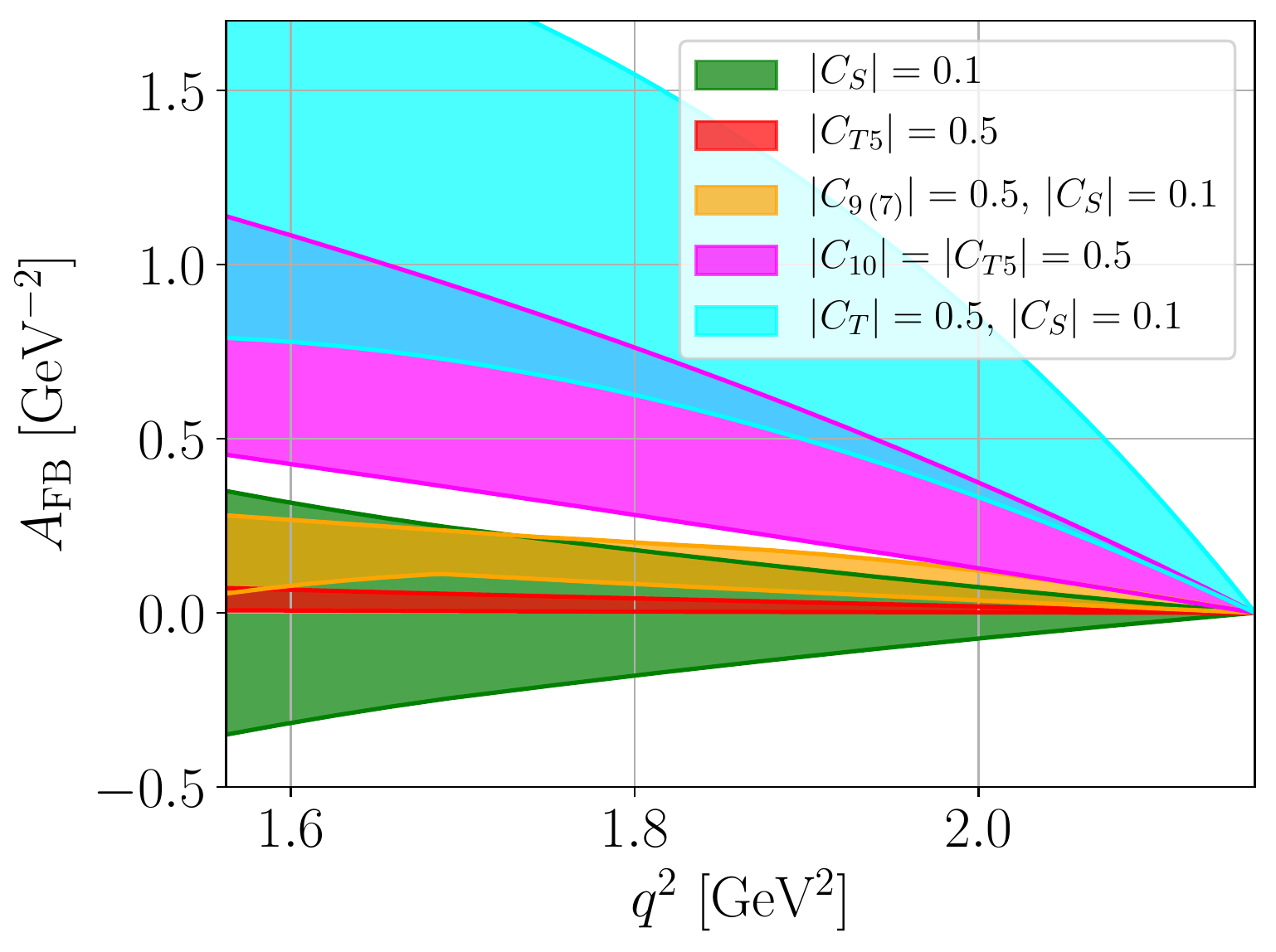} \\ \hspace{3mm}
\includegraphics[width=0.47\textwidth]{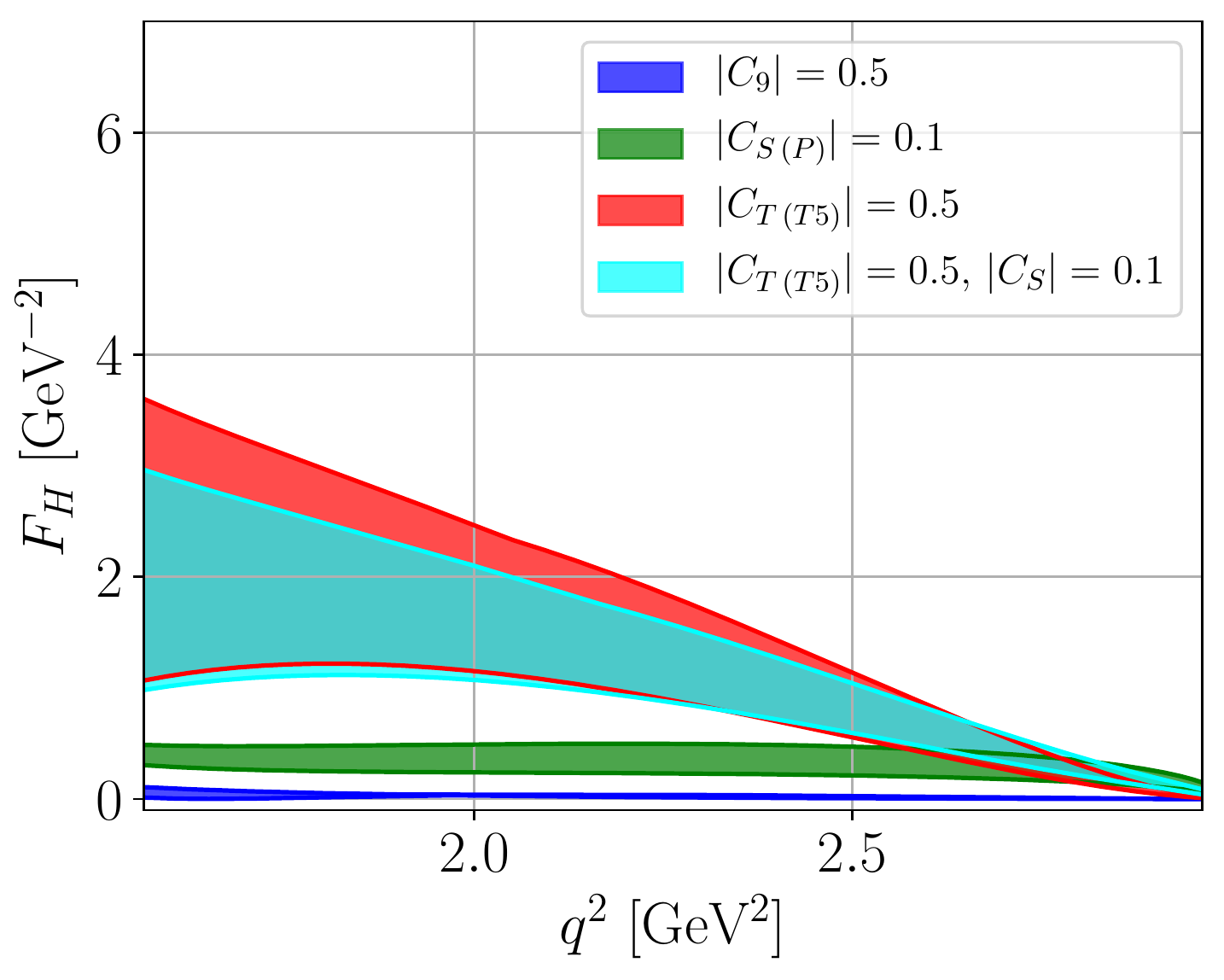} \hspace{3mm}
\includegraphics[width=0.47\textwidth]{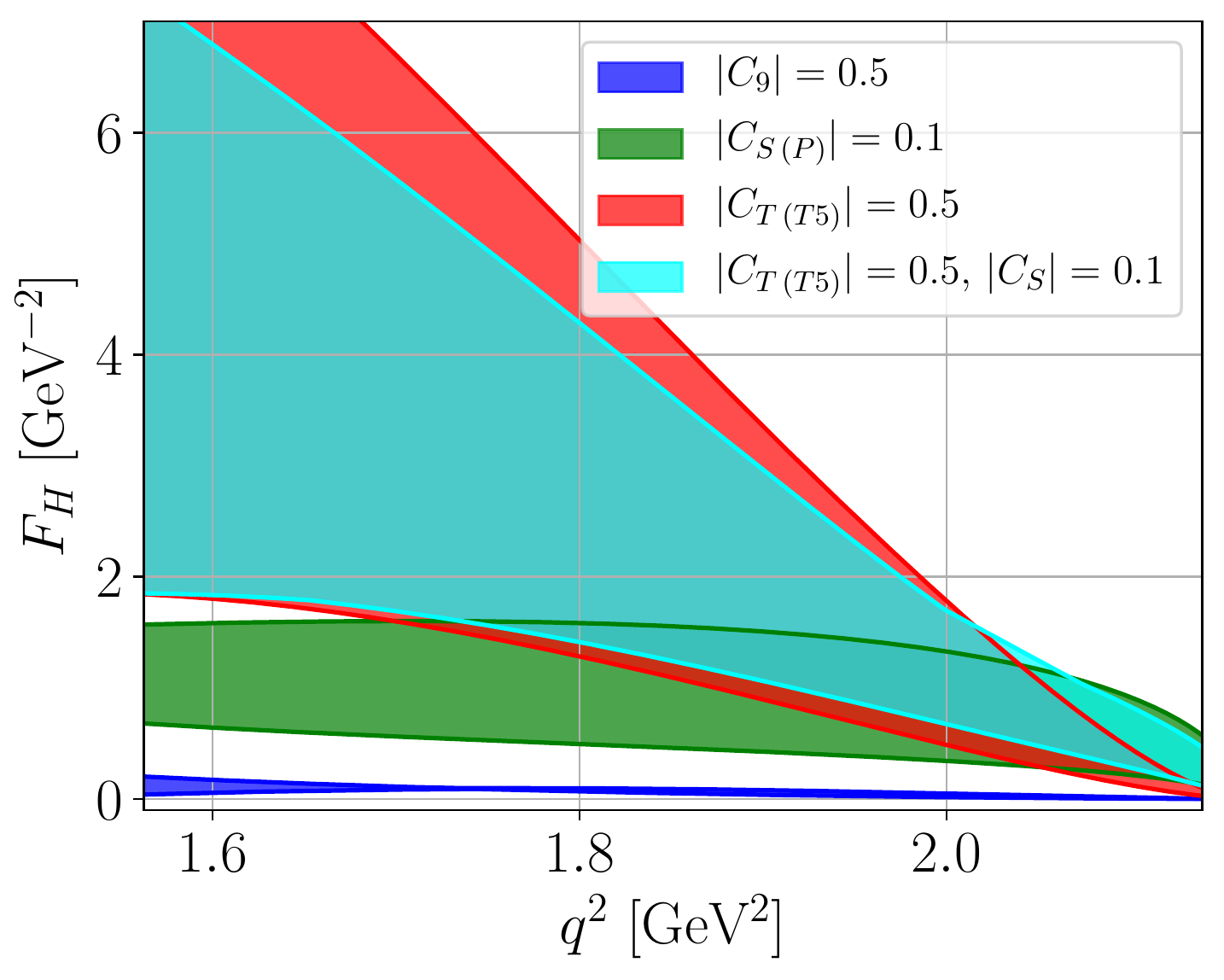}
\caption{The forward-backward asymmetry $A_{\rm FB}$  (upper plots) and $F_H$ (lower plots) in the high--$q^2$ region in different BSM scenarios for $D^+\to\pi^+\mu^+\mu^-$ (plots to the left) and $D_s^+\to K^+\mu^+\mu^-$ (plots to the right).}
\label{fig:AFB-FH}
\end{figure}

In Fig.~\ref{fig:AFB-FH} we show $A_{\rm FB}$ and $F_H$ in different NP scenarios for $D^+\to\pi^+\mu^+\mu^-$ (plots to the left) and $D_s^+\to K^+\mu^+\mu^-$ (plots to the right) at high $q^2$, that is, $q^2_{\rm min}=(1.25 \, \mbox{GeV})^2$ and  $q^2_{\rm max}=(m_D-m_P)^2$.
The band width represents theoretical uncertainties.
One can see that $A_{\rm FB}$ is mostly sensitive to the combinations of tensor and (pseudo)scalar operators, while it is significantly suppressed if only scalar or pseudotensor structure is present. 
The latter effect comes from interference terms, $C_9^RC_S^*$ and $C_P^RC_{T5}^*$, which are suppressed by the light lepton mass and small $\eta$-resonance contribution to $C_P^R$, respectively. 
The flat term turns out to be sensitive to the (pseudo)scalar and especially to the (pseudo)tensor operators. 
The effect of simultaneous presence of scalar and tensor contributions (in cyan) is essentially indistinguishable from the one induced by pure (pseudo)tensor (in red) structure. 
However, large effects in both $F_H$ and $A_{\rm FB}$ would be a signal of tensor and scalar nature of NP.
In Sec.~\ref{sec:BSM_models} we zoom into  Fig.~\ref{fig:AFB-FH},  and discuss in more detail possible signals of NP induced predominantly by (axial-)vector contributions.
The normalization of angular observables has significant impact on the shape of the distributions.
Choosing  $d \Gamma/dq^2$, as in \cite{Sahoo:2017lzi}, instead of $\Gamma$ (\ref{eq:G}), results in markedly different behavior for $F_H$ near the endpoint $q^2 \to (m_D-m_P)^2$, where cancellations can occur.

\subsection{CP--asymmetry}\label{sec:BSM_cpasym}

Another promising observable to probe NP is the CP--asymmetry, defined as~\cite{Fajfer:2012nr,deBoer:2015boa}
\begin{equation}
A_{\rm CP}(q^2) = {1 \over \Gamma + \overline\Gamma} \left( {\text{d}\Gamma \over \text{d}q^2} - {\text{d}\overline\Gamma \over \text{d}q^2} \right) \,,
\end{equation}
where $\overline\Gamma$ denotes the decay rate of the CP--conjugated mode, defined with $q^2$-bin dependence as $\Gamma$ in Eq.~\eqref{eq:G}. The difference of the differential rates can be written as
\begin{equation}
\begin{split}
{\text{d}\Gamma \over \text{d}q^2} - {\text{d}\overline\Gamma \over \text{d}q^2} &= {G_F^2 \alpha_e^2 \over 256\pi^5 m_D^3} \sqrt{\lambda_{DP} \biggl( 1 - {4m_\ell^2 \over q^2} \biggr)}\, \biggl\{\biggr. \\
& \quad {2\over3} \,\text{Im}\left[ C_9 + 2C_7 {m_c \over m_D + m_P} {f_T \over f_+} \right] \text{Im}\left[ C_9^R \right] \left( 1 + {2m_\ell^2 \over q^2} \right) \lambda_{DP} f_+^2 \\
& \quad + \text{Im}\left[ C_P \right] \text{Im} \left[ C_P^R \right] {q^2 \over m_c^2} (m_D^2-m_P^2)^2 f_0^2 \\
& \quad +  4\,\text{Im}\left[ C_T \right] \text{Im}\left[ C_9^R \right]  {m_\ell \over m_D + m_P} \lambda_{DP} f_+ f_T\\
& \quad + 2\,\text{Im}\left[ C_{10} \right] \text{Im}\left[ C_P^R \right] {m_\ell \over m_c} (m_D^2-m_P^2)^2 f_0^2 \biggr.\biggr\} \,.
\end{split}
\label{eq:ACP}
\end{equation}
$A^{\rm SM}_{\rm CP}$ is determined by the first term in Eq.~\eqref{eq:ACP} and remains tiny due to small phases of the CKM factors in $C_9$, see Eq.~\eqref{eq:CSM}.
Existing bounds from $A_{\rm CP}(D^0 \to \rho^0 \gamma)=0.056\pm0.152\pm0.006$~\cite{Abdesselam:2016yvr} do not provide further constraints on $C_7^{(\prime)}$ beyond the ones from branching ratio measurements~\cite{deBoer:2017que}.
Na\"ive T-odd CP-asymmetries from angular distributions in $D \to \pi \pi \mu^+ \mu^-$ decays can probe CP-phases even for vanishing strong phases.
First experimental studies are at the ${\cal{O}}(10\,\%)$ level~\cite{Aaij:2018fpa} which is about where sensitivity to BSM physics starts~\cite{deBoer:2018buv}.

\begin{figure}[!t]\centering
\includegraphics[width=0.49\textwidth]{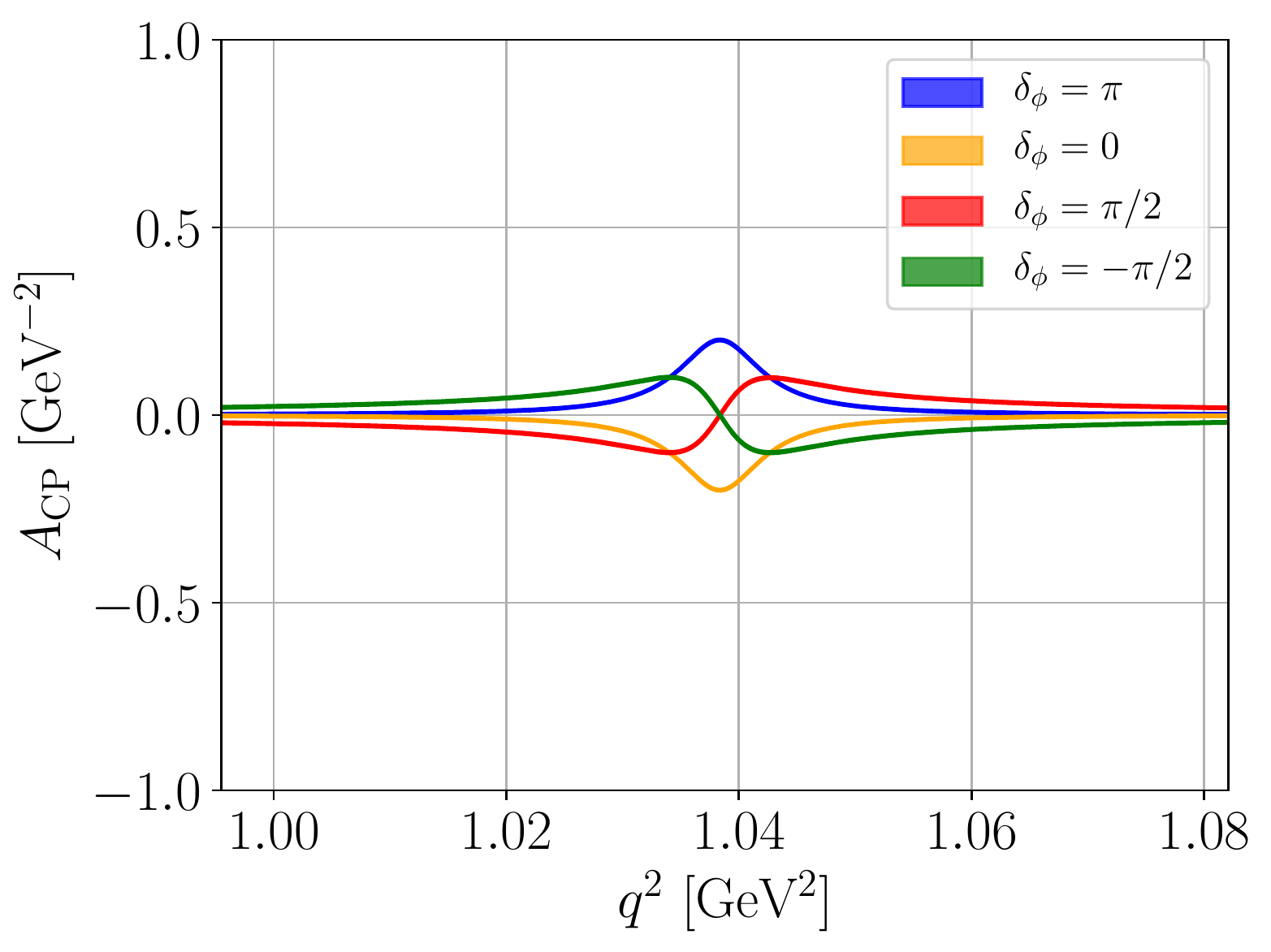}
\includegraphics[width=0.49\textwidth]{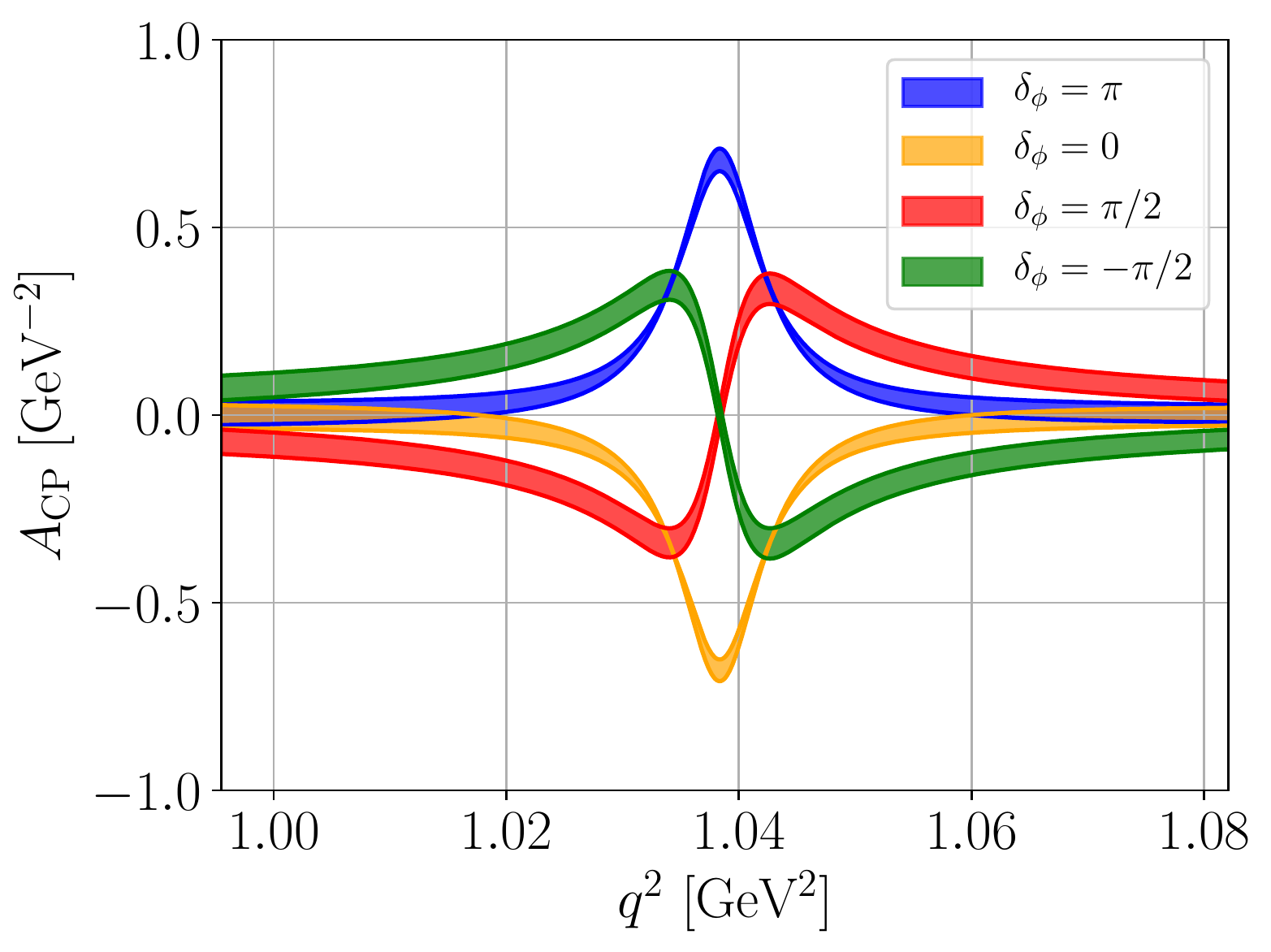}
\includegraphics[width=0.49\textwidth]{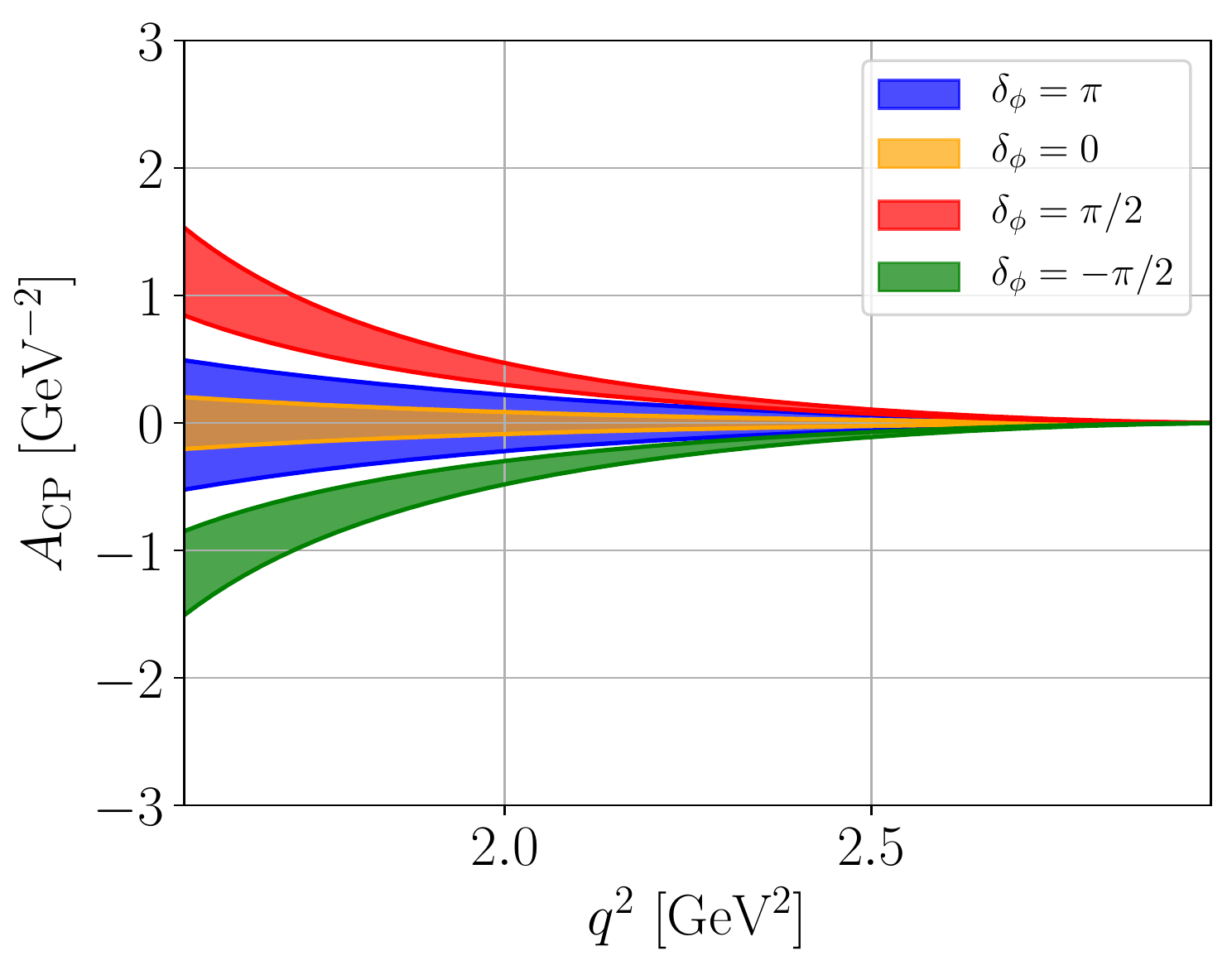}
\includegraphics[width=0.49\textwidth]{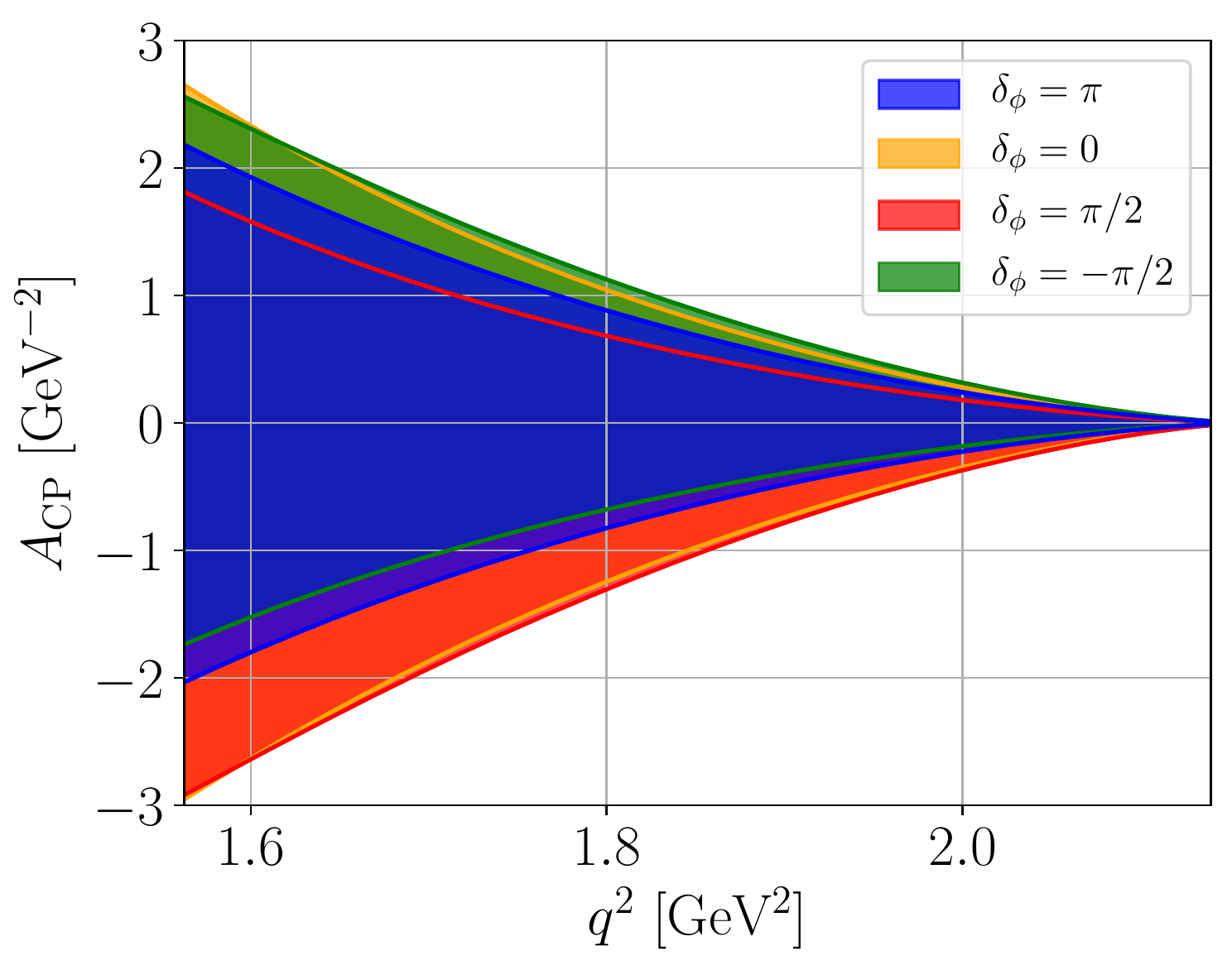}
\caption{The CP--asymmetry in $D^+\to\pi^+\mu^+\mu^-$ (plots to the left) and $D_s^+\to K^+\mu^+\mu^-$ (plots to the right) decays around the $\phi$ resonance $[(m_\phi-5\Gamma_\phi)^2,(m_\phi+5\Gamma_\phi)^2]$ (upper plots) and in  the high--$q^2$ region (lower plots) for different  values of $\delta_\phi =0, \pm \pi/2, \pi$ and $C_9=0.1\exp(\text{i}\,\pi/4)$. The uncertainties are due to the other strong phases $(\delta_\rho,\, \delta_\eta)$, the form factors, as well as the charm quark mass $m_c$. }
\label{fig:ACP}
\end{figure}

For a sizable (T-even) CP--asymmetry, enhanced strong phases and therefore resonance effects are instrumental~\cite{Fajfer:2012nr}. In Fig.~\ref{fig:ACP} we show $A_{\rm CP}$  in the $\phi$--region (upper plots) and at high $q^2$ (lower plots) for $C_9=0.1\exp(\text{i}\,\pi/4)$ and different values of $\delta_\phi$. 
The band width corresponds to the  $1\sigma$ uncertainty due to $m_c$, form factors and resonance parameters ($\delta_{\rho,\eta}$, varied within $-\pi$ and $\pi$). 
NP effects in $C_T$ and $C_{P,10}$  are suppressed by the light lepton mass and the completely negligible $\text{Im}[C_P^R(q^2\simeq m_\phi^2)]$ respectively.
We learn from Fig.~\ref{fig:ACP} that irrespective of the value of $\delta_\phi$, sizable BSM effects occur making this observable promising for NP searches.
Except for $D_s \to K \ell^+ \ell^-$ at high $q^2$, $A_{\rm CP}$ has rather small uncertainties and is useful to extract strong and weak parameters. 
Note that $A_{\rm CP}$ can change its sign around  $q^2 \sim m_\phi^2$ and hence, to avoid a vanishing integrated asymmetry, binning is required.
Due to the way Eq.~\eqref{eq:U} with which leptonic vector contributions enter $D \to P \ell^+ \ell^-$ decays, $A_{\rm CP}$ has similar sensitivity to 
BSM effects from $C_7^{(\prime)}$ than from $C_9^{(\prime)}$.

Similar to the behavior observed in the angular observables $F_H$ and $A_{\rm FB}$, BSM effects in $A_{\rm CP}$ in the $D_s \to K \ell^+\ell^-$ mode are enhanced relative to the $D \to \pi \ell^+ \ell^-$ one, due to the smaller decay rate of the former, caused predominantly by kinematics.

\subsection{Lepton flavor violation}\label{sec:BSM_lfv}

To discuss LFV in $c\to u\ell^-\ell^{\prime+}$ ($\ell\neq\ell^\prime$) decays we introduce the following effective Hamiltonian,
\begin{equation} \label{eq:LFV}
\mathcal H_\text{eff}^{\rm LFV} = -{4G_F \over \sqrt2} {\alpha_e \over 4\pi} \sum_i \left( K_i^{(\ell\ell^\prime)} O_i^{(\ell\ell^\prime)} + K_i^{\prime\,(\ell\ell^\prime)} O_i^{\prime\,(\ell\ell^\prime)} \right) \,,
\end{equation}
where the $K_i^{(\prime)}$ denote Wilson coefficients and the operators $O_i^{(\prime)}$ read as
\begin{equation}
O_9^{(\ell\ell^\prime)} =  (\ubar_L \gamma_\mu c_L) (\lbar \gamma^\mu \ell^\prime) \,, \quad\quad
O_9^{\prime\,(\ell\ell^\prime)} = (\ubar_R \gamma_\mu c_R) (\lbar \gamma^\mu \ell^\prime) \,,
\label{eq:Z9_phenomenological}
\end{equation}
with other operators from Eq.~\eqref{eq:operators} defined in similar way by changing flavor in lepton currents. 
Note that there is no $O_7^{(\prime)}$ contribution since the photon does not couple to different lepton flavors.

The differential distribution for the LFV decays $D \to P e^\pm\mu^\mp$ is given as
\begin{equation}
\begin{split}
{\mathrm d\Gamma (D \to P e^\pm\mu^\mp) \over \mathrm dq^2} &= {G_F^2\alpha_e^2 \over 1024\pi^5m_D^3} \sqrt{\lambda_{DP}}   \bigg\{ \frac23 \left(|K_9|^2 + |K_{10}|^2\right) \lambda(m_D^2,m_P^2,q^2) f_+^2 \\
& +\left(|K_S|^2 + |K_P|^2\right) {q^2 \over m_c^2} (m_D^2 - m_P^2)^2 f_0^2 \\
& +\frac43 \left(|K_T|^2 + |K_{T5}|^2\right) {q^2 \over (m_D+m_P)^2} \lambda_{DP} f_T^2 \\
& +2\mathrm{Re}\left[\pm K_9 K_S^* + K_{10} K_P^*\right] {m_\mu \over m_c} (m_D^2 - m_P^2)^2 f_0^2 \\
& +4\mathrm{Re}\left[K_9 K_T^* \pm K_{10} K_{T5}^*\right] {m_\mu \over m_D+m_P} \lambda_{DP} f_+ f_T \bigg\}
+\mathcal O\left(m_\mu^2\right)\,,
\end{split}
\label{eq:dGamma_LFV}
\end{equation}
where $K_i=K_i^{(\mu e)}+K_i^{\prime\,(\mu e)}$ for $D \to P e^+\mu^-$ and $K_i=K_i^{(e\mu)}+K_i^{\prime\,(e\mu)}$ for $D \to P e^-\mu^+$. 
Here we neglected the electron mass.

Similarly to Eq.~\eqref{eq:NP_constraints_full}, we obtain the following constraints on the LFV Wilson coefficients using the 90\% C.L.~upper limits $ \mathcal{B}(D^+\to\pi^+e^+\mu^-)<2.9\times10^{-6}$ and $\mathcal{B}(D^+\to\pi^+e^-\mu^+)<3.6\times10^{-6}$~\cite{Lees:2011hb}:
\begin{equation}
\begin{split}
& 1.3\,\big(|K_9|^2 + |K_{10}|^2\big) + 2.6\,\big( |K_S|^2 + |K_P|^2 \big) + 0.4\,\big( |K_T|^2 + \big|K_{T5}\big|^2 \big) + \\
& 0.5\,\mathrm{Re}\big[ K_{10} K_P^* \pm K_9 K_S^* \big]  + 0.3\,\mathrm{Re}\big[ K_9 K_T^* \pm K_{10} K_{T5}^* \big] \lesssim 50 \, .
\end{split}
\label{eq:constraint_D-pimue}
\end{equation}

Tighter constraints on $K_{9,10,S,P}^{(\prime)}$ can be obtained from data on leptonic decays, 
\begin{equation}
\begin{split}
\mathcal{B}(D^0\to e^\pm\mu^\mp) &= \tau_{D^0} {G_F^2 \alpha_e^2 m_D^5 f_D^2 \over 64\pi^3 m_c^2} \left(1 - {m_\mu^2 \over m_D^2}\right)^2 \bigg\{ \left| K_S - K_S^\prime \pm {m_\mu m_c \over m_D^2} \left(K_9 - K_9^\prime \right) \right|^2 \\
& \quad + \left| K_P - K_P^\prime + {m_\mu m_c \over m_D^2} \left(K_{10} - K_{10}^\prime \right) \right|^2 \bigg\} \,,
\end{split}       
\label{eq:Br_D-emu} 
\end{equation}
with $K_i^{(\prime)}=K_i^{(\prime)(\mu e)}$ for $D^0\to e^+\mu^-$ and $K_i^{(\prime)}=K_i^{(\prime)(e\mu)}$ for $D^0\to e^-\mu^+$. Using $\mathcal{B}(D^0\to e^\pm\mu^\mp)<1.3\times10^{-8}$ @ 90\% C.L.~\cite{Aaij:2015qmj}, we obtain
\begin{equation}
\big|K_S - K_S^\prime \pm 0.04\,\big( K_9- K_9^\prime \big)\big|^2 + \big|K_P - K_P^\prime + 0.04\,\big( K_{10} - K_{10}^\prime \big)\big|^2 \lesssim 0.01 \, . \\
\label{eq:constraint_D-mue}
\end{equation}

\begin{figure}[!t]\centering
\includegraphics[width=0.49\textwidth]{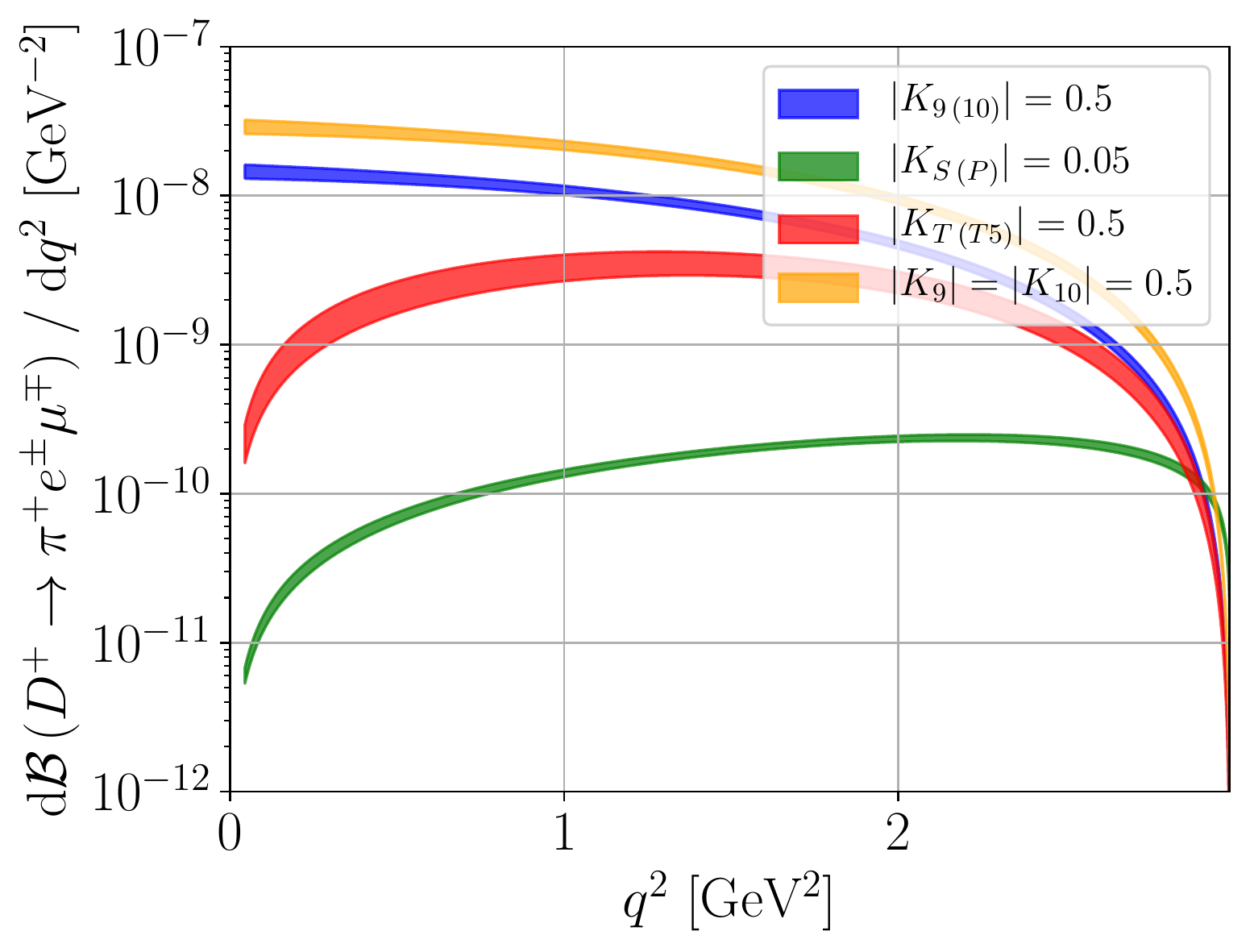}
\includegraphics[width=0.49\textwidth]{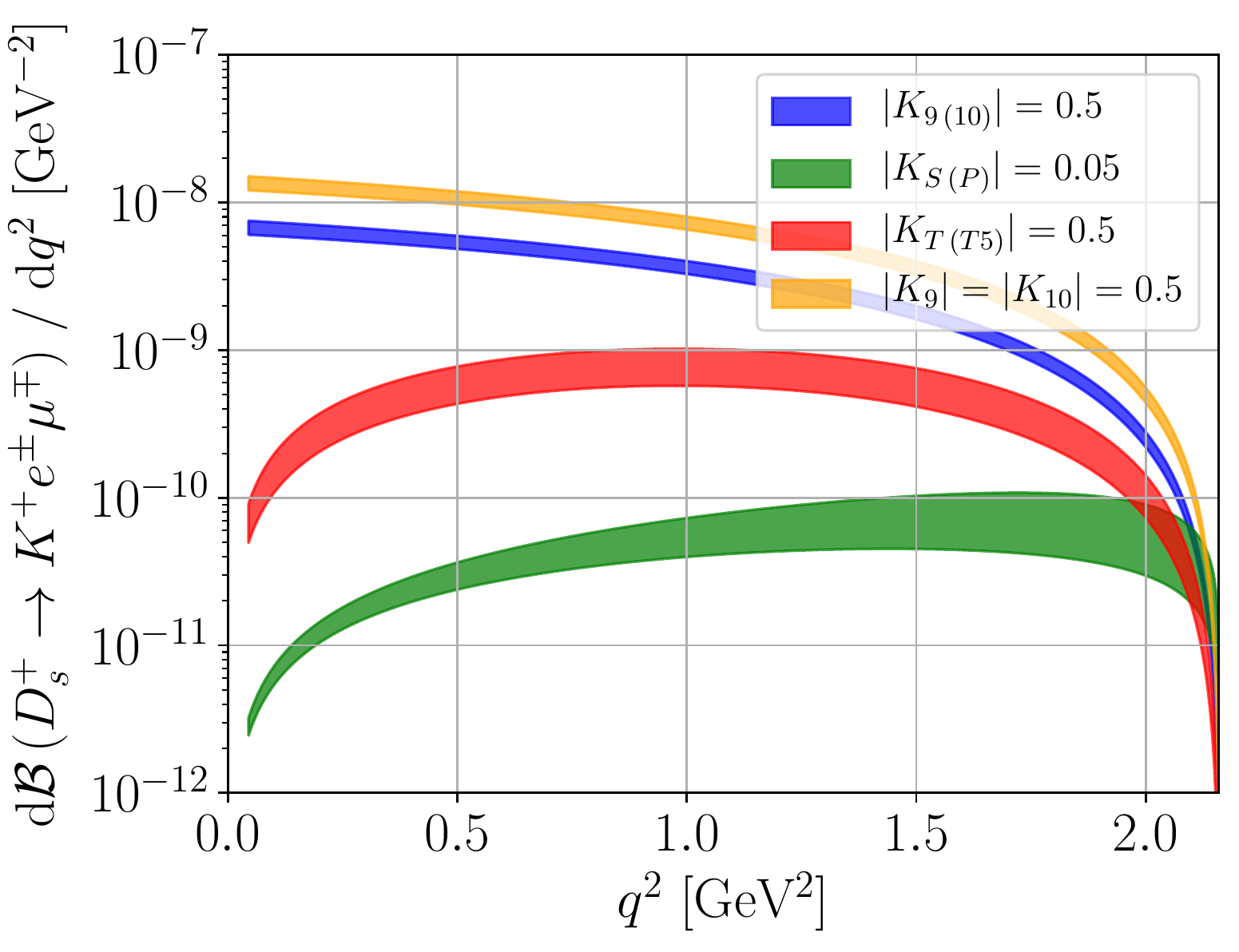}
\caption{The differential branching ratio of $D^+\to\pi^+e^\pm\mu^\mp$ (plot to the left) and $D_s^+\to K^+e^\pm\mu^\mp$ (plot to the right) decays in different LFV-BSM scenarios.}
\label{fig:LFV}
\end{figure}

In Fig.~\ref{fig:LFV} we present the differential branching fractions for LFV decays $D^+\to\pi^+e^\pm\mu^\mp$ and $D_s^+\to K^+e^\pm\mu^\mp$ for various NP Wilson coefficients allowed by Eqs.~\eqref{eq:constraint_D-pimue},\eqref{eq:constraint_D-mue}. 
Since the resonant contributions are absent in the LFV processes, the uncertainties are due to the form factors and $m_c$, making the band widths in Fig.~\ref{fig:LFV} significantly smaller than in Fig.~\ref{fig:dGamma_SM}. 
The difference in the shapes for vector and (pseudo)tensor/scalar structure allow to experimentally distinguish different operators and BSM scenarios.

Note that there exists one opportunity to study $\tau$-couplings in leptonic charm decays,
\begin{equation}
\mathcal{B}(D^0\to e^\pm\tau^\mp) = 1.2\times10^{-8}\, \left( \big|K_S - K_S^\prime \pm 0.7\,\big( K_9- K_9^\prime \big)\big|^2 + \big|K_P - K_P^\prime + 0.7\,\big( K_{10} - K_{10}^\prime \big)\big|^2 \right) \,. \\
\label{eq:constraint_D-tau}
\end{equation}

\section{New Physics models for \texorpdfstring{$c\to u\ell\ell$}{ctoull}\label{sec:BSM_models}}

We discuss signatures of leptoquarks in Sec.~\ref{sec:LQ}, and supersymmetry (SUSY) in Sec.~\ref{sec:susy}.
Contributions to $c \to u$ FCNCs in flavorful $Z^\prime$-models including explicit, anomaly-free realizations are worked out in Sec.~\ref{sec:Z'}.

In general, the models' reach in rare charm decays is limited by kaon data, $D$-mixing and high-$p_T$ searches.
Constraints from the down-sector can be escaped with contributions from singlet up-type quarks, giving rise to primed operators.
Decoupling the  doublets from kaon constraints requires assumptions on flavor.
While this introduces model-dependence, it highlights
the importance of joint charm and kaon studies and their impact on flavor model building.
$D$-mixing constraints are most severe for the $Z^\prime$-models  because, unlike in SUSY or leptoquark scenarios, meson mixing arises at tree-level, see App.~\ref{sec:D_mixing} for details.
We point out that the simultaneous presence of left-handed and right-handed couplings to quarks allows to avoid such constraints.

At the end of each model section  we summarize how the model can signal NP in which observable from Sec.~\ref{sec:BSM}, and refer to corresponding figures and tables.

\subsection{Leptoquark signatures \label{sec:LQ}}

In order to bypass  the constraints from the kaon sector in a most straightforward manner, we consider only the scalar  leptoquarks $S_{1(2)}$ with right(left)-handed couplings and the vector ones $\widetilde{V}_{1,2}$, {\it e.g.}~\cite{deBoer:2015boa,Hiller:2016kry}. The subscript "$1$" denotes  $SU(2)_L$-singlets, whereas "$2$" denotes doublets.
This precludes any (pseudo-)scalar or tensor operators, which otherwise would be induced by $S_{1,2}$, as well as (axial-)vector ones with left-handed quarks.
We stress, however, that even small  Wilson coefficients involving doublet quarks can signal NP in $A_{\rm CP}$, if CP-violating~\cite{deBoer:2015boa}.
The interaction Lagrangian contributing to  $c\to u\ell^-\ell^{\prime+}$ processes then reads~\cite{Buchmuller:1986zs}
\begin{equation} \label{eq:LLQ}
\mathcal{L}_{\rm LQ} \supset 
\lambda_{S_1}^{ij}\, \overline{u}_{iR}^c l_{jR}\, S_1 +
\lambda_{S_2}^{ij}\, \overline{u}_{iR} L_{jL}\, S_2 +
\lambda_{\widetilde{V}_1}^{ij}\, \overline{u}_{iR} \gamma^\mu l_{jR}\, \widetilde{V}_{1\mu} +
\lambda_{\widetilde{V}_2}^{ij}\, \overline{u}_{iR}^c \gamma^\mu L_{jL}\, \widetilde{V}_{2\mu} + {\rm c.c.} \,,
\end{equation}
where $L_L$ denote  lepton doublet and $l_R,u_R$  lepton and up-type quark singlets; $i,j$ are the generation indices.
Hypercharge-assignments of the leptoquarks can be read-off from (\ref{eq:LLQ}); all leptoquarks are color-triplets.
Charm signatures of $S_2$ have been studied in \cite{Fajfer:2015mia,Sahoo:2017lzi} and $\widetilde{V}_1$ in \cite{Fajfer:2015mia}.
Only $S_2^{5/3}$ and $\widetilde{V}_{2}^{1/3}$, and $S_1^{1/3}$, $\widetilde{V}_{1}^{5/3}$ mediate an interaction between up-type quarks and charged leptons,
where  superscripts indicate the electric charge. 
Using Fierz identities  the Wilson coefficients of the $O_9^{\prime\,(\ell\ell^\prime)}$ and $O_{10}^{\prime\,(\ell\ell^\prime)}$ operators can be written as,
{\it e.g.}~\cite{deBoer:2015boa},
\begin{equation}
\begin{split}
K_9^{\prime\,(\ell\ell^\prime)} &= {\sqrt2 \pi \over G_F \alpha_e} \left[ 
 { \lambda_{S_1}^{c\ell^\prime} \lambda_{S_1}^{u\ell*} \over 4 M_{S_1}^2 }
-{ \lambda_{S_2}^{u\ell^\prime} \lambda_{S_2}^{c\ell*} \over 4 M_{S_2}^2 } 
-{ \lambda_{\widetilde{V}_1}^{u\ell^\prime} \lambda_{\widetilde{V}_1}^{c\ell*} \over 2 M_{\widetilde{V}_1}^2 }
+{ \lambda_{\widetilde{V}_2}^{c\ell^\prime} \lambda_{\widetilde{V}_2}^{u\ell*} \over 2 M_{\widetilde{V}_2}^2 } 
\right] \,, \\
K_{10}^{\prime\,(\ell\ell^\prime)} &= {\sqrt2 \pi \over G_F \alpha_e} \left[ 
 { \lambda_{S_1}^{c\ell^\prime} \lambda_{S_1}^{u\ell*} \over 4 M_{S_1}^2 }
+{ \lambda_{S_2}^{u\ell^\prime} \lambda_{S_2}^{c\ell*} \over 4 M_{S_2}^2 } 
-{ \lambda_{\widetilde{V}_1}^{u\ell^\prime} \lambda_{\widetilde{V}_1}^{c\ell*} \over 2 M_{\widetilde{V}_1}^2 }
-{ \lambda_{\widetilde{V}_2}^{c\ell^\prime} \lambda_{\widetilde{V}_2}^{u\ell*} \over 2 M_{\widetilde{V}_2}^2 } 
\right] \, ,
\end{split}
\label{eq:K_LQ}
\end{equation}
where $M_X$, $X=S_{1,2},\widetilde{V}_{1,2}$ denotes the leptoquark mass.
For the  singlets ($S_1$, $\widetilde{V}_1$) $K_9^\prime=K_{10}^\prime$, while for the doublets ($S_2$, $\widetilde{V}_2$) $K_9^\prime=-K_{10}^\prime$. 
The corresponding  lepton flavor conserving contributions to $C_{9,10}^\prime$ can be obtained from Eq.~\eqref{eq:K_LQ} for $\ell^\prime=\ell$.

Using Eqs.~\eqref{eq:Br_D-ll},\eqref{eq:Br_D-emu} and neglecting the SM contribution, we obtain the following constraints  from the upper limits on $\mathcal{B}(D^0\to\mu^+\mu^-)$~\cite{Aaij:2013cza} and $\mathcal{B}(D^0\to e^\pm\mu^\mp)$~\cite{Aaij:2015qmj},
\begin{equation} 
\begin{split}
&\bigl|\lambda_{S_{1,2}}^{c\mu} \lambda_{S_{1,2}}^{u\mu*}\bigr| \lesssim 0.07\, \left({M_{S_{1,2}} \over 1~{\rm TeV}}\right)^2 \,, 
\quad\quad\quad
\bigl|\lambda_{\widetilde{V}_{1,2}}^{c\mu} \lambda_{\widetilde{V}_{1,2}}^{u\mu*}\bigr| \lesssim 0.03\, \left({M_{\widetilde{V}_{1,2}} \over 1~{\rm TeV}}\right)^2 \,, \\
&\bigl|\lambda_{S_{1,2}}^{ce(c\mu)} \lambda_{S_{1,2}}^{u\mu(ue)*}\bigr| \lesssim 0.13\, \left({M_{S_{1,2}} \over 1~{\rm TeV}}\right)^2 \,, \quad
\bigl|\lambda_{\widetilde{V}_{1,2}}^{ce(c\mu)} \lambda_{\widetilde{V}_{1,2}}^{u\mu(ue)*}\bigr| \lesssim 0.07\, \left({M_{\widetilde{V}_{1,2}} \over 1~{\rm TeV}}\right)^2 \,.
\end{split}
\end{equation}
Constraints in a similar ballpark are obtained from $\mathcal{B}(D\to\pi\mu^+\mu^-)$ using Eq.~\eqref{eq:NP_constraints_high},
\begin{equation} 
\bigl|\lambda_{S_{1,2}}^{c\mu} \lambda_{S_{1,2}}^{u\mu*}\bigr| \lesssim 0.09\, \left({M_{S_{1,2}} \over 1~{\rm TeV}}\right)^2 \,, 
\quad
\bigl|\lambda_{\widetilde{V}_{1,2}}^{c\mu} \lambda_{\widetilde{V}_{1,2}}^{u\mu*}\bigr| \lesssim 0.05\, \left({M_{\widetilde{V}_{1,2}} \over 1~{\rm TeV}}\right)^2 \,.
\end{equation}
\begin{figure}[!t]\centering
\includegraphics[width=0.49\textwidth]{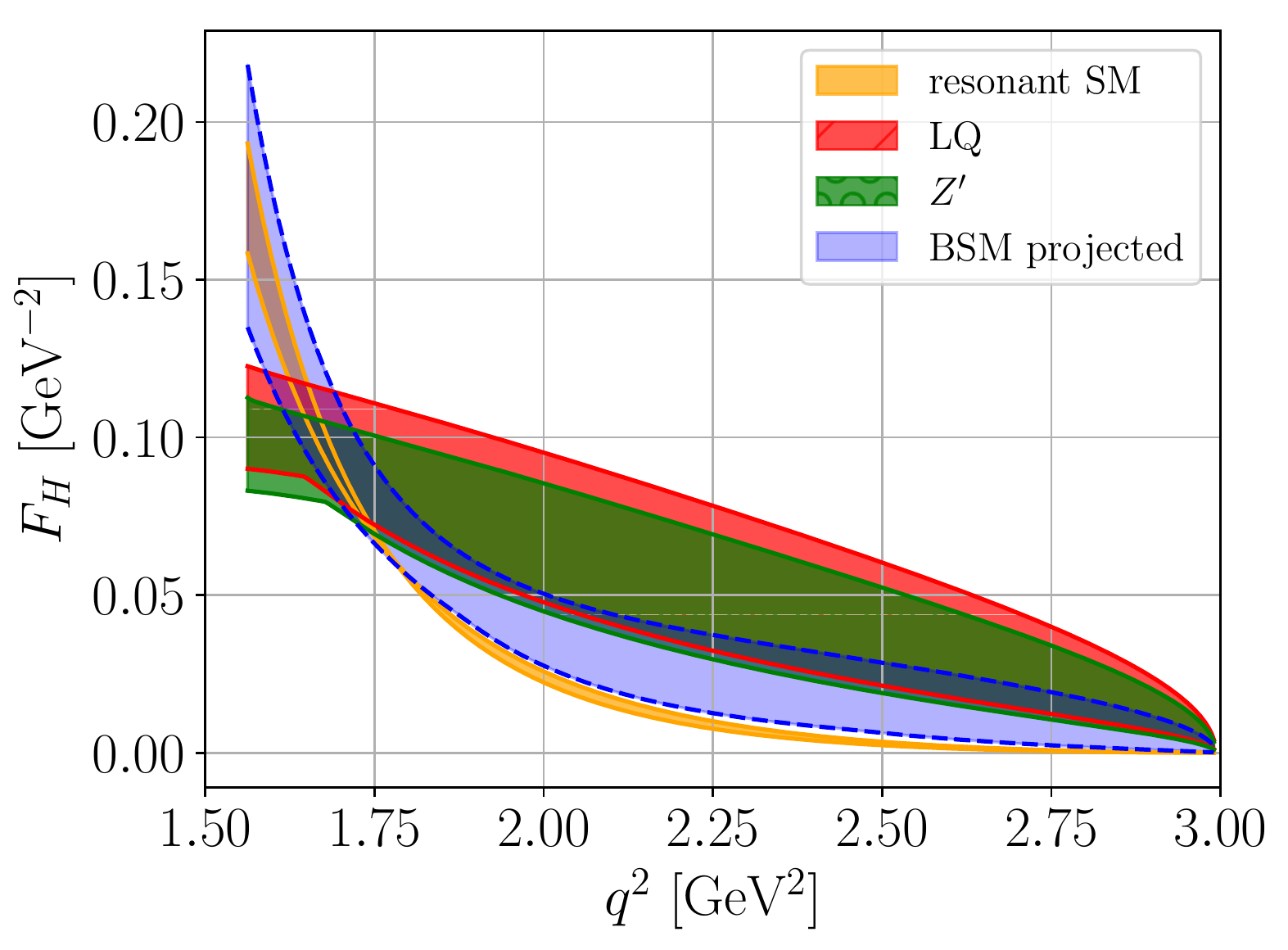}
\includegraphics[width=0.48\textwidth]{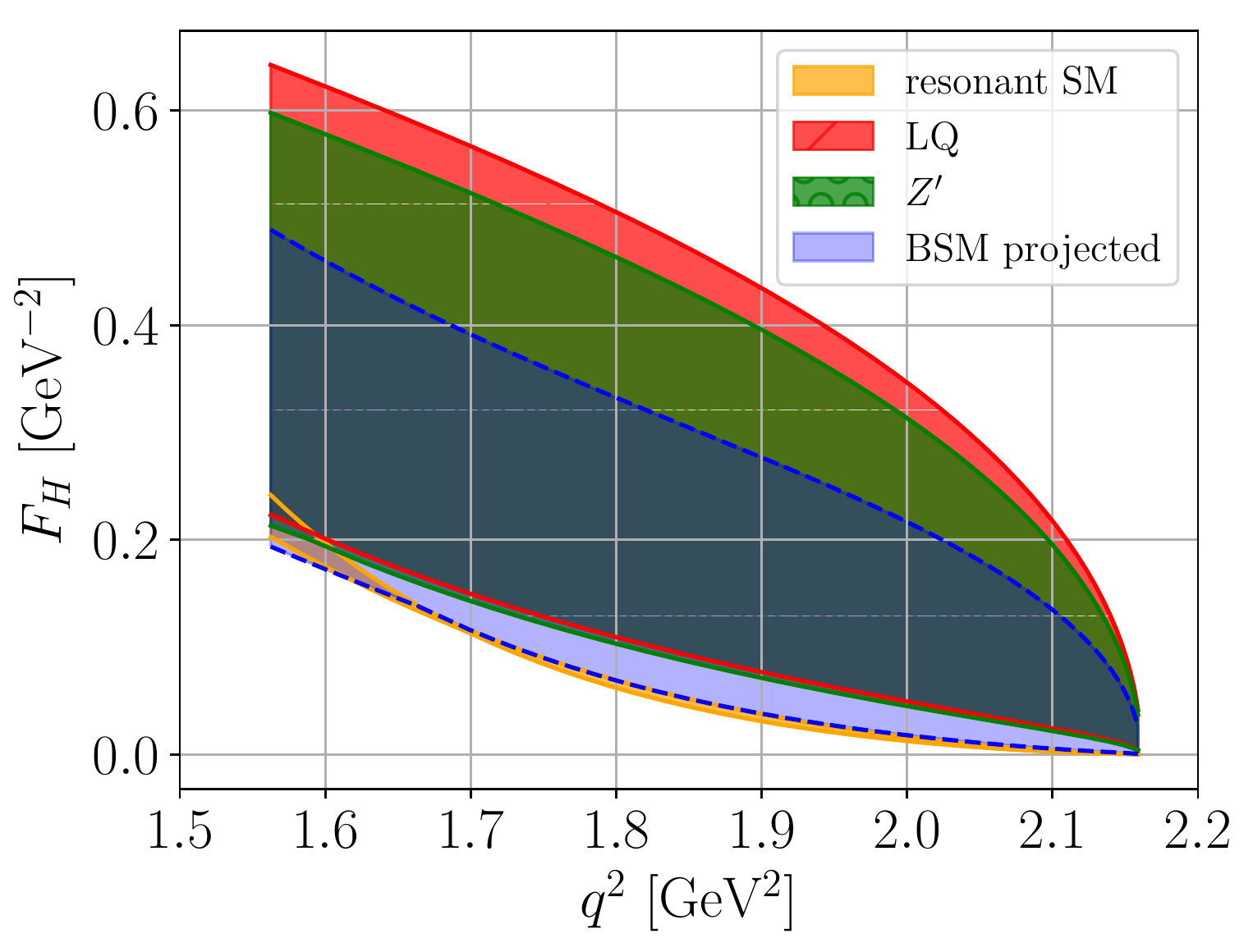}
\caption{$F_H$  in $D^+\to\pi^+ \mu^+ \mu^-$ (plot to the left) and $D_s^+\to K^+ \mu^+ \mu^-$ (plot to the right) decays in the high--$q^2$ region in the SM (thin yellow band).
Also shown are maximal effects in a $Z^\prime$-model (green  bands) for  $C_9+C_9^\prime=-(C_{10}+C_{10}^\prime)=0.63$  and a leptoquark scenario (red bands)
 with $C^\prime_9=-C^\prime_{10}=0.63$, $C_S=C_P=11\,C_T=11\,C_{T5}=-0.0056$, in concordance with kaon and high $p_T$-data and RG-effects, see text for details.
 The lilac bands illustrate the impact of  a less maximal BSM scenario from leptoquarks, or in $Z^\prime$-models, with $C_9+C_9^\prime=-(C_{10}+C_{10}^\prime)=0.1$. }
\label{fig:F_H_models}
\end{figure}
Leptoquark  contributions to  $D^0-\overline{D}^0$ mixing are induced by one-loop box diagrams.
Yet, data on $\Delta m_{D^0}$  results in a somewhat more stringent constraint than the rare decays, \cite{deBoer:2015boa},
\begin{equation}
\big| \lambda_{S_{1,2}}^{ce} \lambda_{S_{1,2}}^{ue*} + \lambda_{S_{1,2}}^{c\mu} \lambda_{S_{1,2}}^{u\mu*} + \lambda_{S_{1,2}}^{c\tau} \lambda_{S_{1,2}}^{u\tau*} \big| \lesssim 0.01 \, \left({M_{S_{1,2}} \over 1~{\rm TeV}}\right) \, ,
\end{equation}
which can be eased for the  individual, lepton-specific terms due to cancellations.
Note that the corresponding mixing bound on the imaginary part is about a factor $0.2$ stronger, see App.~\ref{sec:D_mixing}. 
Also dipole operators are induced by leptoquark-lepton loops \cite{deBoer:2017que}. As we do not consider leptoquarks with couplings to 
left-handed quarks, no chirality-flipping $\tau$-loops are  available, and the resulting $C_7^{(\prime)}$ is below $\mathcal{O}(10^{-3})$ for $S_{1,2}$,
but could reach ${\cal{O}}(10^{-2})$ for $\widetilde{V}_{1,2}$, see \cite{deBoer:2017que} for details.

Note that scenarios with larger coupling $\lambda^{c \mu}_{S_2} \simeq 3.5$   \cite{Fajfer:2015mia,Sahoo:2017lzi}, that could induce significant
$C_{S,P,T,T5}$ together with even a suppressed coupling to doublet quarks,  are excluded by high-$p_T$ studies, 
$\lambda^{c \mu}_{S_2} \lesssim 0.4 M_{S_2} /\mbox{TeV}$ \cite{Greljo:2017vvb}. Taking this constraint into account leptoquark effects in $D \to \pi \mu^+ \mu^-$ decays 
($D_s \to K \mu^+ \mu^-$ decays) in $A_{\rm FB}(q^2)$ do not exceed (few) permille-level. On the other hand, 
$F_H(q^2) \lesssim 0.1 \,   (\mbox{few}  \times 0.1)$  for $D \to \pi \mu^+ \mu^-$ decays ($D_s \to K \mu^+ \mu^-$ decays) in  the high $q^2$-region 
for  $C^\prime_9=-C^\prime_{10}=0.63$ and small, however viable $C_S=C_P=11\,C_T=11\,C_{T5}=-0.0056$, with RG-factors between $C_{S,P}$ and $C_{T,T5}$ from \cite{Hiller:2016kry}, see Fig.~\ref{fig:F_H_models} (red bands). 
For the  $Z^\prime$-model (green bands) with $C_{S,P,T,T5}=0$ but comparable maximal (axial-)vector contributions as the leptoquark one exhibits a similar small window for NP
above the SM prediction (thin yellow band).
The impact of the (pseudo-)scalar and tensor contributions on $F_H$ from viable leptoquark scenarios  is thus very small.

It is apparent from Fig.~\ref{fig:F_H_models} that the SM prediction of the angular observable $F_H$  (\ref{eq:FH})  at high $q^2$ normalized to the high-$q^2$ integrated decay rate  -- unlike in Fig.~\ref{fig:FH-SM}, where the normalization is to the full decay rate -- has very small theory uncertainties. The reason is that the high-$q^2$ region is dominated by a single resonance in $C_9^R$, the $\phi$, whose phase drops out in $F_H$. In the BSM curves the uncertainties, which are given by the band width, arise from interference with the SM, which prohibits the dropping of strong phases.

\begin{figure}[!t]\centering
\includegraphics[width=0.49\textwidth]{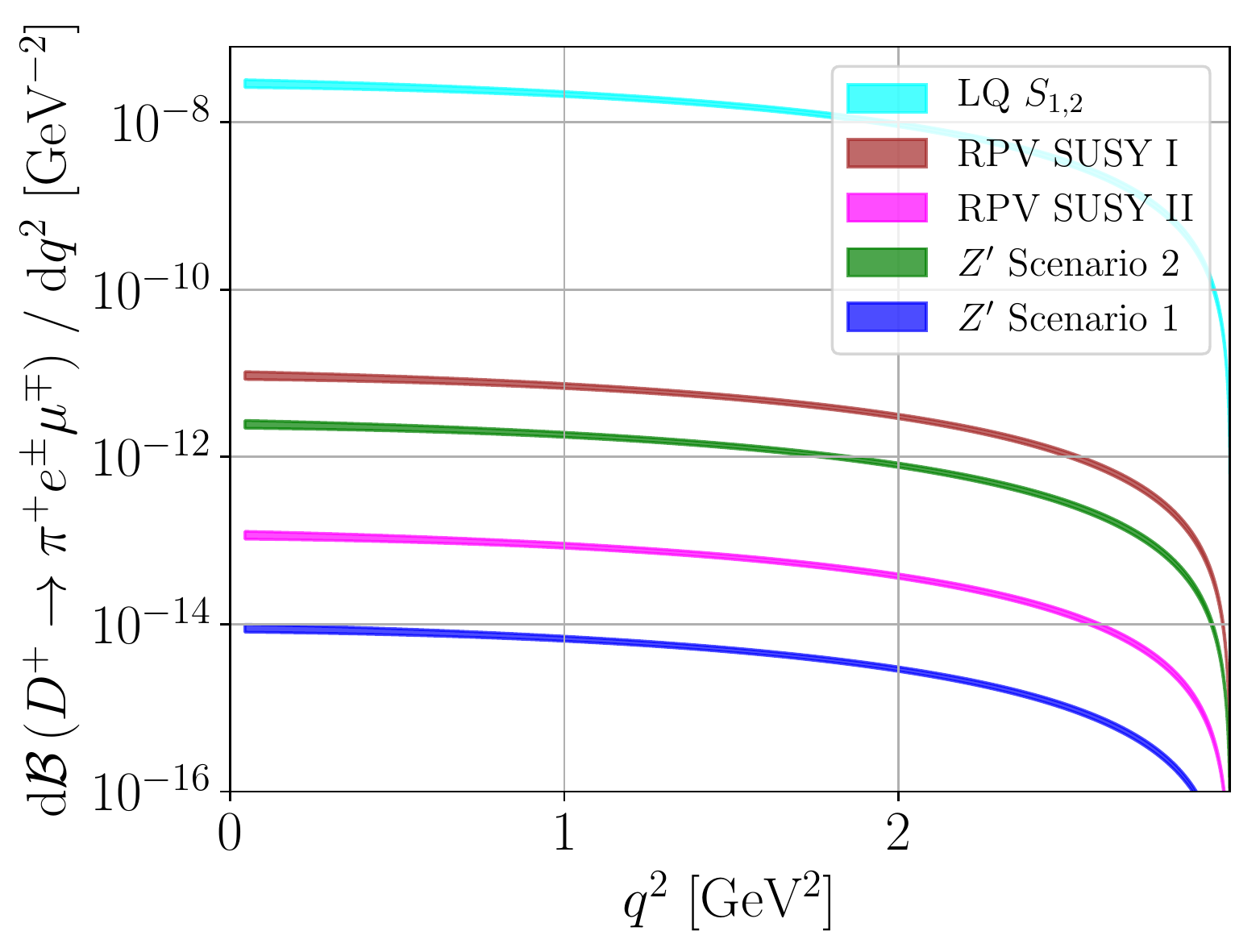}
\includegraphics[width=0.49\textwidth]{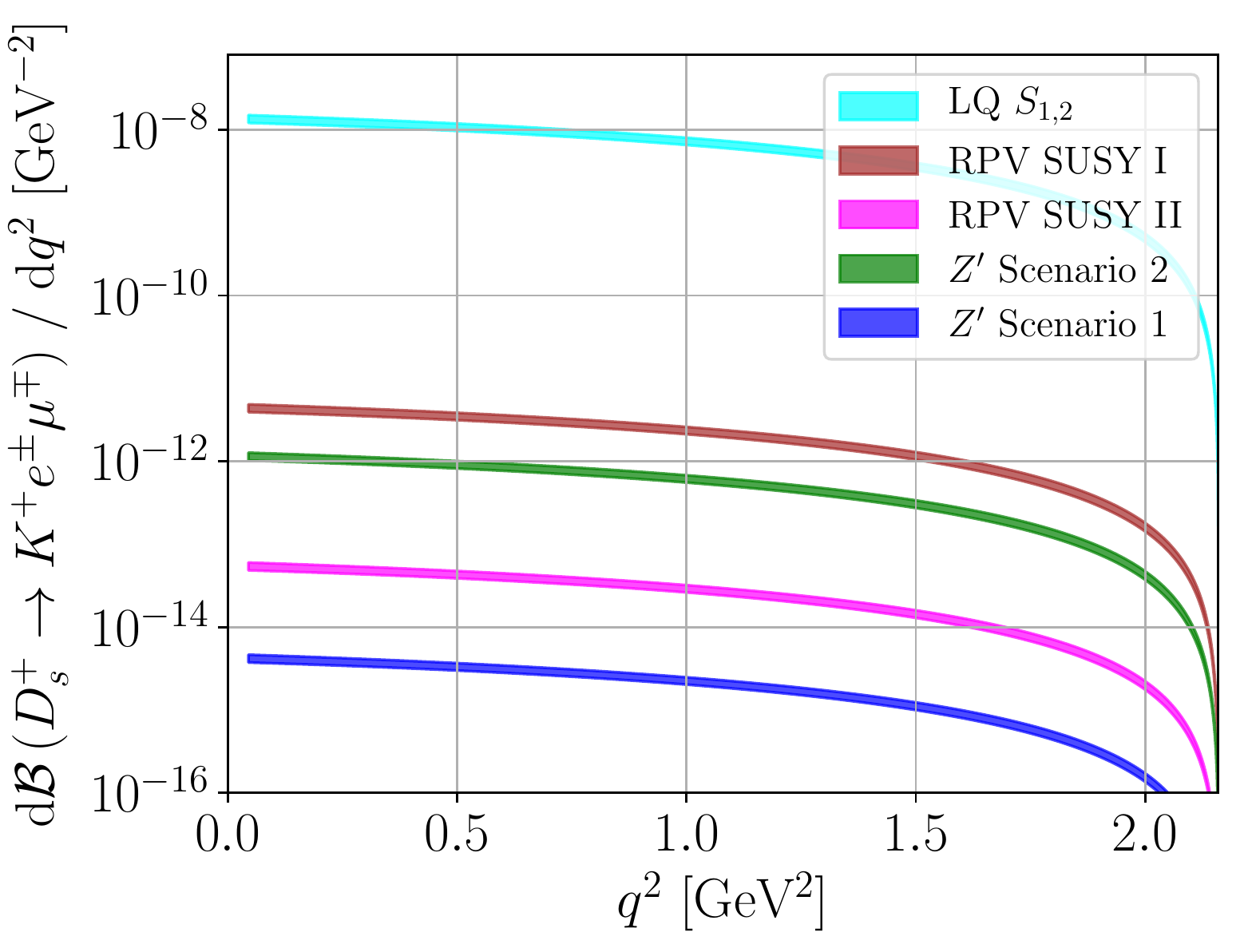}
\caption{The differential branching ratio of $D^+\to\pi^+e^\pm\mu^\mp$ (plot to the left) and $D_s^+\to K^+e^\pm\mu^\mp$ (plot to the right) decays in models
with leptoquarks $S_{1,2}$, $|K_9^\prime|=|K_{10}^\prime|=0.5$  (cyan), for two R-parity violating SUSY benchmarks, $K_9=-K_{10} = 0.009$ (brown) and $K_9=-K_{10}= 0.001$ (magenta), and in  $Z^\prime$-model solution $2$, $K_9+K_9^\prime=-(K_{10}+K_{10}^\prime) = 4.6\times 10^{-3}$ (green), and solution $1$, 
$K_9+K_9^\prime=-(K_{10}+K_{10}^\prime) = 2.8\times 10^{-4}$ (blue).
Solutions $3$-$6,8$ (not shown) give rates near the magenta RPV SUSY II band. See text for details.}
\label{fig:modelLFV}
\end{figure}

Leptoquark signals can hence be observed in $D \to P \ell \ell^{(\prime)}$ decays in lepton-universality tests, with corresponding effects from 
$C_{9,10}^{(\prime)}$ given in Tabs.~\ref{tab:R_ratios} and \ref{tab:R_ratios_Ds},
LFV searches and, if CP-violating, in $A_{\rm CP}$, illustrated in Fig.~\ref{fig:ACP}.
LFV branching ratios from $S_{1,2}$  (cyan) together with SUSY predictions (brown, magenta) and in $Z^\prime$-models (green, blue) are shown in Fig.~\ref{fig:modelLFV}.
There is a small window for NP signals from leptoquarks in the angular observable $F_H$ in the muon modes, shown in Fig.~\ref{fig:F_H_models} by the red bands.

\subsection{Charm reach of SUSY models\label{sec:susy}}

Supersymmetric SM extensions offer two ways for BSM signals in rare charm decays.
One is through  enhanced dipole couplings $C_7^{(\prime)}$ from scalar quark mixing; corresponding contributions to $c \to u \gamma$  decays
have been studied recently in~\cite{deBoer:2017que}, to which we refer for details.
The sensitivity of rare semileptonic $D \to P \ell^+ \ell^-$ decays to $C_7^{(\prime)}$ can be read-off from Fig.~\ref{fig:Br_high} for the branching ratio and from Fig.~\ref{fig:ACP} for the CP-asymmetry using Eq.~\eqref{eq:U}. 
Note that the $D \to P \ell^+ \ell^-$  distributions are sensitive to the sum of dipole coefficients, $C_7 +C_7^\prime$, only. 
This ambiguity can be resolved with polarization studies in $D \to V \gamma$, $V=\rho, K^*, \phi, K_1$ \cite{deBoer:2018zhz,Adolph:2018hde},  
which probe  the fraction of right-handed to left-handed photons.
While the sensitivity window to NP in the branching ratio is rather small, there is large room  in $A_{\rm CP}$ for CP-violating dipole contributions  from supersymmetric flavor violation.

Another possibility to probe SUSY in $c \to u$ FCNCs is with R-parity violating terms $\lambda^\prime LQD$, which induce (axial)-vector couplings with
$C_9=-C_{10}$ through tree level 
exchange of a scalar partner of a singlet down quark, see \cite{Burdman:2001tf} for details; also \cite{Wang:2014uiz}.
However, this unavoidably involves doublet up-type quarks, hence is subject to constraints from rare kaon decays, specifically $K \to \pi \nu \bar \nu$. 
This situation resembles the one of the coupling to left-handed leptons of the scalar leptoquark $S_1$, which has the same quantum numbers as a singlet down quark. 
In SUSY additionally even stronger constraints from $K_L \to e^\pm  \mu^\mp$  apply if the squark doublet masses are smaller or close to the down-singlet ones.
 We illustrate maximal LFV  branching ratios in Fig.~\ref{fig:modelLFV} for $K_9=-K_{10} \simeq 0.009$ (benchmark I) from $K \to \pi \nu \bar \nu$ \cite{Deandrea:2004ae} for decoupled
 doublet squarks and 
 $K_9=-K_{10} \simeq 0.001$ (benchmark II)  from $K_L \to e^\pm  \mu^\mp$ assuming degenerate squarks \cite{Allanach:1999ic,Tanabashi:2018oca}.

To summarize, the only realistic opportunity for SUSY  to signal NP in  $D \to P \ell^+ \ell^-$  decays is
with  CP-violating contributions  in $A_{\rm CP}$, illustrated in Fig.~\ref{fig:ACP},  and in LFV branching ratios in R-parity violating models, shown in Fig.~\ref{fig:modelLFV}.

\subsection{Flavorful \texorpdfstring{$Z^\prime$}{Zprime}--models \label{sec:Z'}}

The effective $Z^\prime$--interaction Hamiltonian part for $c\to u\ell^-\ell^{\prime+}$ processes can be written as 
\begin{equation}
\mathcal{H}_{Z^\prime} \supset \left( g_L^{uc}\, \ubar_L \gamma_\mu c_L + g_R^{uc}\, \ubar_R \gamma_\mu c_R +
g_L^{\ell\ell^\prime}\, \lbar_L \gamma_\mu \ell_L^\prime + g_R^{\ell\ell^\prime}\, \lbar_R \gamma_\mu \ell_R^\prime \right) Z^{\prime\,\mu} + {\rm h.c.} \,.
\end{equation}
The following Wilson coefficients are induced at tree-level
\begin{align}
C_{9/10}^{(\ell\ell)} &= -{\pi \over \sqrt{2} G_F \alpha_e}\, {g_L^{uc} \bigl( g_R^{\ell\ell} \pm g_L^{\ell\ell} \bigr) \over M_{Z^\prime}^2} \,, \quad
C_{9/10}^{\prime\,(\ell\ell)} = -{\pi \over \sqrt{2} G_F \alpha_e}\, {g_R^{uc} \bigl( g_R^{\ell\ell} \pm g_L^{\ell\ell} \bigr) \over M_{Z^\prime}^2} \, ,  \\
K_{9/10}^{(\ell\ell^\prime)} &= -{\pi \over \sqrt{2} G_F \alpha_e}\, {g_L^{uc} \bigl( g_R^{\ell\ell^\prime} \pm g_L^{\ell\ell^\prime} \bigr) \over M_{Z^\prime}^2} \,,  \quad
K_{9/10}^{\prime\,(\ell\ell^\prime)} = -{\pi \over \sqrt{2} G_F \alpha_e}\, {g_R^{uc} \bigl( g_R^{\ell\ell^\prime} \pm g_L^{\ell\ell^\prime} \bigr) \over M_{Z^\prime}^2} \,.
\label{eq:K_Z'}
\end{align}
Using Eqs.~\eqref{eq:Br_D-ll},\eqref{eq:Br_D-emu} and \eqref{eq:NP_constraints_high}, respectively, we obtain the following constraints on the $Z^\prime$--model
\begin{align}
\nonumber
& \bigl|(g_L^{uc} -g_R^{uc}) (g_L^{\mu\mu} - g_R^{\mu\mu}) \big| \lesssim 0.03\, \left({M_{Z^\prime} \over 1~{\rm TeV}}\right)^2 \,, \\ \label{eq:constr_z}
& \bigl|g_L^{uc} -g_R^{uc}\bigr| \sqrt{ \bigl|g_L^{\mu e}\bigr|^2 + \bigl|g_R^{\mu e}\bigr|^2 } \lesssim 0.07\, \left({M_{Z^\prime} \over 1~{\rm TeV}}\right)^2 \,, \\
&  \bigl|g_L^{uc} +  g_R^{uc}\bigr|\,\sqrt{ \bigl|g_L^{\mu \mu}\bigr|^2 + \bigl|g_R^{\mu \mu}\bigr|^2 } \lesssim 0.04\, \left({M_{Z^\prime} \over 1~{\rm TeV}}\right)^2 \,,
\nonumber
\end{align}
where $M_{Z^\prime}$ denotes the mass of the $Z^\prime$.
Stronger bounds, however,  arise from the $Z^\prime$ tree level contribution to $D^0-\overline{D}^0$ mixing, see
App.~\ref{sec:D_mixing} for details (Here and in the following the superscript '$uc$' has been dropped.),
\begin{align}
\label{eq:quadraticform}
 g_L^2+g_R^2-X\,g_L\,g_R \simeq (4.2\pm1.5)\times 10^{-7} \, \left({M_{Z^\prime} \over 1~{\rm TeV}}\right)^2 \,,
\end{align}
with $X\sim20$ for $1 \, {\rm TeV}\lesssim M_{Z^\prime}\lesssim10 \, {\rm TeV}$. 
Main uncertainties are due to the experimental limit on the $D^0-\overline D^0$ mixing parameter and the hadronic matrix elements. 
In the case of only one non-zero coupling present ($g_{L}\neq0$ and $g_{R}=0$ or $g_{R}\neq0$ and $g_{L}=0$), constraints are severe, 
$|g_{L/R}| \lesssim 8 \times 10^{-4} \, (M_{Z^\prime}/ {\rm TeV})$; 
assuming order one muon couplings in Eq.~\eqref{eq:constr_z}, the resulting bounds on the BSM Wilson coefficients read $C_{9/10}\lesssim\mathcal{O}(10^{-2})$ for $M_{Z^\prime}\gtrsim1 \,$TeV, consistent with~\cite{Fajfer:2015mia}. 
However, the presence of both couplings $g_L\neq0$ and $g_R\neq0$ allows to satisfy the mixing constraints even with arbitrarily large values, bounded only by 
Eq.~\eqref{eq:constr_z}, see App.~\ref{sec:D_mixing} for details, if 
\begin{equation}
g_L \approx X g_R \quad \text{or} \quad g_L \approx \frac{1}{X} g_R \,.
\label{eq:linearrelation}
\end{equation}
In the following we show that these conditions can be met in flavorful $Z^\prime$--models by flavor rotations without introducing unnatural hierarchies.
Here, FCNC $c \to u$--transitions arise from non-universal charges, denoted by $F$. Specifically, inducing $g_L$ ($g_R$) requires the charges of the doublet (singlet) charm and up-quarks  to be different from each other
\begin{equation}\label{eq:DeltaF}
\Delta F_{L} =F_{Q_2}- F_{Q_1} \, , \quad 
\Delta F_{R} =F_{u_2}- F_{u_1} \, , 
\end{equation}
\textit{i.e.}, $\Delta F_{L (R)} \neq 0$.
Another requisite ingredient is flavor mixing. Four such rotations between flavor and mass bases exist, those  for up-singlets, $U_u$, for down-singlets $U_d$, and  those for up- and down-doublets, $V_u$ and $V_d$, respectively.
$U_d, U_u,V_u$ and $V_d$ are unitary matrices; in absence of a theory of flavor they are unconstrained, except for $V_u^\dagger V_d=V_{\rm CKM}$.
Here we consider rotations residing  in the up-sector, $U_d=1, V_d =1$, hence $V_u=V_{ \rm CKM}^\dagger$, as flavor rotations in the down sector are subject to strong constraints from the rare kaon decays, \textit{e.g.}~\cite{Buras:2012jb}.
Small mixings, $(V_d)_{12}/(V_u)_{12} \lesssim 10^{-3}$, however, would still be consistent with sizable effects in charm.
To understand the interplay between rotations and charges in order to satisfy Eq.~\eqref{eq:linearrelation}, we simplify the analysis by assuming the third generations to be sufficiently decoupled. 
This allows us  to work with 2 by 2 orthogonal matrices, parametrized by a single angle each, $\theta_u$ and $\Phi_u$ for the up-singlets and -doublets, respectively.
Then,
\begin{equation}
g_L=g_4\,\Delta F_L \cos \Phi_u \sin \Phi_u \, , \quad
g_R=g_4\,\Delta F_R \cos \theta_u \sin \theta_u \, , 
\end{equation}
with $g_4$ the $U(1)^\prime$ gauge coupling strength and, since $V_u=V_{ \rm CKM}^\dagger$, we obtain
\begin{equation}
\frac{g_R}{g_L}  \simeq \frac{  \Delta F_R \cos\theta_u \sin\theta_u}{\Delta F_L  \lambda   }  \, , 
\end{equation}
where $\lambda \sim 0.2$ denotes the Wolfenstein parameter.

We obtain $g_R/g_L=X$ for 
\begin{equation}\label{eq:Xratio}
 \frac{\Delta F_R}{\Delta F_L}\sin 2\theta_u \simeq 8  \,,
\end{equation}
and $g_R/g_L=1/X$ for 
\begin{equation}\label{eq:1overXratio}
 \frac{\Delta F_R}{\Delta F_L} \sin\theta_u \simeq 1/100\,.
\end{equation}
The former case, in the following termed right-handed (RH)-dominated, requires some mild hierarchy between the left versus right charges, whereas the latter case, termed left-handed (LH)-dominated, can be accommodated with mixing alone $\theta_u \sim 10^{-2}$.
In either case, the ratio of left- and right-handed couplings is fixed, 
\begin{align}
\frac{C_{9/10}}{C_{9/10}^\prime},  \frac{K^{(\ell\ell^\prime)}_{9/10}}{K^{(\prime)\,(\ell\ell^\prime)}_{9/10}} \sim X ~(\mbox{LH}) ~~\mbox{or} ~~ \sim 1/X ~(\mbox{RH})\, .
\end{align}

In Tab.~\ref{tab:realistic_Z} we present sample scenarios of $Z^\prime$--models with sizable BSM coefficients, which obey Eq.~\eqref{eq:linearrelation}. 
The models differ in their charge assignments, all of which fulfill anomaly cancellation conditions, see App.~\ref{sec:anomalies} for details. 
For the models presented $ \Delta F_R / \Delta F_L $ ranges within $\sim[0.9,\,35]$. 
Also given are the values of the mixing angle $\theta_u$ for each model. 
Models with $ \Delta F_R / \Delta F_L \geq 8$ can be either RH or LH-dominated, depending on the flavor rotation $\theta_u$.

\begin{table}[!t]
 \centering
  \caption{Scenarios of anomaly-free $Z^\prime$--models and the mixing angle $\theta_u$ for different charge assignments taken from Tab.~\ref{tab:ZprimeModels}. 
  	The primed solutions are RH-dominated, whereas the unprimed ones are LH-dominated.}
  \label{tab:realistic_Z}
  \begin{tabular}{c||c|cc|cc|cc}
 sol. $\#$ & $\Delta F_R$ && $\Delta F_L$ && $g_R/g_L$ case&& $\theta_u$ \\
 \hline
 1 & 3 && 2 && $1/X$ && 0.008 \\
 2 & 12 && 9 && $1/X$ && 0.009 \\
 $3^\prime$ & 35 && 1 && $X$ && 0.122 \\
 3 & 35 && 1 && $1/X$ && 0.0003 \\
 4 & 3 && 3 && $1/X$ && 0.011 \\
 5 & 3 && 3 && $1/X$ && 0.011 \\
 6 & 15 && 16 && $1/X$ && 0.012 \\
 7 & 0 && 0 && - && - \\
 $8^\prime$ & 18 && 1 && $X$ && 0.244  \\
 8 & 18 && 1 && $1/X$ && 0.0006\\
 \end{tabular}
\end{table}

\begin{figure}[!t]\centering
\includegraphics[width=0.7\textwidth]{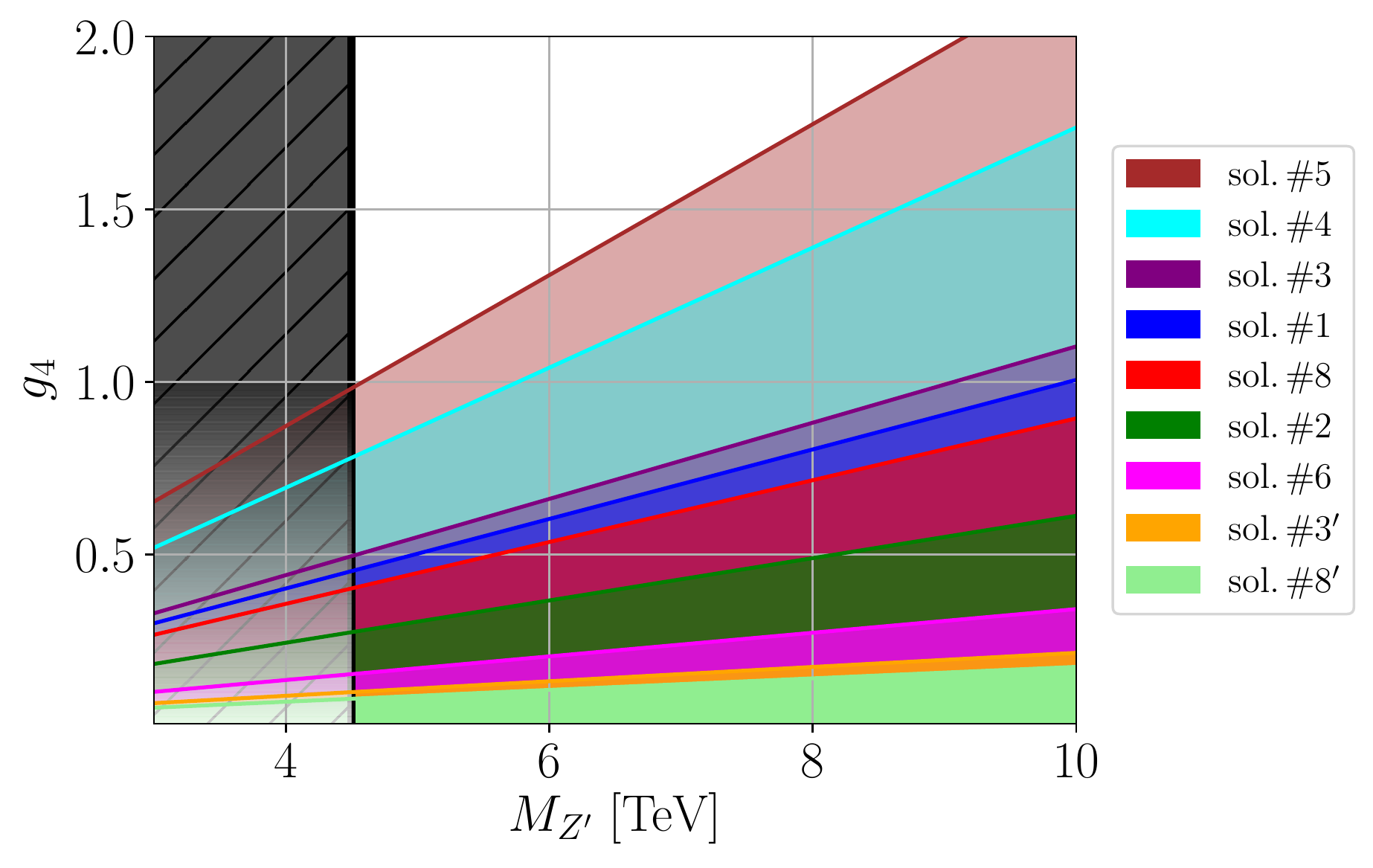}
\caption{Upper limits  on the $U(1)^\prime$ gauge coupling  $g_4$ as a function of the $Z^\prime$ mass (\ref{eq:g4}) for the models in Tab.~\ref{tab:realistic_Z}. 
The black region is excluded by direct searches in dimuon and dielectron spectra~\cite{Tanabashi:2018oca}. 
For lower values of $g_4$, the bounds become model-dependent as indicated by the lighter color, see main text.}
\label{fig:zmodels}
\end{figure}

We work out constraints on the parameters of the concrete $Z^\prime$--models.
The constraints from $D \to \pi \mu^+ \mu^-$ branching ratios in Eq.~\eqref{eq:NP_constraints_high},  using $g_L^{\ell\ell}=g_4 F_{L_\ell}, \, g_R^{\ell\ell}=g_4 F_{e_\ell}$, can be written as
\begin{equation}
g_4^4 \left( \lambda \, \Delta F_L \right)^2 \left\{ 1 + \left( \frac{ \Delta F_R \sin 2 \theta_u }{ \Delta F_L 2 \lambda} \right) \right\}^2 \left(F_{L_2}^2 + F_{e_2}^2\right)  \lesssim 1.8\times10^{-3}\,
\, \left({M_{Z^\prime} \over 1~{\rm TeV}}\right)^4\,.
\end{equation}
Imposing in addition the constraints from $D^0-\overline{D}^0$ mixing fixes $g_4/M_{Z^\prime}$ for each model, with two solutions for those with large $ \Delta F_R / \Delta F_L $. 
We obtain
\begin{equation} \label{eq:g4}
g_4^2  \lesssim \frac{0.21}{\Delta F_L\,\sqrt{F_{L_2}^2 + F_{e_2}^2}}\,\left(\frac{M_{Z^\prime}}{1\,\text{TeV}}\right)^2 \times \begin{cases} (1+X)^{-1} ~~(\mbox{RH}) \\ (1+1/X)^{-1} ~~(\mbox{LH}) \end{cases} \quad .
\end{equation}
In Fig.~\ref{fig:zmodels} the BSM coupling $g_4$ is shown as a function of the $Z^\prime$ mass.
Each line corresponds to an upper limit in the scenarios from Tab.~\ref{tab:realistic_Z} with one specific choice for the charges $F_{L_{1}}\,(F_{L_{2}})$ and $F_{e_{1}}\,(F_{e_{2}})$ for electrons (muons). 
Constraints for RH cases are stronger than for corresponding LH ones in Eq.~\eqref{eq:g4}. 
The black region shows the exclusion region due to resonance searches in dilepton searches of $M_{Z^\prime}>4.5\,$TeV~\cite{Tanabashi:2018oca}. 
The true lower bound for the $Z^\prime$ mass is different for every solution due to the different quark charge assignments and overall coupling strength and in general different for electrons and muons. 
Therefore, part of the region $g_4<0.5$ and $M_{Z^\prime}< 4.5\,$TeV might still be viable or  constrained by other searches~\cite{Smolkovic:2019jow}.

$Z^\prime$-contributions to charged LFV arise by a similar misalignment between flavor and mass bases as for the quarks.
Likewise, there is no left-handed charged LFV if the PMNS-matrix is only due to rotations in the neutrino sector.
In the presence of charged lepton rotations, the decays $(i)$\,$\tau \to (\mu,e)  \ell  \ell$ with $\ell=e,\mu$\,, as well as $(ii)$\,$\mu \to eee$ and $\mu \to e\gamma$  
provide  the most stringent constraints~\cite{Tanabashi:2018oca}.
Charm constraints, shown in Fig.~\ref{fig:zmodels}, continue to be more restrictive for the scenarios in Tab.~\ref{tab:realistic_Z} if the charged lepton rotations are ${\cal{O}}(10^{-2}$--$10^{-3})$ and ${\cal{O}}(10^{-4}$--$10^{-5})$ or smaller for the modes $(i)$ and $(ii)$, respectively.  Upper limits on  $D \to \pi e^\pm \mu^\mp$ and $D_s \to K e^\pm \mu^\mp$ branching ratios consistent with leptonic LFV data are shown in Fig.~\ref{fig:modelLFV}.

As we work in the limit $V_u=V_{\text{CKM}}^\dagger$ we automatically avoid $Z^\prime$-FCNCs with down-type doublet quarks and hence are unable to address present discrepancies in semileptonic rare $B$-decays between the data and the SM.
We would like to point out, however, that simultaneous left-handed contributions to $|\Delta c|=|\Delta u|=1$ and $|\Delta b|=|\Delta s|=1$ 
arise if $V_{us}$ stems from up-rotations and $V_{cb}$ from down-rotations. 
A detailed investigation is beyond the scope of this work. 

To summarize, flavorful $Z^\prime$-models can signal NP in lepton-universality tests, with corresponding effects from 
$C_{9,10}^{(\prime)}$ 
given  in Tabs.~\ref{tab:R_ratios} and \ref{tab:R_ratios_Ds}. As we neglect in the $U(1)^\prime$-study CP-phases in the flavor rotations there are no CP-asymmetries.
We illustrate maximal effects in $F_H$ in the muon modes  in Fig.~\ref{fig:F_H_models}  for $C_9+C_9^\prime=-(C_{10}+C_{10}^\prime)=0.63$ (green bands), leaving 
a small window above the SM at high $q^2$. Charged LFV effects in $Z^\prime$-models are model-dependent, with branching ratios  $\mathcal{B}\left(D \to  \pi \mu^\pm e^\mp \right) \lesssim \text{few}\,10^{-12}$ and $\mathcal{B}\left(D_s \to K \mu^\pm e^\mp \right) \lesssim 10^{-12}$, see Fig.~\ref{fig:modelLFV}.

\section{Conclusions}\label{sec:summary}

We worked out the NP sensitivity of $|\Delta c|=|\Delta u|=1$ SM null tests based on lepton-universality in Sec.~\ref{sec:BSM_lfu}, 
angular distributions in Sec.~\ref{sec:BSM_nulltests}, approximate CP--symmetry in Sec.~\ref{sec:BSM_cpasym} 
and lepton flavor conservation in Sec.~\ref{sec:BSM_lfv} in $D \to \pi \ell \ell^{(\prime)}$ and $D_s \to K \ell \ell^{(\prime)}$ decays into electrons or muons.
These studies provide a rich arena to test the SM and look for different kinds of BSM phenomena:

For $q^2$--values above the $\phi$, a region with lesser resonance pollution, see Fig.~\ref{fig:dGamma_SM}, branching ratios are about $\mathcal{O}(10^{-9}$--$10^{-8})$.
Here, lepton universality tests $R_\pi^D$ and $R_K^{D_s}$ can be sizable, up to ${\mathcal{O}}(10)$, and cleanly signal
BSM physics which exhibits more differences between electrons and muons than their mass, see Tabs.~\ref{tab:R_ratios} and \ref{tab:R_ratios_Ds}.
The angular observables $A_{\rm FB}$ and $F_H$ and CP--asymmetries, the latter around the $\phi$ as well, can be of the order one, see Figs.~\ref{fig:AFB-FH} and \ref{fig:ACP}.
Due to the somewhat smaller $q^2$--integrated decay rates, which appear in the null test observables $F_H, A_{\rm FB}$ and $A_{\rm CP}$ in the denominator,
NP effects in $D_s \to K \ell^+\ell^-$ decays are pronounced relative to the ones in $D \to \pi \ell^+ \ell^-$.
Semileptonic LFV branching ratios can reach $\mathcal{O}(10^{-7})$ model-independently, see Fig.~\ref{fig:LFV}.

Concrete models with potentially large (pseudo)scalar and tensor contributions, which could be signaled with angular observables, 
are scalar leptoquarks with representations $S_1$ and $S_2$. 
However, it is exactly those representations that allow for chirality flipping operators that are unavoidably subject to tight constraints from the kaon sector.
The largest effects in  charm decays within leptoquark models
are therefore from couplings to right-handed up-quarks, inducing primed $\text{(axial-)vector}$ operators.
Both right-handed and left-handed {(axial-)vector} couplings $C_{9,10}^{(\prime)}$ are generically induced in $Z^\prime$--models. 
We give explicit examples with generation-dependent $U(1)^\prime$-charges that are anomaly-free (App.~\ref{sec:anomalies}).

Sizable BSM {(axial-)vector} couplings, as in leptoquarks $S_{1,2}, \widetilde{V}_{1,2}$ and flavorful $U(1)^\prime$-models, can be probed in all null test observables analyzed in this work, except for $A_{\rm FB}$.
There is a small window for NP in  $F_H$ in the muon modes at high $q^2$, shown in Fig.~\ref{fig:F_H_models}.
SUSY models with flavor mixing are prime suspects for chirally-enhanced dipole contributions. 
CP-violating BSM dipole operators in semileptonic modes can be probed with $A_{\rm CP}$. 
The sensitivity to the dipole operators is similar to the one of the four-fermion ones since both types appear in the combination~\eqref{eq:U}, which contains the sum of $C_7 $ and $C_7^\prime$. 
Polarization studies in $D \to V \gamma$, $V=\rho, K^*, \phi, K_1$~\cite{deBoer:2018zhz,Adolph:2018hde}, on the other hand,
are sensitive to the fraction of right-handed to left-handed photons. R-parity violating SUSY LFV branching ratios can reach  $\mathcal{O}(10^{-11})$, see Fig.~\ref{fig:modelLFV}.
Each BSM model's section, Sec.~\ref{sec:LQ} (leptoquarks), Sec.~\ref{sec:susy} (SUSY)
and Sec.~\ref{sec:Z'} (flavorful $Z^\prime$-models) ends with a brief summary of possible "smoking gun"-observables for new physics.

The analysis in $U(1)^\prime$--models explicitly highlights the complementarity between charm and $K$ and $B$--physics. 
Tree level $Z^\prime$--induced FCNCs arise only if the quark flavor basis is sufficiently misaligned with the up quark mass basis, and CKM stems predominantly from up-rotations. 
Looking for BSM physics in charm can therefore provide unique insights into the origin of flavor structure of fundamental matter.

\section*{Acknowledgments}
We would like to thank Stefan de Boer, Javier Fuentes-Martin, Vittorio Lubicz, Olcyr Sumensari and Xinshuai Yan for useful discussions. 
A.T. has been supported by the DFG Research Unit FOR 1873 ``Quark Flavour Physics and Effective Field Theories'' 
and M.G. by the ``Studienstiftung des deutschen Volkes''.
G.H. gratefully acknowledges the hospitality of the theory group at SLAC, where parts of this work have been done.

\appendix
\section{Form Factors}\label{app:FF}

The hadronic matrix elements for $D\to P$ transitions are parametrized in terms of three form factors, $f_{+\,,0\,,T}(q^2)$, defined as~\footnote{The normalization of the tensor matrix element in Ref.~\cite{deBoer:2015boa} is not $m_D+m_P$ but $m_D$. This in turn introduces different denominators in Eqs.~\eqref{eq:dGamma},\eqref{eq:AFB} and \eqref{eq:FH} in the terms involving $f_T$ with respect to Eqs.~(D1)--(D3) of Ref.~\cite{deBoer:2015boa}.}
\begin{align}
& \langle P(k)|\bar u\gamma^\mu c|D(p)\rangle = \left[ (p+k)^\mu - {m_D^2-m_P^2 \over q^2} q^\mu \right] f_+(q^2) + q^\mu {m_D^2-m_P^2 \over q^2} f_0(q^2) \label{eq:me_vector_current} \,, \\
& \langle P(k)|\bar u\sigma^{\mu\nu} (1\pm\gamma_5) c|D(p) \rangle = -\text{i}\,(p^\mu k^\nu - k^\mu p^\nu \pm \text{i}\,\epsilon^{\mu\nu\rho\sigma}p_\rho k_\sigma) {2f_T(q^2) \over m_D+m_P} \,. \label{eq:me_tensor_current}
\end{align}

\begin{figure}[!t]\centering
\includegraphics[width=0.5\textwidth]{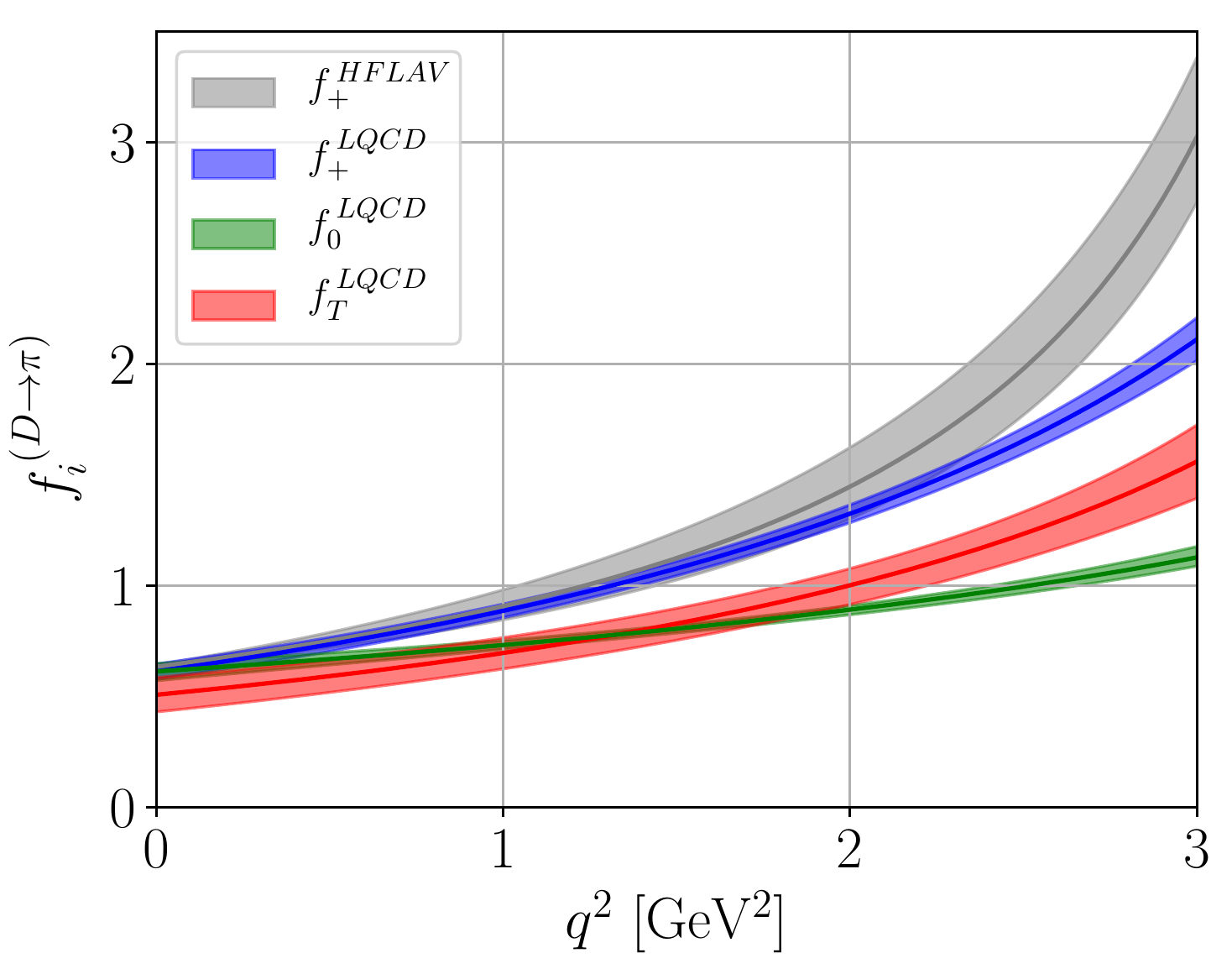}
\caption{The $D\to\pi$ form factors $f_+,f_0,f_T$ from lattice QCD~\cite{Lubicz:2017syv, Lubicz:2018rfs} and $f_+$ from HFLAV~\cite{Amhis:2016xyh} (gray band) with  $1\sigma$ theoretical uncertainties.}
\label{fig:FF}
\end{figure}

In this work we use the $D\to\pi$ form factors computed recently by Lubicz et al.~\cite{Lubicz:2017syv, Lubicz:2018rfs} using lattice QCD (LQCD), parametrized in terms of the $z$--expansion as, $i=+,0,T$,
\begin{align}
\label{FF_para}
f_i(q^2) = {1 \over 1-P_i\,q^2} \biggl[ f_i(0) + c_i\, (z(q^2)-z(0)) \, \biggl( 1 + {z(q^2) + z(0) \over 2} \biggr) \biggr]\,,
\end{align}
where
\begin{align}
z(q^2) = {\sqrt{t_+-q^2} - \sqrt{t_+-t_0} \over \sqrt{t_+-q^2} + \sqrt{t_+-t_0}}\,, \quad
t_\pm=(m_D\pm m_P)^2\,, \quad
t_0 = (m_D+m_P)(\sqrt{m_D}-\sqrt{m_P})^2 \,. 
\end{align}
The numerical values of $f_i(0)$, $c_i$ and $P_i$ parameters together with their uncertainties and covariance matrices are given in~\cite{Lubicz:2017syv, Lubicz:2018rfs}~\footnote{By comparing the diagonal covariance matrix elements of Tab.~7 in Ref.~\cite{Lubicz:2017syv} with $\sigma_i^2$ from Tab.~6, we notice that the ordering of parameters in rows/columns of Tab.~7 must be  $\{f(0),\,c_0,\,c_+,\,P_S,\,P_V\}$ rather than $\{f(0),\,c_+,\,P_V,\,c_0,\,P_S\}$. 
We thank Vittorio Lubicz for confirmation. Here, $P_{V/S}$ corresponds to $P_{+/0}$ in Eq.~\eqref{FF_para}.}. 
In particular, $P_{0,\,T}$ and $c_T$ have large uncertainties, about 40\% and 90\%, respectively, such that na\"ive variation of parameters independently produces larger(huge) uncertainties for $f_{+,0}$($f_T$) at high $q^2$. 
Taking into account correlations, we present in Fig.~\ref{fig:FF} the $D\to\pi$ form factors within $1\sigma$ uncertainties.
Also shown (gray band) is $f_+$ obtained from the HFLAV collaboration~\cite{Amhis:2016xyh}, parametrized as
\begin{equation}
f_+(q^2) = {1 \over \phi(q^2)} \sum_{k=0}^2 a_k\,z^k(q^2) \,,
\end{equation}
where the parameters $r_k=a_k/a_0$ are extracted from data on semileptonic $D\to \pi\ell\overline\nu$ decays from several experiments.
Both results for $f_+$ are in good agreement with each other up to $q^2  \gtrsim  2.3 \, \mbox{GeV}^2$. 
Towards the low recoil end point, $f_+^{\rm HFLAV}/f_+^{\rm LQCD}\simeq1.4$ at $q_{\rm max}^2=(m_D-m_\pi)^2$.
Symmetry breaking effects in the form factor relation $f_T=  f_+(1+\mathcal{O}(\alpha_s,\Lambda_{\rm QCD}/m_c))$ amount to about ${\cal{O}}(30 \%)$ in the LQCD results.
We assume the $D_s\to K$ and $D\to\pi$ form factors to be the same, supported by the lattice study~\cite{Koponen:2012di}. The $D^0\to \pi^0$  form factors receive an 
additional  isospin 
factor $f_i\to f_i/\sqrt{2}$.

\begin{figure}[!t]\centering
\includegraphics[width=0.5\textwidth]{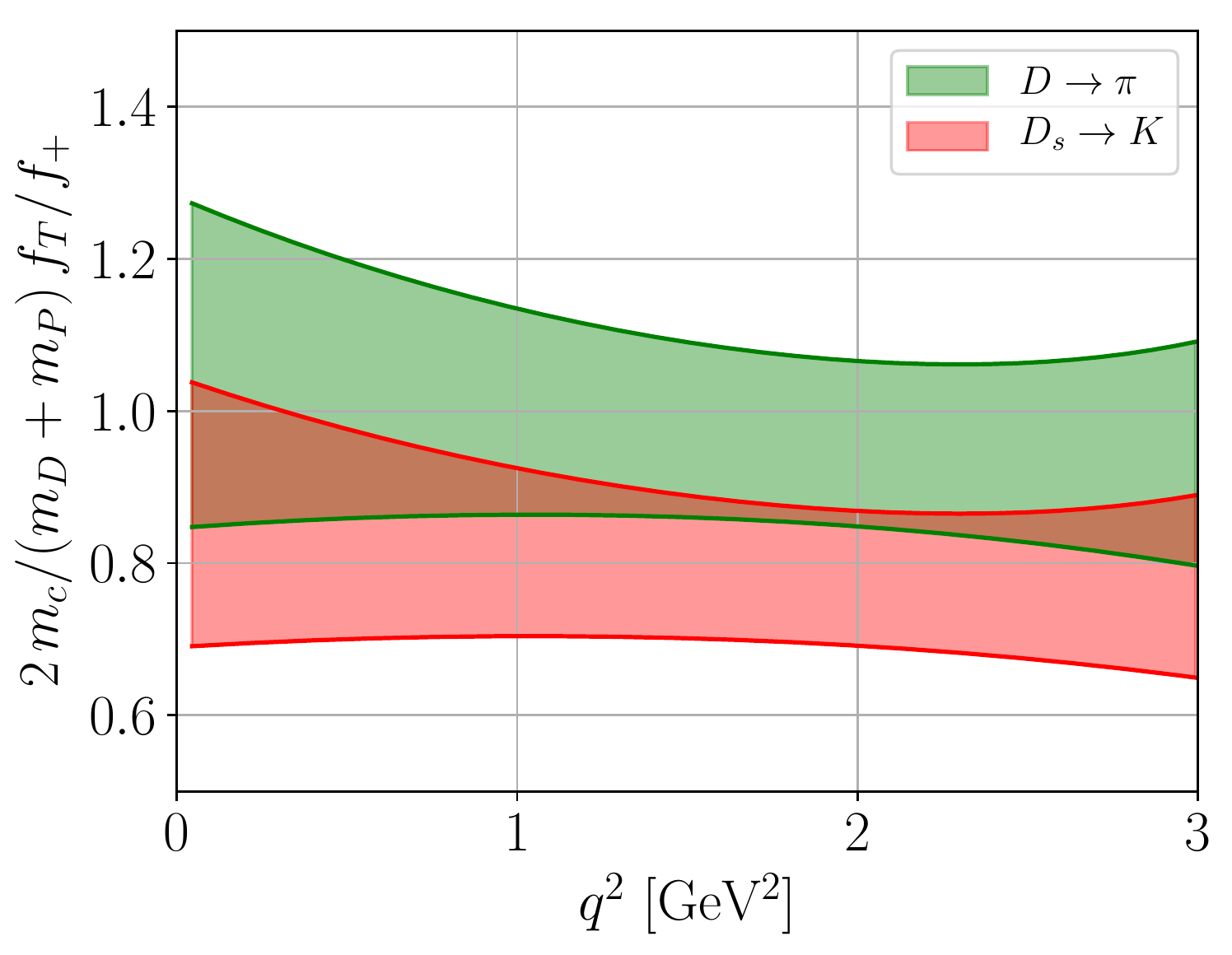}
\caption{The $q^2$--dependence of the factor $\gamma=\frac{2m_c}{m_D+m_P}\frac{f_T}{f_+}$, defined in Eq.~\eqref{eq:U} for $D\to \pi$ (green) and $D_s\to K$ (red).}
\label{fig:c7c9}
\end{figure}

In Fig.~\ref{fig:c7c9} we show the factor $\gamma=\frac{2m_c}{m_D+m_P}\frac{f_T}{f_+}$, defined in Eq.~\eqref{eq:U} for the $D\to \pi$ ($D_s\to K$) transition in green (red).
As noted previously, $\gamma \approx 1$ is a good approximation in the full $q^2$--range.

\section{Constraints from \texorpdfstring{$D^0-\overline{D}^0$}{D-Dbar} mixing \label{sec:D_mixing}}

We compute constraints from the current world average of the $D^0-\overline{D}^0$ mixing parameter $x_D$~\cite{Amhis:2016xyh}
\begin{align}
\label{eq:mixingconstr}
x_D^{\text{exp}}=(4.1^{+1.4}_{-1.5})\times 10^{-3} \,, 
\end{align}
defined as
\begin{align}
\begin{split}
x_D=\frac{\Delta m_{D^0}}{\Gamma_{D^0}}=\frac{2\,|M_{12}|}{\Gamma_{D^0}}=\frac{2}{\Gamma_{D^0}}\,\frac{1}{2m_{D^0}}\,\langle D^0\vert \mathcal{H}_{\rm eff}^{\Delta c=2}\vert\overline{D}^0\rangle\,. 
\end{split}
\end{align}
Here, $\mathcal{H}^{\Delta c=2}_{\rm eff}=\sum_i c_i Q_i$ and~\cite{Golowich:2009ii}
\begin{align}
\begin{split}
Q_1&=(\overline u_L\gamma_\mu c_L)(\overline u_L\gamma^\mu c_L)\,,\quad Q_5=(\overline u_R\sigma_{\mu\nu} c_L)(\overline u_R\sigma^{\mu\nu} c_L)\,,\\
Q_2&=(\overline u_L\gamma_\mu c_L)(\overline u_R\gamma^\mu c_R)\,,\quad Q_6=(\overline u_R\gamma_\mu c_R)(\overline u_R\gamma^\mu c_R)\,,\\
Q_3&=(\overline u_L c_R)(\overline u_R c_L)\,,\quad\quad\quad \,Q_7=(\overline u_L c_R)(\overline u_L c_R)\,,\\
Q_4&=(\overline u_R c_L)(\overline u_R c_L)\,,\quad\quad\quad \,Q_8=(\overline u_L\sigma_{\mu\nu} c_R)(\overline u_L\sigma^{\mu\nu} c_R)\,.
\end{split}
\end{align}
In the $Z^\prime$--model the following $\Delta C=2$ Wilson coefficients are induced at the scale $\mu =M_{Z^\prime}$
\begin{align}
\label{eq:zmodelckoeffs}
 c_1(M_{Z^\prime})  &= \frac{g_L^2}{2M_{Z^\prime}^2}\,, ~
 c_2(M_{Z^\prime})  = \frac{g_L\,g_R}{M_{Z^\prime}^2}\,,~
 c_6(M_{Z^\prime})  = \frac{g_R^2}{2M_{Z^\prime}^2}\,.
\end{align}
The operator $Q_3$ is radiatively induced and needs to be taken into account as well. At the scale $\mu=3$~GeV the $Z^\prime$--contribution is given as~\cite{Golowich:2007ka, Golowich:2009ii}
\begin{align}
\begin{split}
\label{eq:xD}
x_D^{Z^\prime}=\frac{1}{\Gamma_{D^0} m_{D^0}}\,&\left[\phantom{\frac12}r_1\,c_1(M_{Z^\prime})\,\langle Q_1\rangle + \sqrt{r_1}\,c_2(M_{Z^\prime})\, \langle Q_2\rangle \right. \\
&\left.  + \,\frac{2}{3}\,c_2(M_{Z^\prime})\,(\sqrt{r_1}-r_1^{-4})\,\langle Q_3 \rangle + r_1\,c_6(M_{Z^\prime})\,\langle Q_6\rangle\phantom{\frac12}\right]\,,
\end{split}
\end{align}
with the renormalization factor
\begin{align}
r_1=\left(\frac{\alpha_s(M_{Z^\prime})}{\alpha_s(m_t)}\right)^{2/7}\left(\frac{\alpha_s(m_t)}{\alpha_s(m_b)}\right)^{6/23}\left(\frac{\alpha_s(m_b)}{\alpha_s(\mu)}\right)^{6/25}\,,
\end{align}
and the hadronic matrix elements computed at $\mu=3\,$GeV~\cite{Bazavov:2017weg}
\begin{align}
\begin{split}
\langle Q_{1} \rangle &= \phantom{-}0.0805(55) =\langle Q_{6} \rangle\,,\\
\langle Q_2 \rangle &= -0.2070(142)\,,\\
\langle Q_3 \rangle &= \phantom{-}0.2747(129)\,.
\end{split}
\end{align}
In terms of the $Z^\prime$--model parameters one obtains from Eq.~\eqref{eq:xD}
\begin{align}
\label{eq:xD2}
x_D^{Z^\prime}=\frac{r_1 \langle Q_1 \rangle }{2\,\Gamma_{D^0}\,m_{D^0} }\, \frac{ g_L^2 +g_R^2 - X g_L g_R }{M_{Z^\prime}^2} \, , 
\end{align}
where
\begin{align}
X&=-2\,\left(\sqrt{r_1}\,\langle Q_2\rangle + \,\frac{2}{3}\,(\sqrt{r_1}-r_1^{-4})\,\langle Q_3 \rangle \right) \left(r_1\langle Q_1\rangle\right)^{-1}\,,
\end{align}  
and recovers Eq.~\eqref{eq:quadraticform}.
Numerically, $X= 19.2,\,24.0,\,26.2$ for $M_{Z^\prime}=1,5,10$~TeV, respectively.
To obtain the approximations~\eqref{eq:linearrelation}, we rewrite Eq.~\eqref{eq:quadraticform} as
\begin{align}
\vert g_L\vert=\vert g_R\vert \,\left(\frac{X}{2}\pm \,\sqrt{\left(\frac{X^2}{4}-1\right)+\frac{\tilde x}{\vert g_R\vert^2}}\right)\,, \quad \tilde{x}=\frac{2\,x_D^{\text{exp}}\,\Gamma_{D^0}\,m_{D^0}\,M^2_{Z^\prime}}{r_1\langle Q_1\rangle}\,,
\end{align} 
and employ $\tilde x \ll \vert g_R\vert^2$ and $4/X^2 \ll 1$.

Experimental constraints on CP-violation  in  \texorpdfstring{$D^0-\overline{D}^0$}{D-Dbar} mixing,  $x_{12} \sin \phi_{12} \lesssim 2 \times 10^{-4}$, are stronger than \eqref{eq:mixingconstr} by about $\sim 0.04$~\cite{Amhis:2016xyh,Gedalia:2009kh}. Both SUSY and leptoquark models contribute to $\Delta C=2$ Wilson coefficients
with the square of the couplings relevant for $\Delta C=1$ coefficients. For instance,  $C_7^{(\prime)}$  in SUSY is proportional to one power of flavor-violating scalar
quark mass terms whereas
mixing is induced at second order. Constraints  on the  $\Delta C=1$ couplings  from mixing allowing for order one phases are hence about $\sim 0.2$ stronger  than
without CP-phases.

\section{Anomaly-free \texorpdfstring{$Z^\prime$}{Zprime}--models \label{sec:anomalies}}

We consider  $Z^\prime$--extensions of the  SM with generation-dependent
$U(1)^\prime$--charges $F_{\psi_i}$  for quarks and leptons  $\psi=Q,u,d,L,e,\nu$ in representations under $SU(3)_C \times SU(2)_L \times U(1)_Y \times U(1)^\prime$ 
\begin{align} 
\begin{split}
Q_i & \sim (3, 2, 1/6,F_{Q_i})\,,  \quad
u_i \sim (3, 1, 2/3, F_{u_i})\,,  \quad 
d_i \sim (3, 1, -1/3, F_{d_i})\,,  \\
L_i & \sim (1, 2, -1/2, F_{L_i})\,,   \quad 
e_i \sim (1, 1, -1,F_{e_i})\,,  \quad 
\nu_i \sim (1, 1, 0,F_{\nu_i})\,,
\end{split}
\end{align}
where we allow for the possibility of having three RH neutrinos $\nu$.
The charges $F_{\psi_i}$ are subject to constraints from gauge anomaly cancellation conditions (ACCs).
An excellent introduction to the subject is given in~\cite{Bilal:2008qx}, for recent phenomenological applications, see, for instance,~\cite{Ellis:2017nrp,Allanach:2018vjg,Rathsman:2019wyk,Costa:2019zzy}. 
The ACCs read~\cite{Allanach:2018vjg}:
\begin{align}
SU(3)_C^2 \times U(1)_F^\prime : \quad \quad & \sum_{i=1}^3 \, \left(2F_{Q_i} - F_{u_i} - F_{d_i} \right)=0 \,, \label{eqn:su3squ1f} \\
SU(2)_L^2 \times U(1)_F^\prime  : \quad \quad & \sum_{i=1}^3 \, \left(3 F_{Q_i} + F_{L_i} \right)=0 \,, \label{eqn:su2squ1f} \\
U(1)_Y^2 \times U(1)_F^\prime  : \quad \quad & \sum_{i=1}^3 \, \left(F_{Q_i} + 3 F_{L_i} - 8 F_{u_i} - 2 F_{d_i} - 6 F_{e_i} \right)=0 \,,\label{eqn:ysqu1f} \\ 
\text{gauge-gravity:}  \quad \quad &\sum_{i=1}^3 \, \left(6 F_{Q_i} + 2 F_{L_i} - 3 F_{u_i} - 3 F_{d_i} - F_{e_i} - F_{\nu_i} \right)=0 \,,  \label{eqn:gravsqu1f} \\
U(1)_Y \times U(1)^{\prime 2}_F: \quad \quad & \sum_{i=1}^3 \,  \left(F^2_{Q_i} - F^2_{L_i} - 2 F^2_{u_i} + F^2_{d_i} + F^2_{e_i} \right)=0 \,,\label{eqn:quad} \\ 
U(1)_F^{\prime 3}: \quad \quad & \sum_{i=1}^3 \, \left(6 F^3_{Q_i} + 2 F^3_{L_i} - 3 F^3_{u_i} - 3 F^3_{d_i} - F^3_{e_i} - F^3_{\nu_i} \right)=0 \,.  \label{eqn:cubic} 
\end{align}
Since they are SM singlets, the RH neutrinos only appear in Eqs.~\eqref{eqn:gravsqu1f},\eqref{eqn:cubic}. 
In the SM+$3\nu_R$ (the SM), the 18 (15) charges are constrained by 6 ACCs.
Important features of the ACCs and their solutions are (see~\cite{Allanach:2018vjg} and references therein for details):
\begin{itemize}
	\item \textit{Rational solutions}: We assume that all $U(1)^\prime$--charges are rational numbers $F_\psi \in \mathbb{Q}$\,.
	
	\item \textit{Rescaling invariance}: Any solution of the ACCs  can be rescaled by any rational number $k \in \mathbb{Q}$\,, $F_\psi \rightarrow kF_\psi, \,\forall \psi \in  \left\lbrace Q_i, u_i, d_i, L_i, e_i, \nu_i \right\rbrace$, which constitutes another solution. 
	As rescaling all charges is equivalent to a rescaling of the $U(1)^\prime$ gauge coupling, these solutions are in the same equivalence class and, hence, are not independent from each other.
	We therefore assume integer solutions $F_\psi \in \mathbb{Z}$ without loss of generality.
	
	\item \textit{Permutation invariance of fermions}: The ACCs are  invariant under the permutation of generation indices within each specific species 
	$\psi$.
\end{itemize}

To extract concrete solutions of the non-linear Eqs.~\eqref{eqn:su3squ1f}-\eqref{eqn:cubic}, we use computational algebraic geometry and perform a Gr\"obner basis computation~\cite{Rathsman:2019wyk} with the aid of \textit{Mathematica} and obtain analytical expressions for the charges $F_\psi$.
We impose phenomenological constraints, for instance, $F_{d_i}=F_{\nu_i}=0$, to further reduce the number of free parameters.
We then search for integer charges for all fermions by employing rescaling invariance, and focus on solutions with the smallest absolute value of the sum of all charges. 
To obtain large $c \to u$ FCNCs we search for solutions with $\Delta F_{L,R} \neq 0$, see Eq.~\eqref{eq:DeltaF}, that are consistent with Eq.~\eqref{eq:Xratio}.

\addtolength{\tabcolsep}{2pt}  
\begin{table}[!t]
	\centering
	\caption{Sample solutions of the ACCs~\eqref{eqn:su3squ1f}-\eqref{eqn:cubic} in a $U(1)^\prime$ extension of the SM$+3\nu_R$. 
	Only solutions 4, 7 require RH neutrinos. The ordering of generations is arbitrary due to permutation invariance, see text.}
	\label{tab:ZprimeModels}
	\begin{tabular}{c||c|c|c|c|c|c}
		solution $\#$ &$F_{Q_i}$& $F_{u_i}$ & $F_{d_i}$ & $F_{L_i}$ & $F_{e_i}$ & $F_{\nu_i}$ \\ 
		\hline
		1&$\begin{matrix} -4 & -2 & \phantom{-}6 \\ \end{matrix}$ & $\begin{matrix} -2 & \phantom{-}1 & \phantom{-}1 \\ \end{matrix}$ & $\begin{matrix} \phantom{-}0 & \phantom{-}0 & \phantom{-}0 \\ \end{matrix}$ & $\begin{matrix} -8 & \phantom{-}3 & \phantom{-}5 \\ \end{matrix}$ & $\begin{matrix} -3 & -3 & \phantom{-}6 \\ \end{matrix}$ & $\begin{matrix} \phantom{-}0 & \phantom{-}0 & \phantom{-}0 \\ \end{matrix}$  \\
		2&$\begin{matrix} -6 & \phantom{-}3 & \phantom{-}3 \\ \end{matrix}$ & $\begin{matrix} -8 & \phantom{-}4 & \phantom{-}4 \\ \end{matrix}$ & $\begin{matrix} -10 & \phantom{-}0 & \phantom{-}10 \\ \end{matrix}$ & $\begin{matrix} -6 & \phantom{-}1 & \phantom{-}5 \\ \end{matrix}$ & $\begin{matrix} \phantom{-}0 & \phantom{-}0 & \phantom{-}0 \\ \end{matrix}$ & $\begin{matrix} \phantom{-}0 & \phantom{-}0 & \phantom{-}0 \\ \end{matrix}$ \\
		3&$\begin{matrix} -20 & \phantom{-}7 & \phantom{-}8 \\ \end{matrix}$ & $\begin{matrix} -29 & \phantom{-}3 & \phantom{-}6 \\ \end{matrix}$ & $\begin{matrix} -19 & \phantom{-}4 &  \phantom{-}25\\ \end{matrix}$ & $\begin{matrix} \phantom{-}0 & \phantom{-}6 & \phantom{-}9 \\ \end{matrix}$ & $\begin{matrix} \phantom{-}3 & \phantom{-}13 & \phantom{-}14  \\ \end{matrix}$ & $\begin{matrix} \phantom{-}0 & \phantom{-}0 & \phantom{-}0  \\ \end{matrix}$ \\	
		4&$\begin{matrix} -1 & -1 & \phantom{-}2  \\ \end{matrix}$ & $\begin{matrix}-1 & -1 & \phantom{-}2  \\ \end{matrix}$ & $\begin{matrix} \phantom{-}0 & \phantom{-}0 & \phantom{-}0 \\ \end{matrix}$ & $\begin{matrix}  -1 & \phantom{-}0 & \phantom{-}1 \\ \end{matrix}$ & $\begin{matrix} -2 &  \phantom{-}0 & \phantom{-}2 \\ \end{matrix}$ & $\begin{matrix} -2 & -1 & \phantom{-}3 \\ \end{matrix}$  \\
		5&$\begin{matrix} -1 & -1 & \phantom{-}2  \\ \end{matrix}$ & $\begin{matrix}-1 & -1 & \phantom{-}2  \\ \end{matrix}$ & $\begin{matrix}-1 & -1 & \phantom{-}2  \\ \end{matrix}$ & $\begin{matrix}  -1 & \phantom{-}0 & \phantom{-}1 \\ \end{matrix}$ & $\begin{matrix} -1 &  \phantom{-}0 & \phantom{-}1 \\ \end{matrix}$ &  $\begin{matrix} \phantom{-}0 & \phantom{-}0 & \phantom{-}0 \\ \end{matrix}$  \\
		6&$\begin{matrix} -10 & \phantom{-}2 & \phantom{-}6  \\ \end{matrix}$ & $\begin{matrix}-13 & \phantom{-}2 & \phantom{-}3  \\ \end{matrix}$ & $\begin{matrix} -11 & \phantom{-}2 & \phantom{-}13 \\ \end{matrix}$ & $\begin{matrix}  -6 & \phantom{-}3 & \phantom{-}9 \\ \end{matrix}$ & $\begin{matrix} \phantom{-}2 &  \phantom{-}4 & \phantom{-}6 \\ \end{matrix}$ & $\begin{matrix} \phantom{-}0 & \phantom{-}0 & \phantom{-}0 \\ \end{matrix}$  \\	
		7&$\begin{matrix} \phantom{-}1 & \phantom{-}1 & \phantom{-}1 \\ \end{matrix}$ & $\begin{matrix} \phantom{-}1 & \phantom{-}1 & \phantom{-}1 \\ \end{matrix}$ & $\begin{matrix} \phantom{-}1 & \phantom{-}1 & \phantom{-}1 \\ \end{matrix}$ & $\begin{matrix} -3 & -3 & -3 \\ \end{matrix}$ & $\begin{matrix} -3 & -3 & -3 \\ \end{matrix}$ & $\begin{matrix} -3 & -3 & -3 \\ \end{matrix}$  \\	
		8&$\begin{matrix} -15 & \phantom{-}6 & \phantom{-}7 \\ \end{matrix}$ & $\begin{matrix} -14 & \phantom{-}2 & \phantom{-}4 \\ \end{matrix}$ & $\begin{matrix} -25 & \phantom{-}9 & \phantom{-}20 \\ \end{matrix}$ & $\begin{matrix} -24 & \phantom{-}11 & \phantom{-}19 \\ \end{matrix}$ & $\begin{matrix} \phantom{-}1 & \phantom{-}3 & \phantom{-}8 \\ \end{matrix}$ & $\begin{matrix} \phantom{-}0 & \phantom{-}0 & \phantom{-}0 \end{matrix}$ \\	
	\end{tabular}
\end{table}
\addtolength{\tabcolsep}{-2pt}   

In Tab.~\ref{tab:ZprimeModels} we present examples of anomaly-free $Z^\prime$--models. 
Due to permutation invariance, the ordering of generations within each fermion species is not fixed. 
Solution 1 has $F_{d_i}=F_{\nu_i}=0$, and solution 2 has $F_{e_i}=F_{\nu_i}=0$.
Solutions 4 and 7, which are also discussed in \cite{Allanach:2018vjg}, are the only ones with non-zero RH neutrino charges. 
Solution 7 has generation-independent couplings, and therefore no  $Z^\prime$-induced FCNCs at tree level.
Solution 4 has $F_{d_i}=0$ and equal charges for two generations of $Q$ and $U$-quarks. To induce both $\Delta F_{L,R} \neq 0$, 
the generations with equal charges could be first and third, or second and third generation. (This assignment does not have to be the same for $Q$ and $U$.)
The "two generations have equal charge"-condition for all quarks $Q,U,D$ is imposed in solution 5.
Solutions 3 and 8 are obtained by searching for large (small) values of $\Delta F_{R}$ ($\Delta F_{L}$) to satisfy Eq.~\eqref{eq:Xratio}.
As shown in Sec.~\ref{sec:Z'}, these solutions can be either RH- or LH-dominated depending on flavor mixing.

Sizable renormalization group coefficients can be induced in models with large $U(1)^\prime$--charges. The study of these effects is beyond the scope of this work.

\end{document}